\pdfoutput=1
\documentclass[iop,revtex4]{emulateapj}
\slugcomment{Accepted for publication 03 Feb 2014 to {\it The Astrophysical Journal}}
\shortauthors{Kirkpatrick et al.}
\shorttitle{AllWISE Motion Survey}

\usepackage{color,amsmath,lscape,longtable}
\begin{document}

\title{The AllWISE Motion Survey and The Quest for Cold Subdwarfs}

\author{J.\ Davy Kirkpatrick\altaffilmark{1},
Adam Schneider\altaffilmark{2},
Sergio Fajardo-Acosta\altaffilmark{1},
Christopher R.\ Gelino\altaffilmark{1,3},
Gregory N.\ Mace\altaffilmark{4}, 
Edward L.\ Wright\altaffilmark{4}, 
Sarah E.\ Logsdon\altaffilmark{4}, 
Ian S.\ McLean\altaffilmark{4},
Michael C.\ Cushing\altaffilmark{2},
Michael F.\ Skrutskie\altaffilmark{5},
Peter R.\ Eisenhardt\altaffilmark{6},
Daniel Stern\altaffilmark{6},
Mislav Balokovi\'c\altaffilmark{7},
Adam J.\ Burgasser\altaffilmark{8},
Jacqueline K.\ Faherty\altaffilmark{9},
George B.\ Lansbury\altaffilmark{10},
J.\ A.\ Rich\altaffilmark{11},
Nathalie Skrzypek\altaffilmark{12},
John W.\ Fowler\altaffilmark{1}, 
Roc M.\ Cutri\altaffilmark{1}, 
Frank J.\ Masci\altaffilmark{1}, 
Tim Conrow\altaffilmark{1}, 
Carl J.\ Grillmair\altaffilmark{1}, 
Howard L.\ McCallon\altaffilmark{1}, 
Charles A.\ Beichman\altaffilmark{1,3}, and
Kenneth A.\ Marsh\altaffilmark{13}
}

\altaffiltext{1}{Infrared Processing and Analysis Center, MS 100-22, California Institute of Technology, Pasadena, CA 91125, USA; davy@ipac.caltech.edu}
\altaffiltext{2}{Department of Physics and Astronomy, MS 111, University of Toledo, 2801 W. Bancroft St., Toledo, OH 43606-3328, USA} 
\altaffiltext{3}{NASA Exoplanet Science Institute, MS 100-22, California Institute of Technology, Pasadena, CA 91125, USA}
\altaffiltext{4}{Department of Physics and Astronomy, UCLA, 430 Portola Plaza, Box 951547, Los Angeles, CA, 90095-1547, USA}
\altaffiltext{5}{Department of Astronomy, University of Virginia, Charlottesville, VA, 22904, USA}
\altaffiltext{6}{NASA Jet Propulsion Laboratory, 4800 Oak Grove Drive, Pasadena, CA 91109, USA}
\altaffiltext{7}{California Institute of Technology, MC 249-17, Pasadena, CA 91125, USA}
\altaffiltext{8}{Department of Physics, University of California, San Diego, CA 92093, USA}
\altaffiltext{9}{Department of Terrestrial Magnetism, Carnegie Institution of Washington, Washington, DC 20015, USA}
\altaffiltext{10}{Department of Physics, Durham University, Durham DH1 3LE, UK}
\altaffiltext{11}{Observatories of the Carnegie Institution of Washington, 813 Santa Barbara Street, Pasadena, CA 91101, USA}
\altaffiltext{12}{Astro Group, Imperial College London, Blackett Laboratory, Prince Consort Road, London SW7 2AZ, UK}
\altaffiltext{13}{School of Physics and Astronomy, Cardiff University, Cardiff CF24 3AA, UK}

\begin{abstract}

The AllWISE processing pipeline has measured motions for all objects detected on WISE images taken between 2010 
January and 2011 February. In this paper, we discuss new capabilities made to the software pipeline in order to make 
motion measurements possible, and we characterize the resulting data products for use by future researchers. Using a 
stringent set of selection criteria, we find 22,445 objects that have significant AllWISE motions, of which 3,525 
have motions that can be independently confirmed from earlier 2MASS images yet lack 
any published motions in SIMBAD. 
Another 58 sources lack 2MASS counterparts and are presented as motion candidates only.
Limited spectroscopic follow-up of this list has already revealed eight new 
L subdwarfs. These may provide the first hints of a ``subdwarf gap'' at mid-L types that would indicate the break 
between the stellar and substellar populations at low metallicities (i.e., old ages). 
Another object in the motion list 
-- WISEA J154045.67$-$510139.3 -- is a bright ($J \approx 9$ mag) object of type M6; both the spectrophotometric distance 
and a crude preliminary parallax place it $\sim$6 pc from the Sun.
We also compare our list of motion objects to the recently published list 
of 762 WISE motion objects from \cite{luhman2014}. 
While these first large motion studies with WISE data have been very successful in revealing 
previously overlooked nearby dwarfs, both studies missed objects that the other found, demonstrating that many other 
nearby objects likely await discovery in the AllWISE data products.

\end{abstract}


\section{Introduction}

The Wide-field Infrared Survey Explorer (WISE; \citealt{wright2010}) was built to survey the sky simultaneously in 
four bands with central wavelengths of 3.4, 4.6, 12, and 22 $\mu$m (hereafter referred to as W1, W2, W3, and W4, 
respectively). The four-band cryogenic mission covered the sky 1.2 times between
the dates 2010 Jan 07 and 2010 Aug 06. After the solid hydrogen cryogen was 
depleted in the outer tank, data covering 30\% of the sky
continued to be acquired in the three-band (W1, W2, and W3) cryogenic mission until 
2010 Sep 29, when cryogen in the inner tank was exhausted. WISE surveyed 70\% of the sky in the two-band (W1 and W2) 
post-cryogenic NEOWISE mission (\citealt{mainzer2011}) until the satellite was placed into hibernation on 2011 Feb 01. 
In total, WISE surveyed the full sky twice in at least two bands and 20\% of the sky a third time, with each
coverage epoch separated by approximately six months.

Data were processed and released separately for the cryogenic and post-cryogenic missions. The aim of the AllWISE 
program was to combine all of the data from the WISE mission to leverage the full depth and time history of the
 W1 and W2 data. Specifically, one of these goals was to use the full WISE data to perform the first all-sky proper 
motion survey at these wavelengths. In addition to the exciting prospects created by doing a kinematic survey in a 
new wavelength regime, AllWISE processing -- because of the short, six-month time baseline between epochs 
-- enables us to search for very high motion objects that may have been overlooked by the longer time baselines of 
earlier all-sky motion surveys. In these surveys, an object could be ``lost'' between consecutive views. A case in 
point is the first motion discovery from AllWISE, the usdM3 star WISEA J070720.50+170532.7 (\citealt{wright2013}), 
which is moving at $1\farcs8$ yr$^{-1}$ and was missed by all previous motion surveys despite being in the more 
heavily studied northern hemisphere, being easily detected at optical wavelengths, and falling in an uncomplicated 
region free of source confusion.

In this paper, we describe how motions are measured in AllWISE (section 2), characterize the resulting 
motions and present caveats to users of the AllWISE data products (section 3), present a catalog of 
new motion discoveries (section 4), compare that to the recent list of WISE motion discoveries published by \cite{luhman2014}
(section 5), discuss our spectroscopic follow-up of some of the discoveries (section 6), and highlight a 
couple of specific science cases -- the hunt for rare, L-type subdwarfs (section 7) and 
the search for missing members of the immediate Solar Neighborhood (section 8) -- that we have already explored using 
the discovery list. In section 9 we provide a summary of our AllWISE results thus far.

\section{Measuring Motions in AllWISE}

AllWISE combines all previous WISE single-exposure images (called Level 1b frames) that satisfy a minimum quality 
standard. These individual frames are match-filtered and co-added on pre-defined $1\fdg56\times1\fdg56$
Atlas Tile footprints on the sky. Pixel 
outlier rejection is used to suppress transient features and objects of extremely high motion (e.g., asteroids and 
satellite streaks).

In processing of the earlier WISE data sets, the following procedure was used for source detection and 
astrometric/photometric measurement. A list of detections from a multiband signal-to-noise ratio map of the
coadded images was first generated for each Atlas Tile. At the position of each detection in the list, a point response
function (PSF) fit was then performed to measure the source position and flux via a $\chi^2$ minimization
procedure on the stack of individual frames that generated the coadd. Because the 
WISE beam size is large (the full-width at half-maximum, FWHM, is 12$\arcsec$ at W4), the deblending of overlapping 
detections -- and the further deconvolution of those detections into possible multiple components -- is handled 
through a $\chi^2$ minimization procedure explained in more detail in the WISE All-Sky Release Explanatory 
Supplement\footnote{See http://wise2.ipac.caltech.edu/docs/release/allsky/expsup/.}. 

During AllWISE processing, the source position measurement model was augmented with linear motion terms
in RA and Dec (the ``motion fit''). In AllWISE processing, the fit excluding these new motion terms (the 
``stationary fit'') is performed first, and the full 
set of photometric parameters is computed. The motion fit begins with the stationary position as the initial position 
estimate at a fiducial time that is the flux-weighted mean observational epoch of each source. Because of the 
effects of noise on the non-linear $\chi^2$ minimization algorithm, it is sensitive to the initial estimates used to 
start the minimization search. Of the several ways to weigh the epoch averaging, flux weighting based on the 
stationary-fit single-frame flux solutions was found to be the most effective in reproducing motions of known moving 
objects. Readers are encouraged to consult the AllWISE Release Explanatory 
Supplement\footnote{See http://wise2.ipac.caltech.edu/docs/release/allwise/expsup/.} for additional details.

As a result of this addition to the AllWISE processing, motions are now measured for all coadd-detected WISE sources, 
giving researchers a powerful new tool with which to study and/or identify nearby stars and brown dwarfs. This paper 
characterizes these new motion measurements, discusses the caveats, and showcases two research areas -- the search 
for low-metallicity stars and brown dwarfs and the hunt for missing members of the Solar Neighborhood -- that are 
particularly well suited for this data set.

\section{Analysis of the AllWISE Motion Products}

\subsection{Apparent Motion vs. Proper Motion}

The AllWISE time baseline is typically between six months and a year for any point on the sky. Except for ecliptic 
polar regions, the time sampling will be confined to discrete epochs separated by roughly six months. Multiple 
measurements, spanning the range of a few days, are taken at each epoch. Most of the sky has two 
epochs covering six months, and $\sim$20\% of the sky has three epochs spanning a time range of one year.

It is important to understand what this time sampling means for motion measurements. In general, it is the closest 
objects to the Sun that exhibit the largest proper motions. Being close, these objects will also exhibit substantial 
parallax. Because data are taken near $90\degr$ solar elongation, WISE observed these objects at their maximum 
parallax factors. As an example, an object seen at two WISE epochs will have motion that is a combination of 
half its yearly proper motion (because the two epochs are only separated by six months) and twice its parallax 
(because the vantage point between epochs has shifted by 2 AU). Having only two epochs, AllWISE cannot disentangle 
the effects of parallax and proper motion and thus measures only {\it apparent motion} on the sky as opposed to 
{\it proper motion}.

\begin{figure}
\figurenum{1}
\includegraphics[scale=0.45,angle=0]{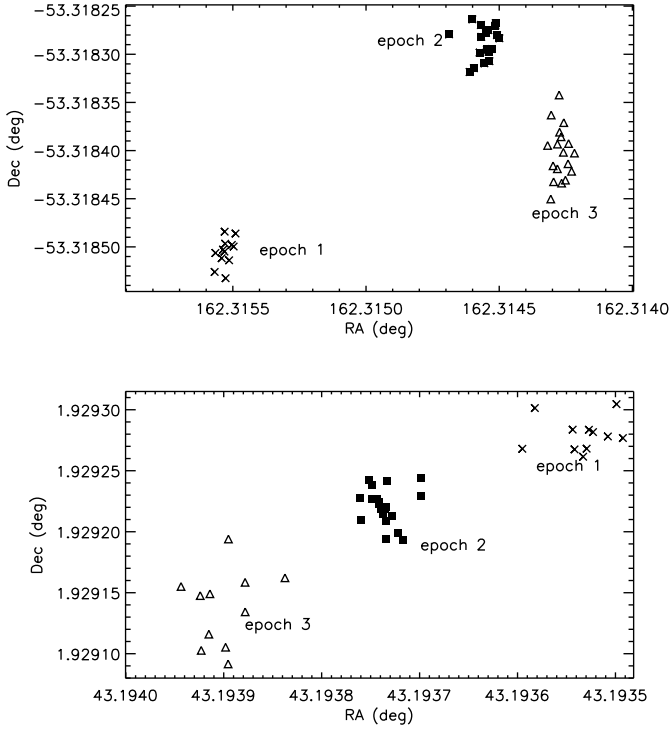}
\caption{Astrometry measured from the individual Level 1b frames for the nearby L+T dwarf binary WISE 
J104915.57-531906.1 (top) and LHS 161 (bottom). The three clusters of points labeled ``epoch 1'', ``epoch 2'', 
and ``epoch 3'' show the WISE data separated by six-month intervals.
\label{individual_epoch_positions}}
\end{figure}

Figure~\ref{individual_epoch_positions} (top panel) shows the individual Level 1b frame measurements of the position 
of the nearby L+T dwarf binary WISE J104915.57-531906.1. This source has three epochs of coverage in AllWISE. The 
motion is not linear because this binary's parallax ($\pi = 0{\farcs}50$) and proper motion ($\mu = 2{\farcs}78$ 
yr$^{-1}$) are comparable in size (\citealt{luhman2013}). It would be possible to measure the proper motion by using 
only the epoch-1 and epoch-3 points, since these clusters of points were taken at the same parallax factor; however, 
AllWISE uses data from all three epochs and thus will measure a motion that is not the actual ``proper'' motion. The 
extreme M subdwarf LHS 161 in Figure~\ref{individual_epoch_positions} (bottom panel) demonstrates a case at the 
opposite extreme, where the parallax is swamped by a much larger proper motion ($\pi$ = 0$\farcs$025 and $\mu = 
1\farcs455$ yr$^{-1}$; \citealt{vanaltena1995}). As expected for this case, the three separate epochs of data more 
closely follow a linear progression in time.

\begin{figure}
\figurenum{2}
\includegraphics[scale=0.5,angle=0]{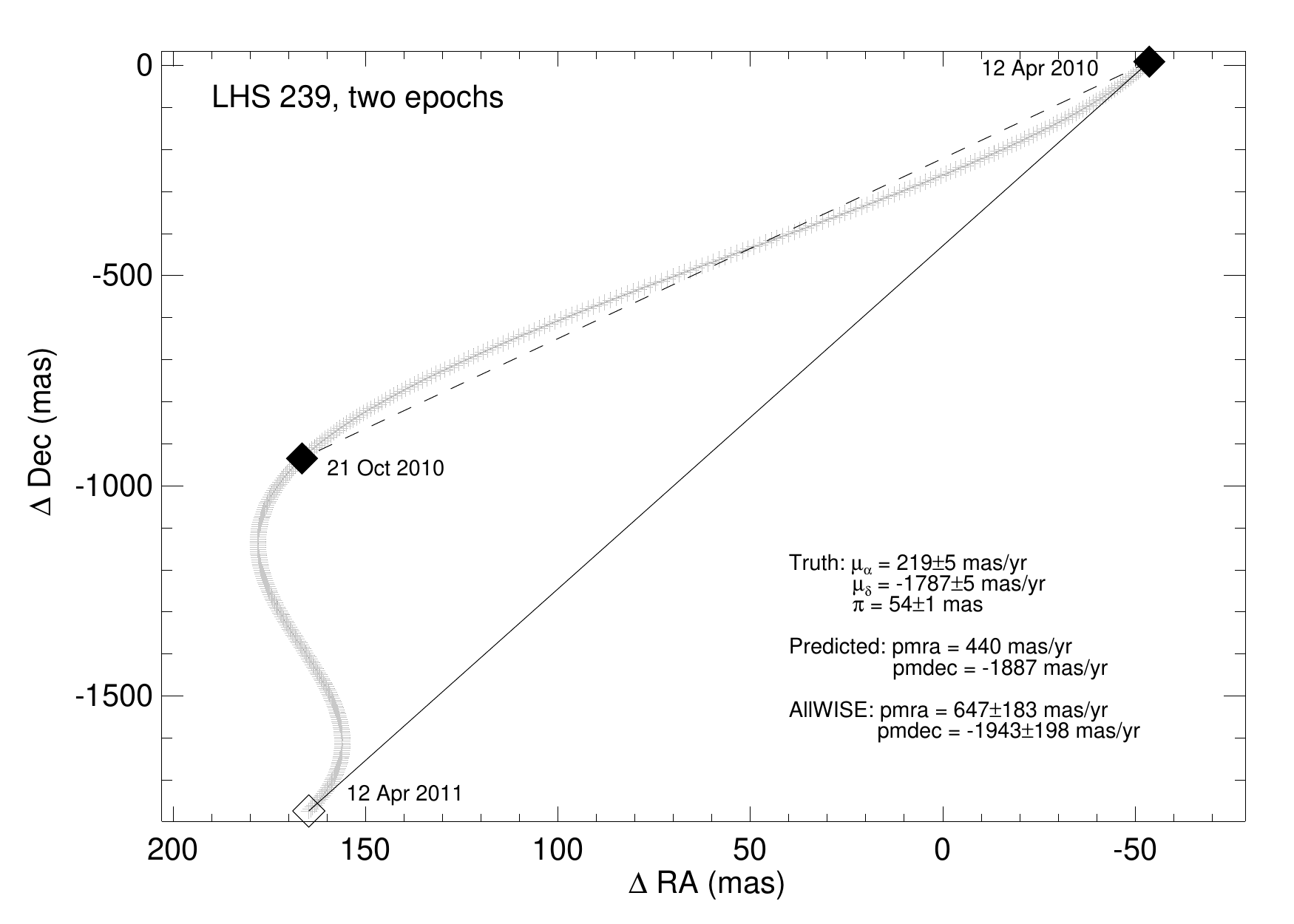}
\caption{The predicted motion track (grey plus signs, one symbol per day) for the DC12 white dwarf LHS 239. Solid 
diamonds show the dates of the WISE observations. The open diamond shows the date twelve months after the initial 
WISE observation. AllWISE measures a motion over six months as noted by the dashed line (that is, the reported yearly
 motion will be twice this long), whereas the true proper motion over a full year is denoted by the solid line.
\label{lhs239_predicted_track}}
\end{figure}

\begin{figure}
\figurenum{3}
\includegraphics[scale=0.5,angle=0]{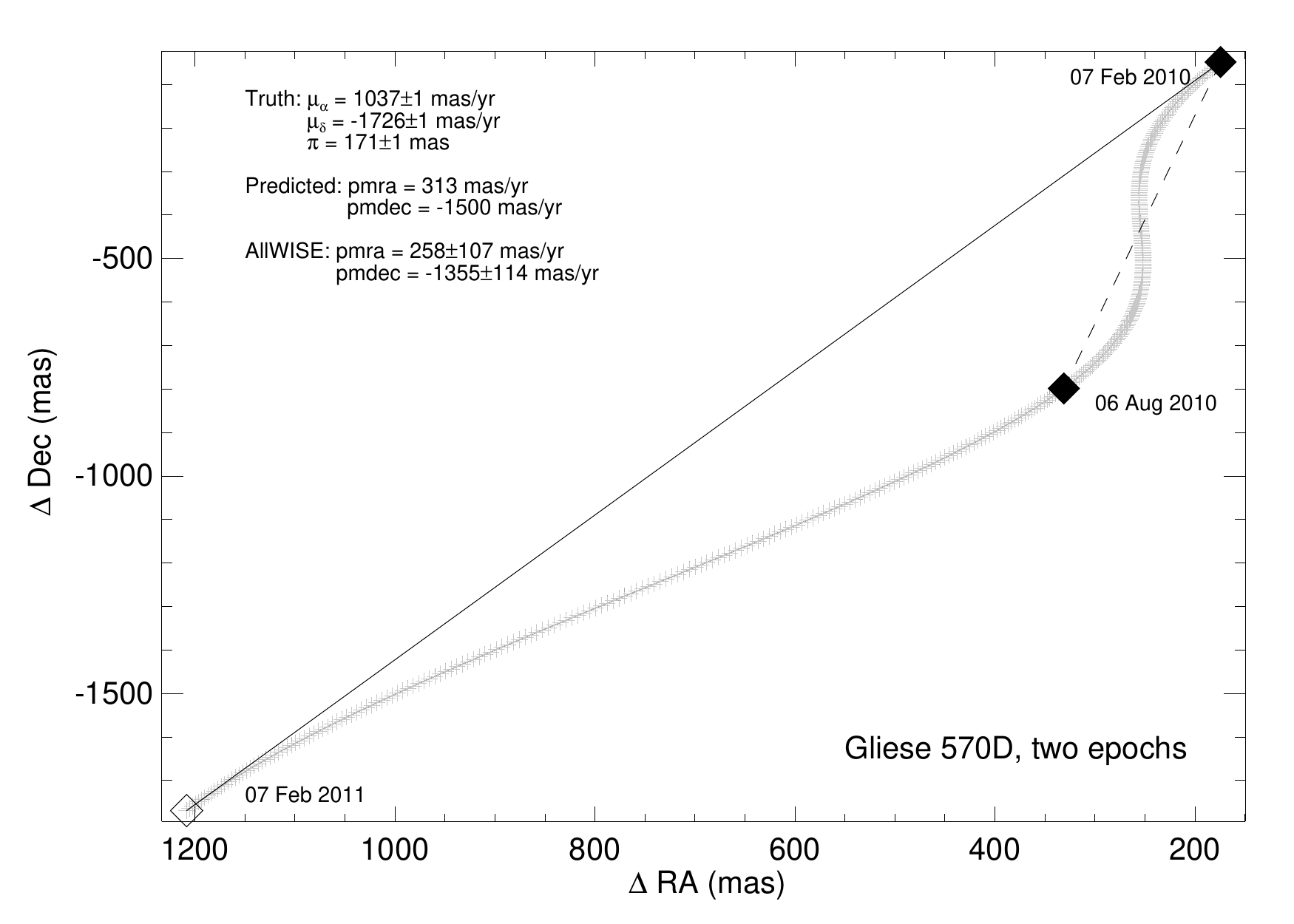}
\caption{The predicted motion track (grey plus signs, one symbol per day) for the T7.5 brown dwarf Gliese 570D. 
Solid diamonds show the dates of the WISE observations. The open diamond shows the date twelve months after the 
initial WISE observation. AllWISE measures a motion over six months as noted by the dashed line, whereas the true 
proper motion over a full year is denoted by the solid line.
\label{gl570d_predicted_track}}
\end{figure}

This can be further demonstrated with a couple of examples drawn from areas with two-epoch coverage, which represents 
the majority of WISE data. Figure~\ref{lhs239_predicted_track} shows the predicted path of the 
white dwarf LHS 239 on the sky, and Figure~\ref{gl570d_predicted_track} shows the path for the T dwarf Gliese 570 D. In each of 
these plots, the grey track shows the predicted path of the object from the date of the first WISE epoch to a date 
one year later. The date of the second (i.e., final) WISE epoch, approximately six months after the first, is also 
noted. The motion that AllWISE measures is the positional difference between the start of the track and the point 
six months later, whereas the true proper motion would be that measured between the start of the track and the point 
twelve months later. The true proper motion and parallax, the predicted motion from AllWISE, and the actual AllWISE 
values are given in the legend of both figures. Note that the actual AllWISE measurements are, within their errors, 
identical to the predicted values, although not necessarily close to the true proper motion measures. As stated 
earlier, these AllWISE measurements should be regarded as {\it apparent motion} values (parallax + proper motion) 
and not as pure {\it proper motion} values. 

\begin{figure}
\figurenum{4}
\includegraphics[scale=0.5,angle=0]{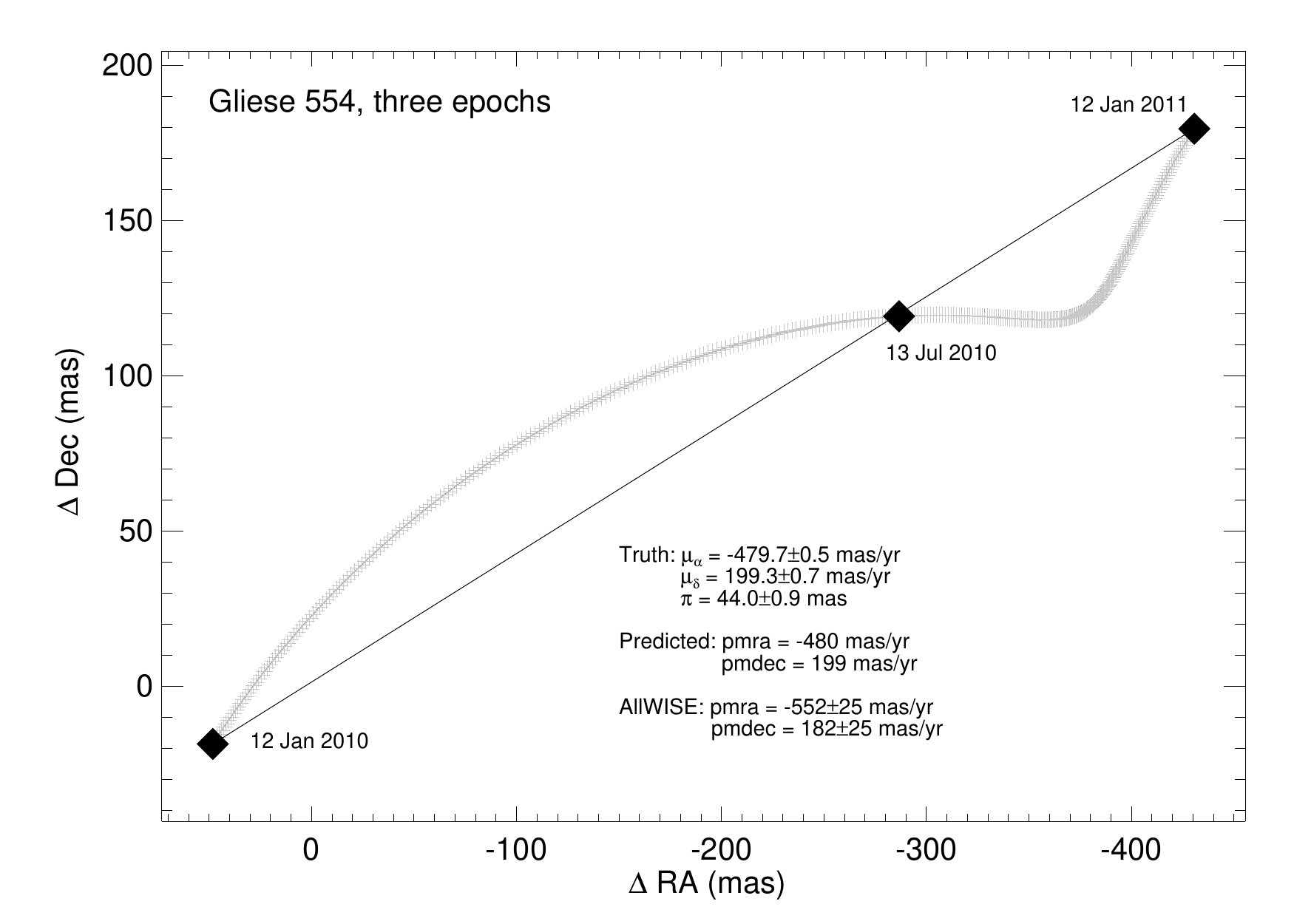}
\caption{The predicted motion track (grey plus signs, one symbol per day) for the K3 dwarf Gliese 554. Dates of 
the WISE data points (denoted by solid diamonds) cover a full year. AllWISE measures a motion that is very close 
to the true proper motion in this case.
\label{gl554_predicted_track}}
\end{figure}

Figure~\ref{gl554_predicted_track} illustrates the predicted sky path for the K dwarf Gliese 554, located in an area 
with three epochs of coverage. In this case, the WISE data fall almost perfectly along the proper motion vector. It 
would be expected that the AllWISE motion measure would be very close to the true proper motion value, which is 
indeed the case. It should be noted that a measurement using only epoch 1+2 data would have a larger motion than a 
measurement using only epoch 2+3 data, although both would have the correct position angle.

Because the AllWISE-measured motions do not decouple proper motion and parallax, these motions cannot be used to 
predict source positions at other epochs. Rather, their utility is in identifying potentially interesting, nearby 
objects for further follow-up.

\subsection{Motion Limits}

\subsubsection{What is the Largest Motion Measurable?}

AllWISE is able to identify objects with very high motion. Table~\ref{highest_motion_stars} lists the eleven 
highest proper motion systems known ($\mu > 4{\farcs}6$ yr$^{-1}$) and compares their published proper motions to 
the motions derived by AllWISE. Most of these sources are extremely bright and heavily saturated on the WISE detectors. 
Agreement between published values and those in AllWISE is not expected due both to saturation effects and parallax. 
Nonetheless, the comparison shows that the highest motion stars can be identified as such in AllWISE data.

\begin{turnpage}
\begin{deluxetable*}{lccccccccc}
\tabletypesize{\tiny}
%\rotate
\tablewidth{8.0in}
\tablenum{1}
\tablecaption{Motion Comparisons for the Eleven Highest Proper Motion Systems Known\label{highest_motion_stars}}
\tablehead{
\colhead{Object Name} &                          
\colhead{Sp.\ Type} &  
\colhead{AllWISE} &
\colhead{AllWISE} &     
\colhead{AllWISE} &
\colhead{AllWISE} &
\colhead{Published $\mu_\alpha$} &
\colhead{Published $\mu_\delta$} &
\colhead{Published $\pi$} &
\colhead{Ref.} \\
\colhead{} &                          
\colhead{} &  
\colhead{Designation} &
\colhead{W1} &     
\colhead{RA Motion} &
\colhead{Dec Motion} &
\colhead{(mas yr$^{-1}$)} &
\colhead{(mas yr$^{-1}$)} &
\colhead{(mas)} &
\colhead{} \\
\colhead{} &                          
\colhead{} &  
\colhead{} &
\colhead{(mag)} &     
\colhead{(mas yr$^{-1}$)} &
\colhead{(mas yr$^{-1}$)} &
\colhead{} &
\colhead{} &
\colhead{} &
\colhead{} \\
\colhead{(1)} &                          
\colhead{(2)} &  
\colhead{(3)} &     
\colhead{(4)} &
\colhead{(5)} &
\colhead{(6)} &
\colhead{(7)} &
\colhead{(8)} &
\colhead{(9)} &
\colhead{(10)}  
}
\startdata
Barnard's Star  &  M5 V  & WISEA J175747.94+044323.8  &  5.02&  -1630$\pm$71&   13703$\pm$65&    -798.58$\pm$1.72&   10328.12$\pm$1.22&  548.31$\pm$1.51&  1,9\\
Kapteyn's Star  &  sdM1p & WISEA J051146.81$-$450204.5&  4.92&   8166$\pm$29&   -8065$\pm$32&    6505.08$\pm$0.98&   -5730.84$\pm$0.96&  255.66$\pm$0.91&  3,9\\
Groombridge 1830&  G7 V+ & WISEA J115302.28+374206.6  &  4.33&   2224$\pm$59&  -14307$\pm$66&    4003.98$\pm$0.37&   -5813.62$\pm$0.23&  109.99$\pm$0.41&  4,9\\
Lacaille 9352   &  M2 V  & WISEA J230557.86$-$355057.2&  2.59&   1866$\pm$56&    -841$\pm$60&    6768.20$\pm$0.59&    1327.52$\pm$0.68&  305.26$\pm$0.70&  5,9\\
CD$-$37 15492   &  M3 V  & WISEA J000529.35$-$372151.0&  4.40&   -471$\pm$77&   -2662$\pm$86&    5634.68$\pm$0.86&   -2337.71$\pm$0.71&  230.42$\pm$0.90&  5,9\\
61 Cygni A      &  K5 V  & WISEAR J210657.70+384530.8\tablenotemark{a} 
                                                      &  0.18&  -4028$\pm$52&    5420$\pm$54&    4168.31$\pm$6.57&    3269.20$\pm$12.08& 286.82$\pm$6.78&  1,9\\
61 Cygni B      &  K7 V  & WISEA J210658.89+384504.5  &  0.97&  -1171$\pm$58&     215$\pm$66&    4106.90$\pm$0.32&    3144.68$\pm$0.44&  285.88$\pm$0.54&  1,9\\
Ross 619        &  M4 V  & WISEA J081158.31+084529.4  &  7.18&   1661$\pm$34&   -5277$\pm$32&    1081.4$\pm$1.6  &   -5087.3$\pm$1.6  &  145.5$\pm$3.9&    2,11\\
Teegarden's Star&  M6.5 V& WISEA J025303.34+165213.2  &  7.15&   3471$\pm$56&   -3666$\pm$25&    3403.8$\pm$2.2  &   -3807.0$\pm$2.2  &  260.63$\pm$2.69&  6,10\\
Lalande 21185   &  M2 V  & WISEA J110319.67+355722.4  &  2.47&    338$\pm$50&     786$\pm$48&    -580.27$\pm$0.62&   -4765.85$\pm$0.64&  392.64$\pm$0.67&  1,9\\
$\epsilon$ Ind A&  K2 V  & WISEA J220326.69$-$564735.5&  0.87&   1218$\pm$32&     -26$\pm$36&    3960.93$\pm$0.24&   -2539.23$\pm$0.17&  276.06$\pm$0.28&  8,9\\
$\epsilon$ Ind BaBb                                           
                &  T1+T6 & WISEA J220415.75$-$564724.4& 10.72&   3132$\pm$41&   -3124$\pm$38&    3960.93$\pm$0.24\tablenotemark{b}&   -2539.23$\pm$0.17\tablenotemark{b}&  276.06$\pm$0.28\tablenotemark{b}&  7,9\\
Wolf 359        &  M6 V  & WISEA J105626.19+070025.0  &  5.84&  -2618$\pm$44&   -4881$\pm$42&   -3838.6$\pm$0.2&     -2697.8$\pm$0.2&    418.9$\pm$2.4&    1,12\\
\enddata
\tablecomments{References for spectral type and high-precision astrometry: (1) \cite{kirkpatrick1991}, 
(2) \cite{henry1994}, (3) \cite{keenan1980}, (4) \cite{keenan1953}, (5) \cite{walker1983}, (6) \cite{teegarden2003}, 
(7) \cite{mccaughrean2004}, (8) \cite{vandekamp1953}, (9) \cite{vanleeuwen2007}, (10) \cite{henry2006}, 
(11) \cite{harrington1980}, (12) \cite{harrington1993}.}
\tablenotetext{a}{61 Cygni A is the only one of these sources failing to satisfy the criteria for inclusion in the 
AllWISE Source Catalog. This object is part of a small-separation same-tile (SSST) source group (see 
section~\ref{false_sources}). Unfortunately, despite best efforts to recover such sources, neither member of the 61 
Cyg A SSST group satisfies all the criteria necessary for Catalog consideration. The entry for 61 Cyg A can, however, 
be found in the AllWISE Reject Table, as the source designation indicates.}
\tablenotetext{b}{The published astrometry for $\epsilon$ Ind A is used here for $\epsilon$ Ind BaBb.}
\end{deluxetable*}
\end{turnpage}
%\clearpage
%\end{landscape}

Objects with motions as high as Barnard's Star (10$\farcs$4 yr$^{-1}$) can be successfully recovered with AllWISE, 
but what is the theoretical limit to the largest motion that AllWISE can measure for a real, astrophysical source? 
There are sources in the combination of the AllWISE Source Catalog and Reject Table that have single-axis motions as 
large as 670$\arcsec$ yr$^{-1}$ (670,000 mas yr$^{-1}$); spot checks of these sources show them to be artifacts 
associated with extended structures like diffraction spikes and optical ghosts. Ultimately, the algorithm of $\chi^2$ 
minimization is only limited by the detection step on the coadd images; that is, as long as a source is identified 
on the coadds, extremely large motions can be recovered in the Level 1b images. The only real sources that will not 
have a coadd detection are those moving so fast that outlier rejection will have eliminated them from the coadd 
entirely. Real examples of this include main belt asteroids and near-earth objects. Outlier rejection is more likely to 
eliminate objects seen less than 50\% of the time in the frame 
stack\footnote{See http://wise2.ipac.caltech.edu/docs/release/allsky/expsup/ sec4\_4f.html\#outrej of the WISE 
All-Sky Release Explanatory Supplement for more details.}. So it is just possible, in the two-epoch case, for an 
object near the ecliptic plane moving at $\sim$2.5$\times$FWHM between the first and last exposures over one epoch, 
or 15$\arcsec$ per $\sim$2 days ($\sim$2700$\arcsec$ yr$^{-1}$), to have survived coaddition and be detected. For 
areas of sky with three epochs of coverage, assuming equal numbers of frames per epoch, a source moving at 
$\sim$2700$\arcsec$ yr$^{-1}$ would be seen within a 2.5$\times$FWHM window only 33\% of the time, and would thus 
be removed from the coadd because it fails to meet the 50\% outlier threshold. For such a three-epoch case, the 
motion limit is more likely to be $\sim$2.5$\times$FWHM between two consecutive epochs, or $\sim$15$\arcsec$ per 
6 months ($\sim$30$\arcsec$ yr$^{-1}$).

\subsubsection{What is the Smallest Motion Measurable?\label{smallest_motion_section}}

\begin{figure}
\figurenum{5}
\includegraphics[scale=0.45,angle=0]{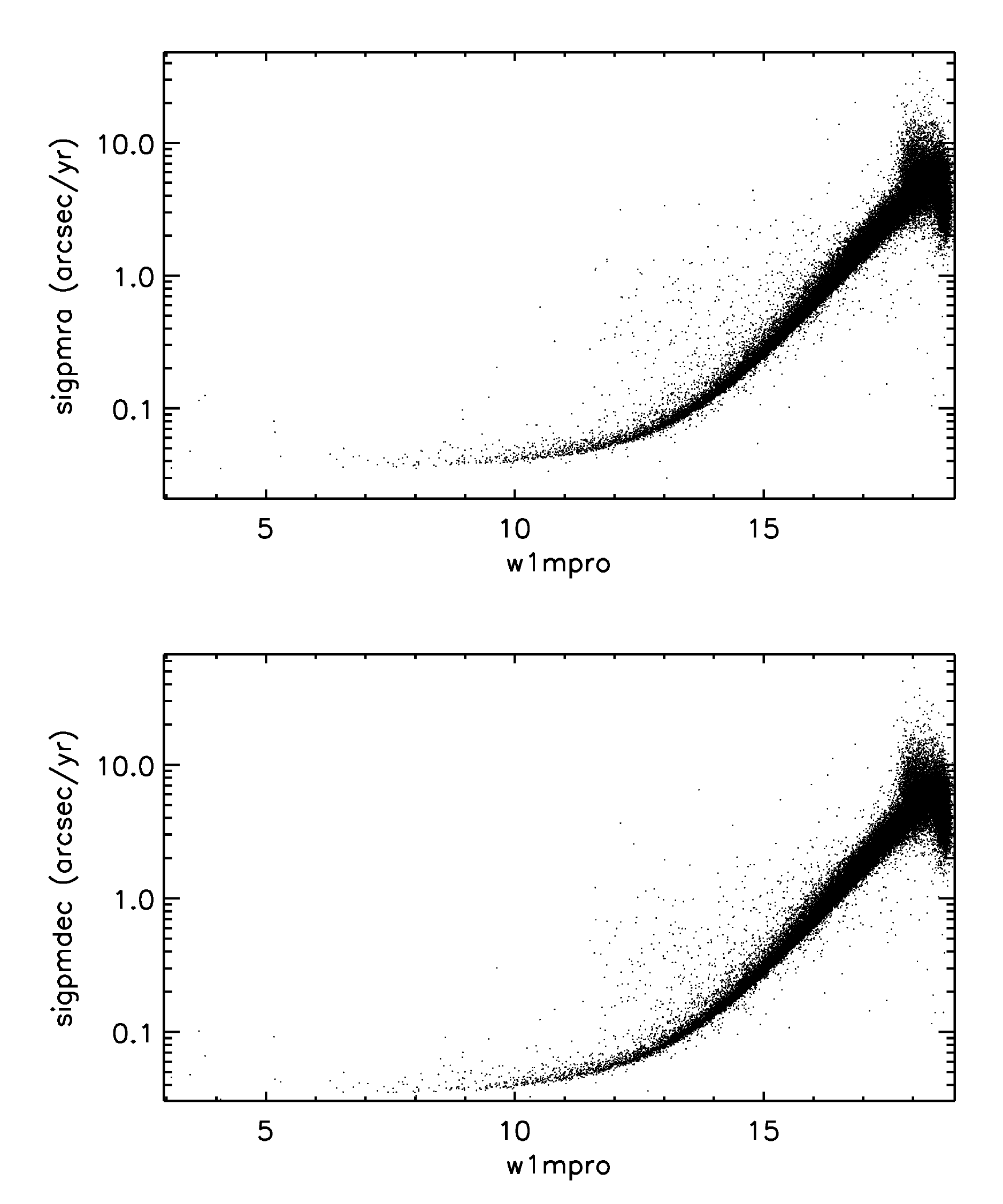}
\caption{Uncertainty in AllWISE motion measures in RA (top), and Dec (bottom) for all sources in coadd 
2243m213\_ac51, plotted as a function of the W1 profile-fit photometry.
\label{smallest_motions}}
\end{figure}

Figure~\ref{smallest_motions} shows the motion measurement uncertainties for 
sources in a typical AllWISE Atlas Image plotted against W1 magnitude. 
Bright sources free of saturation effects (8 $<$ W1 $<$ 10 mag) approach an asymptote of 
$\sim$0$\farcs$035 yr$^{-1}$ in their per-axis errors. Requiring a 3$\sigma$ detection per axis means that the 
smallest significant motion\footnote{Requiring the stricter ${\chi^2}_{motion} > 27.63$ criterion from 
section~\ref{motions_real_false} results in a smallest motion of $\sim$0$\farcs$18 yr$^{-1}$.} that is 
measurable is $\sim$0$\farcs$15 yr$^{-1}$. At fainter magnitudes, the smallest 
measurable significant motion increases dramatically as the minimum motion errors themselves increase. At W1 = 15 
mag, for example, the motion errors are roughly 10$\times$ larger, meaning that the smallest measurable significant 
motion is $\sim$1$\farcs$5 yr$^{-1}$.

The Atlas Image used in this example, 2243m213\_ac51, has an average depth of coverage of $\sim$24 framesets, as is typical of Atlas Tiles
located near the ecliptic plane and observed at two WISE epochs. Atlas Tiles farther from the ecliptic will have
greater coverage depths. For these cases, and for any Tiles with three WISE epochs, motion limits will be even
smaller than those quoted above.

\subsection{Comparisons to Truth}

Despite the fact that AllWISE measures an apparent motion over only a short time baseline, these measurements can 
still be compared to proper motion values determined for stars specifically targeted in astrometric monitoring 
campaigns as long as the sample size is statistically significant and scattered over the entire sky so that the 
influence of parallax is suppressed. In the first subsection below, the AllWISE motion measurements are compared 
to published proper motion values for an all-sky collection of late-type dwarfs. In the second subsection, the 
internal repeatability of the AllWISE measurements is checked using the individual components of widely separated 
common-proper-motion binary systems.

\subsubsection{External Comparisons Using Previously Known Motion Objects}

\begin{figure*}
\figurenum{6}
\includegraphics[scale=0.9,angle=90]{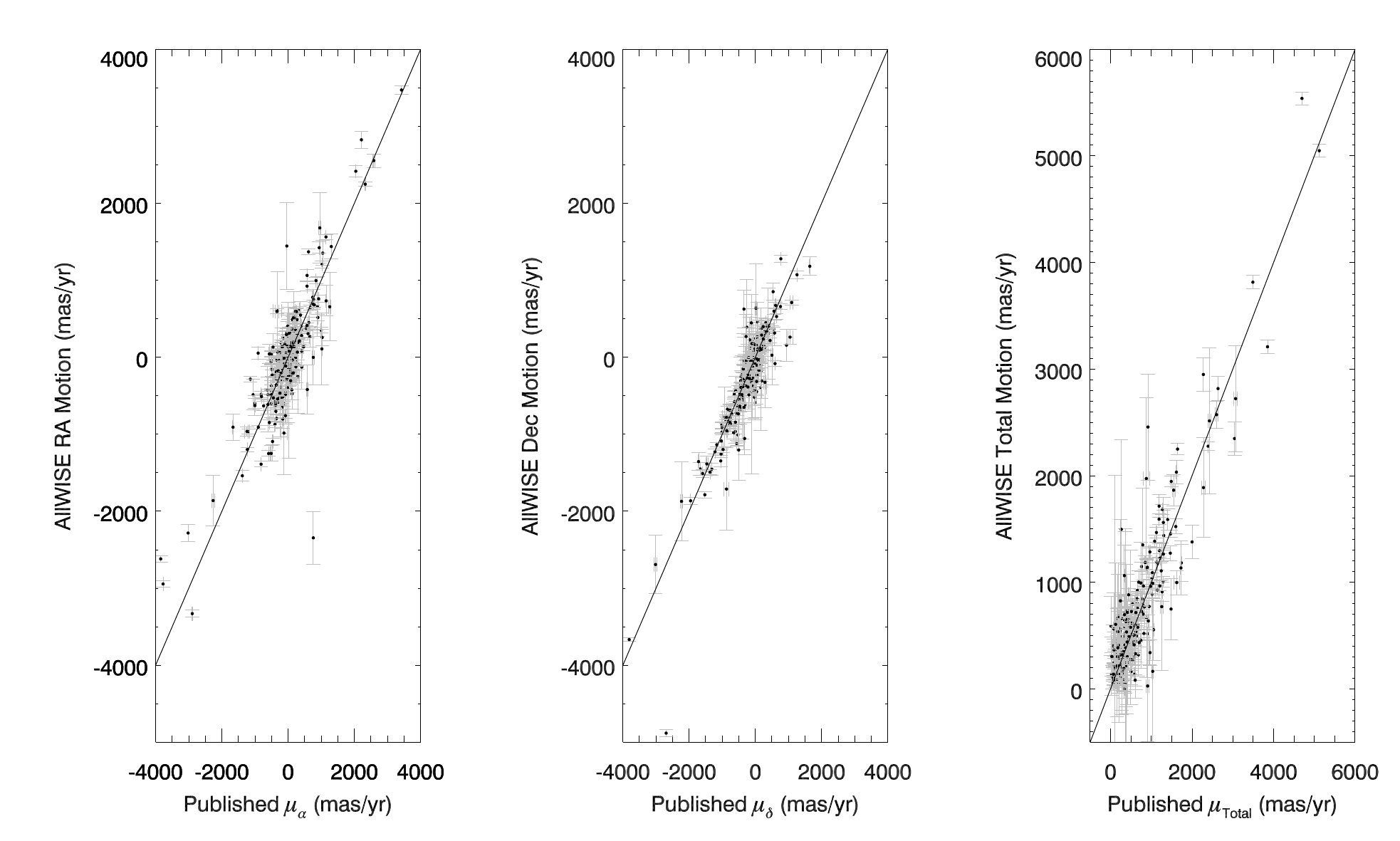}
\caption{Comparison of the AllWISE RA, Dec, and total motion measurments to published RA, Dec, and total proper 
motion measurements of late-M, L, T and Y dwarfs in the literature. The straight line corresponds to perfect 
one-to-one correspondence. The discrepant point near (800, -2350) in the left panel is Wolf 940B ($cc\_flags = 
`Hh00'$), which is contaminated by the halo of Wolf 940A, and the discrepant point near (-2700, -4900)
in the middle panel is the saturated star Wolf 359 (see Table~\ref{highest_motion_stars}). 
\label{truth_vs_allwise}}
\end{figure*}

A typical user of the AllWISE data products may wish to check how the AllWISE measurements compare to published 
values from a favorite motion catalog. As demonstrated in section 3.1, parallax contributes to the AllWISE 
motion measurements, making a direct comparison difficult. Are other statistical effects seen when a large 
sample is used? To this end,
the AllWISE measurements have been compared to published astrometry 
for a set of known low-mass stars and brown dwarfs.  These objects were selected because they have high 
signal-to-noise ratios in one or more WISE bands but are generally not saturated. Specifically, objects in the 
Database of Ultracool Parallaxes\footnote{See 
https://www.cfa.harvard.edu/$\sim$tdupuy/plx/Database\_of\_ Ultracool\_Parallaxes.html.}, compiled as of October 
2013, were used (\citealt{dupuy2012}). This list contains proper motions for all late-M, L, T, and 
Y dwarfs that have published parallaxes. After excluding close binaries and blends, 233 objects were available for 
comparison. Figure~\ref{truth_vs_allwise} shows a comparison of the published values of $\mu_\alpha$, $\mu_\delta$, 
and $\mu_{Total}$ to the RA, Dec, and total motions measured by AllWISE. The overall agreement is excellent, with 
the trend very closely following the line of one-to-one correspondence.

\begin{figure*}
\figurenum{7}
\includegraphics[scale=0.9,angle=90]{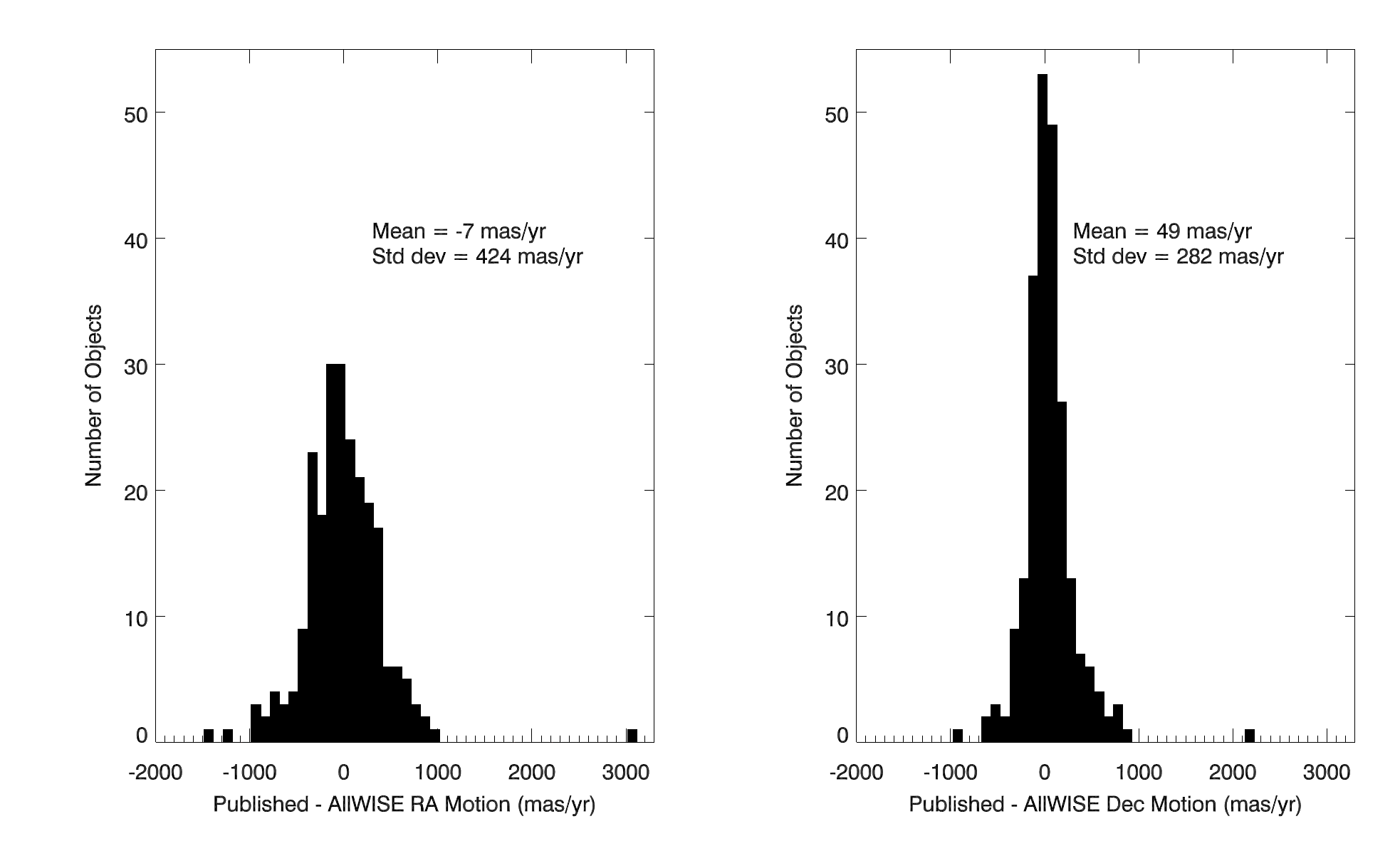}
\caption{Histograms of the difference between published and AllWISE motions in RA (left panel) and Dec (right panel) 
for late-M, L, T, and Y dwarfs.   
\label{truth_histograms}}
\end{figure*}

\begin{figure*}
\figurenum{8}
\includegraphics[scale=0.9,angle=90]{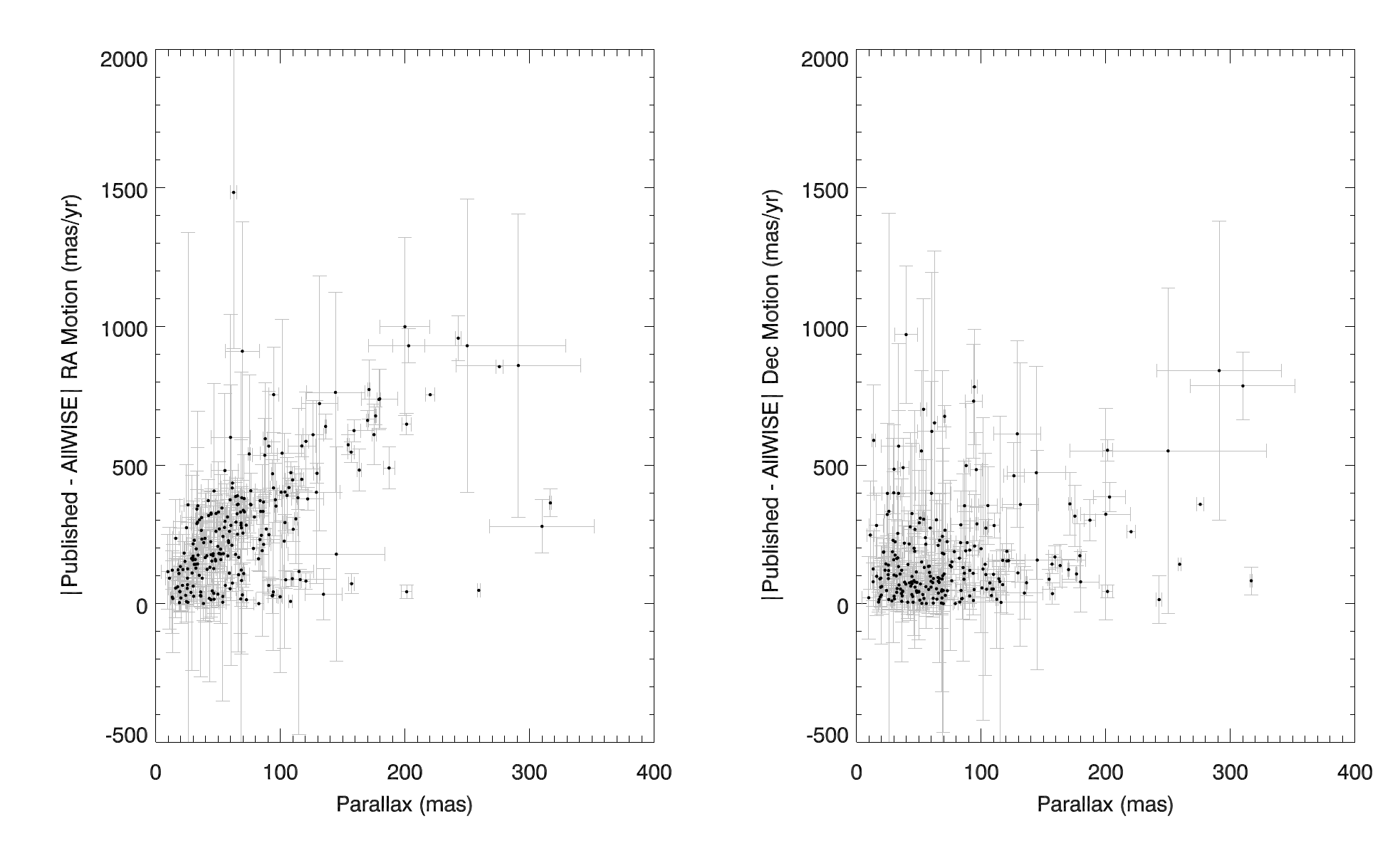}
\caption{The absolute value of the difference between published proper motion and AllWISE motion in RA (left) and 
Dec (right) as a function of parallax for late-M, L, T, and Y dwarfs.
\label{truth_vs_parallax}}
\end{figure*}

The influence of parallax on the AllWISE motion measurements can be seen in Figure~\ref{truth_histograms}.
In this figure, the difference between the published proper motions and 
the AllWISE motions shows a larger dispersion in RA than in Dec, and both dispersions are larger than would 
be expected based on the average AllWISE motion errors, which are 122.3 mas and 129.4 mas for RA and Dec, respectively. 
As shown in 
Figure~\ref{truth_vs_parallax}, which illustrates the motion differences in RA and Dec plotted as a function of 
parallax, the RA differences have a 
striking dependence on parallax (i.e., objects with the largest differences tend to have the largest 
parallax values) while the Dec differences show only a modest increase at the largest parallaxes.
Because the ecliptic and equatorial systems are tilted with respect
to each other by only $\sim$23 degrees, parallactic motion, which is primarily manifested in ecliptic longitude, 
affects equatorial longitude (RA) more severely than equatorial latitude (Dec). Other than this effect, no other
issues are seen, as the RA and Dec 
differences show a mean near zero and have no appreciable skew.

\begin{figure*}
\figurenum{9}
\includegraphics[scale=0.9,angle=90]{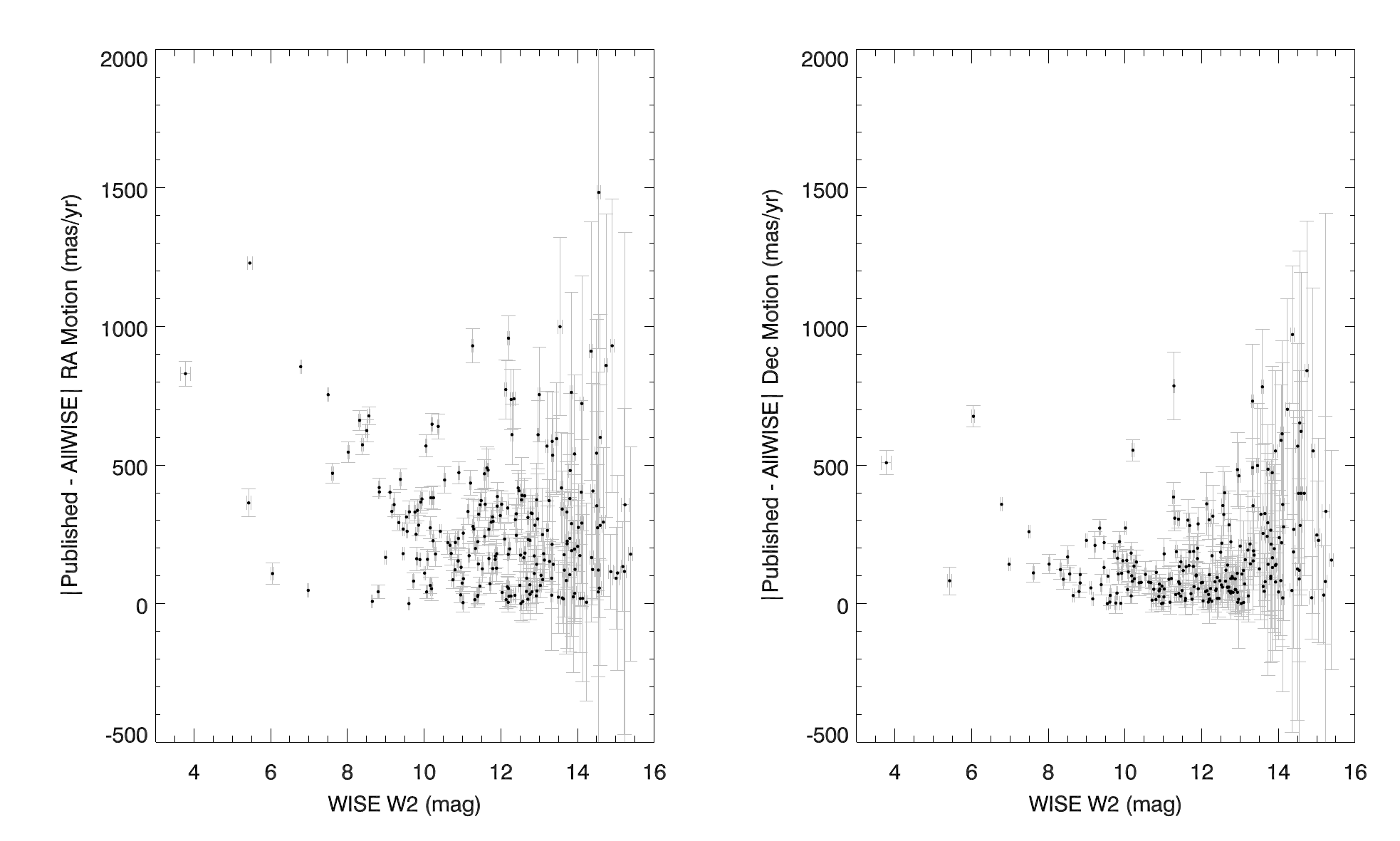}
\caption{The absolute value of the difference between published proper motion and AllWISE motion in RA (left) and 
Dec (right) as a function of WISE W2 profile-fit magnitude for late-M, L, T, and Y dwarfs.
\label{truth_vs_w2}}
\end{figure*}

Figure~\ref{truth_vs_w2} shows the motion difference plotted as a function of W2 mag, which for most L, T, and 
Y dwarfs is the WISE band with the highest signal-to-noise ratio and thus the band driving the astrometry. 
Agreement is best in the range 7 $<$ W2 $<$ 12 mag. Astrometric precision is degraded at brighter magnitudes 
(W2 $<$ 7 mag) because such objects have saturated image cores, and it quickly degrades at fainter magnitudes (W2 $>$ 
12 mag) due to poorer photon counting statistics.

\subsubsection{Internal Consistency Checks Using Common-proper-motion Binaries}

As part of AllWISE Quality Assurance checks, the software reported sources in each Atlas Tile that had a high likelihood 
of being real motion stars. (See section~\ref{motions_real_false} below.) During the course of quality assessment, 
reviewers noted fifty-five pairs of objects identified as having significant motions and located within 
$\sim15\arcsec$ - $400\arcsec$ of each other. These pairs were checked in SIMBAD and were confirmed to be known 
common-proper-motion systems, found to be new ones verified through independent means, or identified as possible 
new pairs based on independent checks. The AllWISE measured motions should be identical for each member of the pair 
because the common parallax between components will affect the motion measures identically. Therefore, we can use 
these systems to perform an internal consistency check to see whether AllWISE motion measurements are the same 
within the errors.

\begin{figure}
\figurenum{10}
\includegraphics[scale=0.45,angle=0]{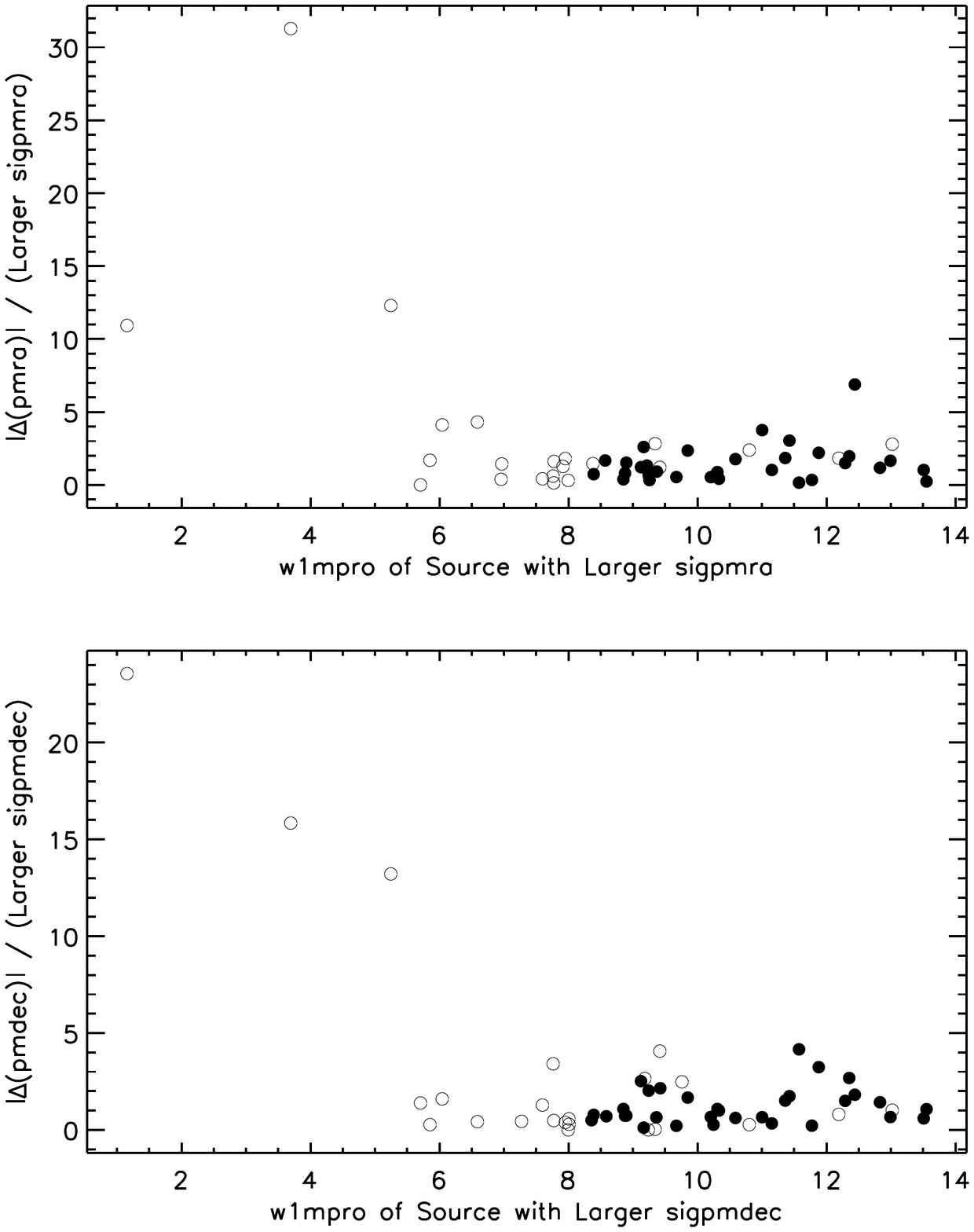}
\caption{Difference in motion in RA (top) and Dec (bottom), expressed in units of the larger motion error within 
the pair, for fifty-five common proper motion binaries noted during quality assessment. These differences are plotted 
as a function of the W1 profile-fit magnitude of the source with the larger motion error. Pairs having at least one 
component in the saturated regime (W1 $<$ 8.1 mag or W2 $<$ 7.0 mag) are shown by open circles. All others are shown 
by solid circles. 
\label{truth_cpm_binaries}}
\end{figure}

Table 2 %~\ref{cpm_pairs} 
(found at the end of this manuscript)
lists these fifty-five common-proper-motion binaries and Figure~\ref{truth_cpm_binaries} 
shows the difference in motion between the two members of each pair. The difference is shown per axis, is expressed 
in units of the larger motion error within the pair, and is plotted as a function of the W1 profile-fit magnitude 
of the source with the larger motion error. Overall, agreement is very good. The majority of unsaturated 
pairs has a difference $<2\times$ the error. 

\notetoeditor{From this point forward, I've had to insert the table numbers by hand because the label tagging
assigns the wrong table number. I can't figure out why this is happening.}

\subsection{Motions, Both Real and False\label{motions_real_false}}

In this section, we examine the distribution of motions for objects in a typical 
$1\fdg56\times1\fdg56$ AllWISE Atlas Image. We then extract sources having 
statistically significant motions and discuss the average 
success rate for finding bona fide motion stars. In the final subsection, we take a broader view and
categorize the sources of false motion seen during our quality assurance checks 
(described in more detail in section 4.1) over the entire sky.

\subsubsection{A Typical Distribution of Motion Measurements}

\begin{figure}
\figurenum{11}
\includegraphics[scale=0.35,angle=90]{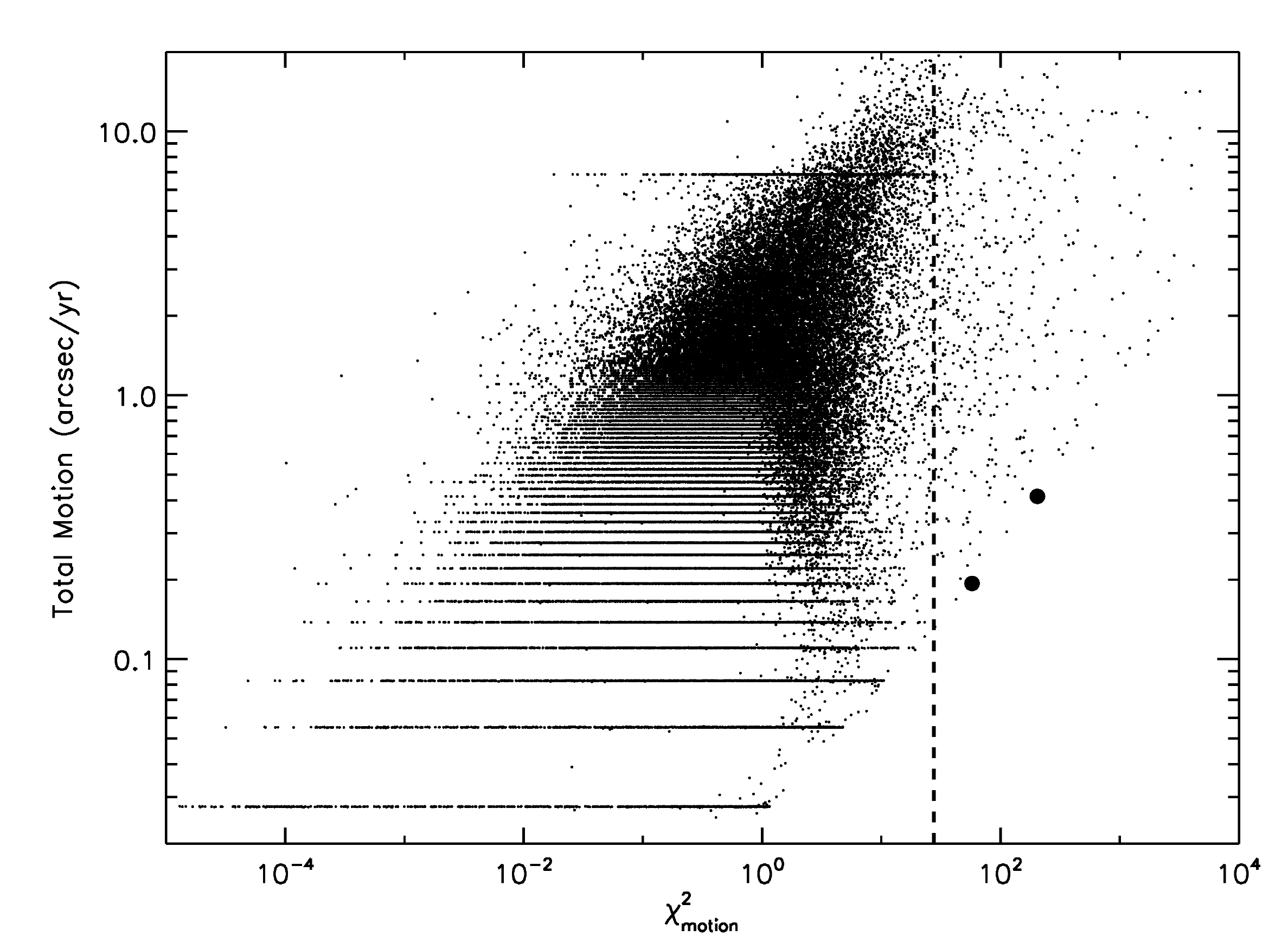}
\caption{Total motions for sources in AllWISE Atlas Tile 0440p166\_ac51 plotted as a function of ${\chi^2}_{motion}$. 
Objects to the right of the dashed line have ${\chi^2}_{motion} > 27.63$ and are considered to have statistically 
significant motions. The previously known motion stars LTT 10950 and NLTT 9223 are shown by the larger dots. The 
quantized motion measures (further illustrated in the upper panel of Figure~\ref{motion_quantization_and_pile-up};
see section~\ref{motion_quantization}) 
are seen for small motion values in the lower portion of the plot, and the motion ``pile-up'' for spurious sources 
(further illustrated in the lower panel of Figure~\ref{motion_quantization_and_pile-up}; see 
section~\ref{motion_pile-up}) 
is seen near the top of the 
plot at a value of 6$\farcs$875 yr$^{-1}$.
\label{typical_motions}}
\end{figure}

Motion measurements for all sources in a typical 
Atlas Image are illustrated in Figure~\ref{typical_motions}. 
The total proper motion is plotted as a function of ${\chi^2}_{motion} = (pmra/sigpmra)^2 + (pmdec/sigpmdec)^2$, where 
$pmra$ and $pmdec$ are the AllWISE-measured RA and Dec motions, respectively, and $sigpmra$ and $sigpmdec$ are their 
associated uncertainties. AllWISE tiles typically have $\sim10^5$ extracted sources, so any source with $Q = 
e^{-{\chi^2}_{motion}/2} < 10^{-6}$ violates\footnote{This threshold was found to produce a sample of potentially 
interesting motion sources for human perusal that was not too small and not overwhelmingly large.}, with a very high 
degree of confidence, the hypothesis of a real object with 
zero motion (and errors correctly characterized by the Gaussian uncertainties). This limit on $Q$ corresponds to 
${\chi^2}_{motion} > 27.63$. Sources in this tile that satisfy this criterion are found to the right of the dashed 
line in Figure~\ref{typical_motions}. Although a few of these objects are motion stars (see 
Section~\ref{real_moving_sources}), other sources have complications that create high values of ${\chi^2}_{motion}$ 
but do not have real motions. Examples are contaminated and spurious objects flagged as such by $cc\_flags$, objects with low 
signal-to-noise in all bands, and blended or extended sources that can often be identified by the $nb$, $w?chi2$, 
and $rchi2$ criteria\footnote{Throughout the paper, we use the shorthand ``$w?chi2$'' to refer to $w1chi2$, 
$w2chi2$, $w3chi2$, and $w4chi2$.}. Other examples of false motion sources are given in Section~\ref{false_sources}. 

\subsubsection{Real Moving Sources\label{real_moving_sources}}

A total of sixteen objects in Figure~\ref{typical_motions} passed additional criteria
-- no blending ($nb = 1$), $w?rchi2$ values indicating a point source ($<$ 2.0 in all 
bands), and other criteria as further explained in section 4.1. 
Only two of these, highlighted by larger dots in the figure,
also passed a visual inspection step showing them to be point-like, unconfused, and moving (if present in
data from earlier surveys). Both are previously published motion stars.
The first, 
WISEA J025517.56+161832.7, has AllWISE measured motions in (RA, Dec) of (367$\pm$29, 190$\pm$29) mas yr$^{-1}$ and 
is the known motion star LTT 10950 with published proper motions in (RA, Dec) of ($203.09{\pm}0.82$, 
$-47.76{\pm}0.58$) mas yr$^{-1}$ (\citealt{vanleeuwen2007}). The discrepancy in measured values is likely due more 
to the fact that this source is heavily saturated in AllWISE (W1 = 5.7 mag) rather than to parallax, which is only 
37.32$\pm$0.66 mas (\citealt{vanleeuwen2007}). The second source, WISEA J025326.12+172429.8, has AllWISE measured 
motions in (RA, Dec) of ($-55{\pm}26$, $-185{\pm}25$) mas yr$^{-1}$ and is the known motion star NLTT 9223 with 
published proper motions in (RA, Dec) of (20, -258) mas/yr (\citealt{lepine2005}). This source is also saturated 
in AllWISE (W1 = 7.7 mag), which is likely the reason for the poor agreement between AllWISE and published values.

This Tile highlights the typical success rate in finding new motion objects. Generally fewer than two dozen motion
candidates are retained by the automated criteria, and only a small number survive the visual inspection step.
Most of those that remain have been previously identified as motion objects by other surveys. Given that we 
identified 3,525 new motion objects over the entire sky and that there are 18,240 Atlas Tiles total, the average 
Tile does not contain a new discovery. 

\subsubsection{False Moving Sources\label{false_sources}}

\begin{figure*}
\figurenum{12}
\includegraphics[scale=0.475,angle=0]{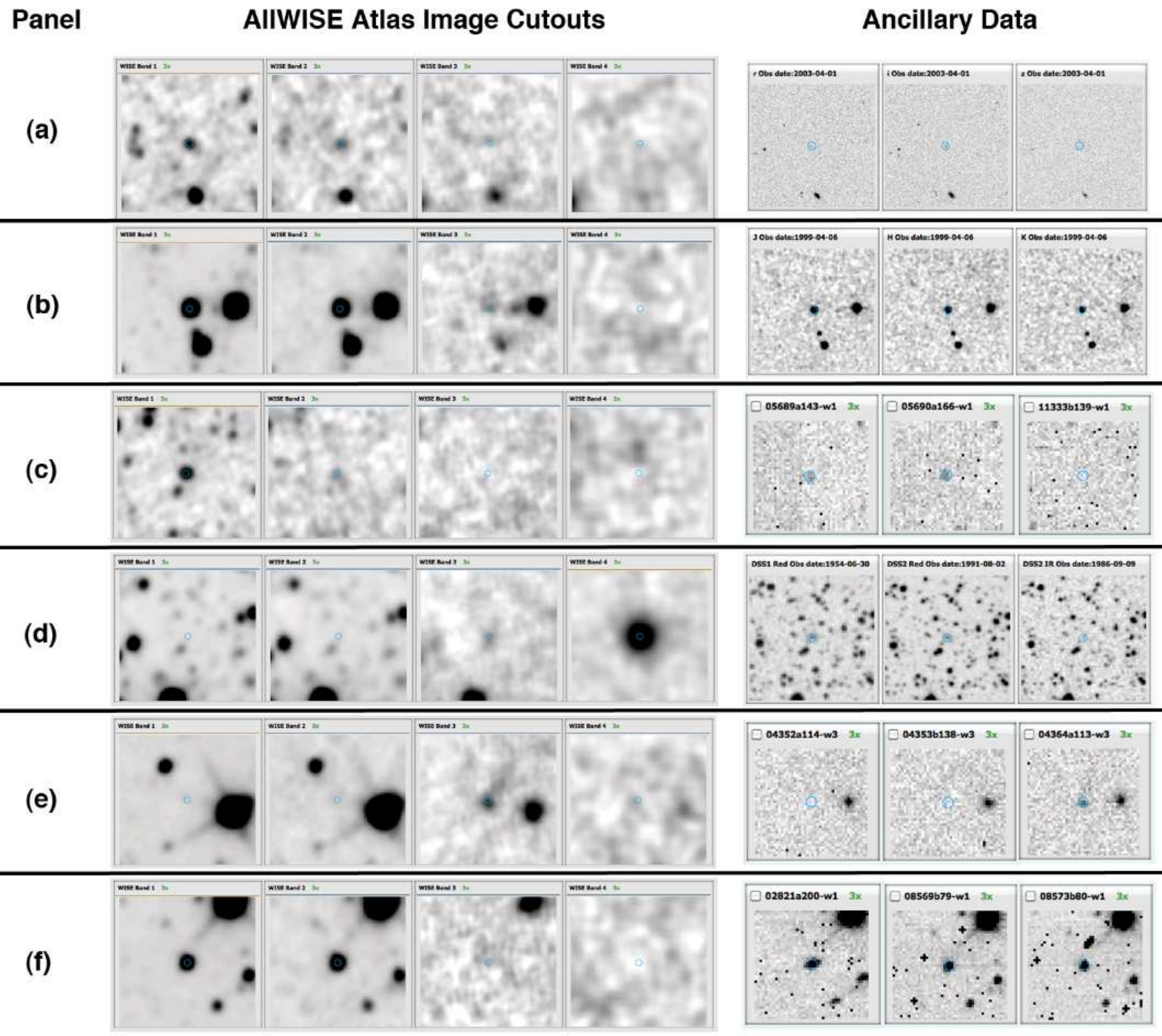}
%Note: I made this figure by doing a PNG screen grab while working in Photoshop, then converted the PNG file to
%eps using http://www.tlhiv.org/rast2vec/ .
\caption{Seven categories of false-motion sources. The W1, W2, W3, and W4 AllWISE Atlas Image fields 
($2\arcmin\times2\arcmin$ with north up and east to the left) along with three ancillary images are shown for 
each example. The faint blue circle denotes the position of the AllWISE source. Ancillary images are identified 
as -- (a) SDSS $r$, $i$, and $z$. Note that a faint source is seen at the AllWISE position. (b) 2MASS $J$, $H$, 
and $K_s$. Note that the blue circle is not well centered on the bright source in either the AllWISE or 2MASS 
images. (c) Representative WISE Level 1b frames at W1; the first two frames are from the first epoch of WISE 
observations and the third frame is from the second epoch. Note that the object is seen only at the first epoch. 
(d) DSS1 $R$ and DSS2 $R$ and $I$. Note that there is a faint source at the AllWISE position. (e) Three WISE 
Level 1b frames at W3. Note that the first two show no source at the location of the AllWISE object; the third 
is the only Level 1b image at this sky location showing a source at this position. (f) The only first-epoch WISE 
Level 1b image at W1 along with two representative second-epoch W1 images. Note that the first-epoch frame is 
significantly smeared due to spacecraft motion.
\label{spurious_movers}}
\end{figure*}

False-motion sources generally fall into one of the following categories. Representative examples and their probable 
causes are discussed below:

\noindent (1) Blended/extended sources: As in any motion survey, blended or extended objects in AllWISE create the 
potential for false motions. One such source is WISEA J130716.06+061032.7, an object whose AllWISE measurements 
indicate a significant motion of 4795$\pm$471 and 2381$\pm$493 mas yr$^{-1}$ in RA and Dec, respectively. A very 
faint, extended source is seen at this position in the SDSS $r$ and $i$ band imagery taken in 2003 
(Figure~\ref{spurious_movers}a), so the source is clearly not moving at the rates indicated by AllWISE. This source 
is extended in the W2 band and the photocenters at W1 and W2 do not align, likely giving rise to the false motion.

\noindent (2) Spurious small-separation, same-tile (SSST) sources\footnote{An SSST source is part of a group of 
detections with unphysically small separations 
extracted on the same Atlas Tile and having nearly identical positions and fluxes. The false sources in these groups 
are created by a coding error in the AllWISE processing pipeline. The photometry routine works on 
detections in descending order of brightness; the intention was to process the brightest sources first so that their 
flux could be subtracted from the individual frames before processing fainter sources. The software error
caused the flux subtraction step to be skipped in most cases. For a faint source in the wings 
of a brighter source, this error can cause the faint source to migrate to the flux from the brighter object because 
that minimizes $\chi^2$ better than a near zero-motion solution near the faint-source detection position itself. 
Post-processing identified sources in these SSST groups, and the {\it rel} parameter was used to flag these in the
Catalog and Reject Table. Sources 
thought to be the original, correct entries have {\it rel} codes of $s$ or $c$ and were considered for inclusion 
in the AllWISE Source Catalog; sources believed to be unreliable members of the SSST group were given {\it rel} = 
$r$ and will be found only in the AllWISE Reject Table. Sources with {\it rel} = {\it null} are not affected by
the SSST phenomenon.}: WISEAR J103558.13-370522.0 is a source in the 
AllWISE Reject Table with a seemingly high and statistically significant motion; AllWISE measures $3527{\pm}350$ 
and $-1209{\pm}119$ mas yr$^{-1}$ in RA and Dec, respectively. This, however, is a spurious small-separation, 
same-tile (SSST) source and is flagged as such by the $rel$ flag in the AllWISE 
Reject Table. As seen in Figure~\ref{spurious_movers}b, this source falls in the wings of a real object, named 
WISEA J103558.04$-$370520.9, that is found in the AllWISE Source Catalog and that has motion measures of $-4{\pm}44$ 
and $27{\pm}44$ mas yr$^{-1}$ in RA and Dec, respectively, indicating no apparent motion. Users are cautioned to 
pay particular attention to the $rel$ flag when using motion measures from the AllWISE Reject Table.

\noindent (3) Flux transients: Flux variables can sometimes trigger false motions. When an object appears at only 
one epoch, a nearby noise blip at the other epoch(s) may trigger a false, generally large motion. Because the 
derived motion is so large, it may appear to be statistically significant despite having rather large errors. 
An example of this is WISEA J122559.53+070005.2, which has AllWISE-reported motions of 24902$\pm$4071 and 
11567$\pm$7922 mas yr$^{-1}$ in RA and Dec, respectively. This object appears only in the first epoch and not the 
second (Figure~\ref{spurious_movers}c), as confirmed by the AllWISE Multi-Epoch Source Table, which gives a mean 
$w1mpro$ value of $\sim$15.3 mag for the first epoch and limits of $>$17.0 mag for the second.

\noindent (4) W3- and W4-dominated sources: Sources that are detected only at W3 and/or W4 may show spurious 
motions. Such sources will be seen only at the earlier epochs before cryogen was exhausted and not at the later 
epochs when those bands were not operational. Nearby noise blips at the later epochs in W1 and/or W2 can trigger 
false motions. One such example is the planetary nebula PN SB 24, also known as WISEA J185716.62$-$175050.4 
(Figure~\ref{spurious_movers}d), which has AllWISE-measured motions of $6243{\pm}1101$ and $-11981{\pm}1162$ mas 
yr$^{-1}$ in RA and Dec, respectively, despite the fact that the nebula is known to have near-zero proper motion 
(\citealt{kerber2008}).

\noindent (5) Cosmic rays in low-coverage areas: Areas having less than five frames of coverage do not reap the 
benefit of outlier rejection\footnote{See http://wise2.ipac.caltech.edu/docs/release/allsky/expsup/ sec4\_4f.html\#outrej 
in the WISE All-Sky Release Explanatory Supplement.} in the coaddition step, so spurious sources can bleed through 
into the coadds. In some cases, these spurious sources have falsely measured, yet apparently
significant, motions in AllWISE. 
One example is WISEA J210220.65$-$083948.6 (Figure~\ref{spurious_movers}e), a cosmic ray that appears in only one 
W3 frame (04364a113) in an area of sky where that W3 frame is the only one of acceptable quality for AllWISE 
processing. A nearby noise blip in another band at a different epoch nonetheless causes AllWISE to measure a large, 
although false, motion for it of 4573$\pm$1094 and 6531$\pm$1213 mas yr$^{-1}$ in RA and Dec, respectively.

\noindent (6) Sources falling in a streaked frame in a low-coverage area: Another consequence of low-coverage 
areas is the increased influence of the occasional poorer quality frame going into the coadd. The most extreme example is 
one in which only a single frameset is available for one of the WISE epochs, and that frameset has non-optimal 
image quality. False motions may result, as is the case for source WISEA J061658.37+701209.5 
(Figure~\ref{spurious_movers}f). This object lies in an area of sky with single-frameset coverage (02821a200) at 
the first epoch, but this frameset is slightly smeared due to momentum 
dumping\footnote{See http://wise2.ipac.caltech.edu/docs/release/allwise/expsup/ sec4\_2.html\#torque in the 
AllWISE Explanatory Supplement.} by the spacecraft. The astrometric solution for the entire frameset is slightly 
biased in RA because of this smearing. Data at the later epoch are not affected. As a result of the mismatch in 
positions across epochs, a false motion of $1547{\pm}129$ and $-5{\pm}127$ mas yr$^{-1}$ in RA and Dec, 
respectively, is created. It should be noted that this is an extremely rare occurrence in AllWISE, as Quality 
Assurance flagged almost all of the smeared frames and eliminated them from consideration for coaddition. 
Nonetheless, users are advised to treat with caution any motion measures in areas of low-coverage at a single epoch.

\subsection{Other Caveats\label{other_caveats}}

\subsubsection{Motion Measurements are Quantized\label{motion_quantization}}

\begin{figure}
\figurenum{13}
\includegraphics[scale=6.5,angle=0]{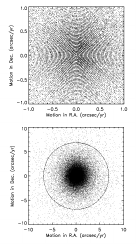}
\caption{Plots of the AllWISE-measured motions in RA plotted against the AllWISE-measured motions in Dec for 
a typical Atlas Tile (1174p075\_ac51). (Upper panel) A zoom-in showing motion quantization at small values. (Lower panel) 
A zoom-out showing the motion ``pile-up'' at a value of $6\farcs875$ yr$^{-1}$.
\label{motion_quantization_and_pile-up}}
\end{figure}

A plot of AllWISE motion measurements in a single Atlas Tile 
(upper panel of Figure~\ref{motion_quantization_and_pile-up}) shows that 
certain, discrete values have a higher frequency of occurrence. The solution for motion values and source 
positions is an iterative process whereby a minimization of $\chi^2$ is sought in the phase space of source 
position and motion. Convergence is rapidly achieved for position estimates, using initial values. On the other 
hand, convergence on motion values is achieved only after one or more iterative steps. The step size used for 
motion estimates is 27.5 mas yr$^{-1}$, and converged motion values are thus quantized at multiples of this 
value\footnote{See http://wise2.ipac.caltech.edu/docs/release/allwise/expsup/ sec5\_3bii.html\#gradient-descent\_algo 
in the AllWISE Explanatory Supplement for further information.}.

\subsubsection{Spurious Motion Measurements show ``Pile-up'' at Large Values\label{motion_pile-up}}

A total motion value of 6875 mas yr$^{-1}$ is often seen for spurious sources in AllWISE. The ``pile-up'' of motion 
measures at this value is illustrated in the lower panel of Figure~\ref{motion_quantization_and_pile-up}. The 
algorithm for minimization of $\chi^2$ has a maximum number of iterations of 250. Each iteration step for motion 
uses, as stated above, a step size of 27.5 mas yr$^{-1}$. A spurious or pure-noise solution that has not converged 
after the maximum number of steps will lead to a motion at the ``pile-up''  
value\footnote{See http://wise2.ipac.caltech.edu/docs/release/allwise/expsup/ 
sec5\_3bii.html\#effect\_noise\_motion\_solution in the AllWISE Explanatory Supplement for more details.} of
6875 mas yr$^{-1}$.

\section{A Catalog of Motion Discoveries}

\subsection{Criteria for Selecting Motion Candidates\label{selection_criteria}}

As part of AllWISE pipeline processing at IPAC, the Quality Assurance code filtered the list of detections from 
each Atlas Tile to identify objects believed to be moving significantly. These criteria were used to select 
only the choicest motion objects from the source lists and to keep the number of candidates manageable on the tight 
schedule allotted for AllWISE Quality Assurance checks. The criteria used are as follows:

(1) To ensure that the object is not flagged as an artifact, the $cc\_flags$ parameter was required not to have a 
capital-letter value at any position in its four-character code, in which the four characters refer to the four 
WISE bands. Capital letters are used when the software has determined in that band that the object was a spurious
detection of either 
a diffraction (`D') spike, a scattered light halo (`H') surrounding a bright star, an optical 
(`O') ghost image, or persistence (`P') -- i.e., a short-term latent image -- left behind by a bright source on 
the detector. Values of $cc\_flags$ having only lower-case versions of these same letters (indicating that the 
object is likely real and only contaminated by an artifact) or zero (indicating that the source is unaffected by known 
artifacts) were allowed. 

(2) To ensure that the object is point-like, a value of $nb=1$ (no blends) was required. Also, the motion-fit, 
reduced $\chi^2$ values in all bands, $w?rchi2\_pm$, were required to be less than two to eliminate sources that 
are obviously extended.

(3) To eliminate more poorly measured sources from consideration, the signal-to-noise ratio, $w?snr$, was required 
to be greater than seven in at least one band.

(4) To ensure that the code selected only those objects with robust and statistically significant motions, the 
following criteria were applied: (a) The ratio of the reduced $\chi^2$ value between the stationary and motion 
fit, $rchi2/rchi2\_pm$, was required to be greater than 1.03. This ensures that the motion-fit solution significantly 
improves the overall reduced $\chi^2$ value compared to the stationary fit. (b) As explained in 
section~\ref{motions_real_false} above, the ${\chi^2}_{motion}$ criterion was required to be greater than 27.63 to 
select only those objects with statistically significant motions. (c) The second and third characters of $pmcode$ 
were required to be `N' and `0', respectively. A value of `N' means that no blend-swapping has occurred and a value 
of `0' means that the offset between the stationary-fit and the motion-fit position at the mean epoch is less than 
one arcsecond. These values were chosen because blended sources often have blend swapping, and spurious sources 
often have unphysically large positional differences at the mean epoch between the non-motion and motion 
solutions\footnote{See http://wise2.ipac.caltech.edu/docs/release/allwise/expsup/ sec2\_1a.html\#pmcode in the 
AllWISE Explanatory Supplement.}.

(5) To ensure that the motion measurement has a solid foundation of W1- and W2-band data underlying it, we 
experimented with various values of $w1m$, $w2m$, and $w1mJDmax - w1mJDmin$. The first two parameters specify the 
number of frames in W1 and W2, and the third parameter specifies the time difference between the earliest and 
latest frames in W1. In the end, these parameters were all set to be greater than zero because regions of very 
sparse coverage were found not to contribute a large number of spurious motion sources, mainly because the vast 
majority of the sky is well covered across two or more epochs.

These criteria were written to eliminate blended/extended sources, spurious objects, and pure noise. Other 
contaminants listed in section~\ref{false_sources} -- the SSST sources, flux transients, W3- and W4-dominated 
sources, and cosmic rays and streaked frames in low coverage areas -- were discovered as a result of these Quality 
Assurance checks. These contaminants were generally easy to recognize via a visual inspection step, which displayed 
images of the field in the AllWISE W1, W2, W3, and W4 coadds along with 2MASS images at $J$, $H$, and $K_s$. For 
WISE objects that were also visible in 2MASS, which represents the vast majority of sources identified, any motion 
should be obvious in the 10+ years between the 2MASS and WISE image sets\footnote{Because the smallest motion 
measureable with AllWISE is $\sim0\farcs15$ yr$^{-1}$ (section~\ref{smallest_motion_section}), this means that 
objects will have moved $>15\arcsec$ in the 10+ years between the 2MASS and WISE imaging sets, which is a 
sufficiently large motion to be obvious by eye.}. The Quality Assurance scientists checked these images for all 
candidate motion objects and retained only those objects having no 2MASS source at the AllWISE position.

\subsection{Vetting the List of Candidates}

The Quality Assurance scientists (JDK and SFA) produced a list of 26,850 candidates as a result of the criteria 
filtering and image checking discussed above. This list was then run (by JDK) through SIMBAD using a two-arcminute 
search radius around the AllWISE position to search for previously known objects. If SIMBAD noted a proper motion 
source within 10$\arcsec$ of the AllWISE position, the object was eliminated from the list\footnote{Roughly speaking, 
a circle of radius 10$\arcsec$ around the AllWISE J2000 position at epoch 2010.54 will capture all SIMBAD motion 
objects, reported using their J2000 positions and epoch 2000.00, moving less than $\sim1\arcsec$ yr$^{-1}$.}. When 
known proper motion sources were found outside of 10$\arcsec$, the AllWISE- and SIMBAD-reported motions were checked 
to see if the object was a very high-motion source, generally $>1\arcsec$ yr$^{-1}$. This further check eliminated 
many known, small-numbered LHS (Luyten Half Second) objects.

The remaining list of objects was then checked (by JDK) against the Fourth US Naval Observatory CCD Astrograph 
Catalog\footnote{Available at http://irsa.ipac.caltech.edu.} (UCAC4; \citealt{zacharias2013}) using a one-arminute 
search radius. Objects with significant UCAC4-reported motions were considered to be possible matches to the AllWISE 
source. The match was considered confirmed if the UCAC4 source appeared to have a motion of the right magnitude and 
in the correct direction, as inferred from visual inspection of the motion source across the DSS2, 2MASS, and WISE 
images, to be the AllWISE-selected object. Those objects matching a source with reliable motions in UCAC4 were 
flagged as such but kept in the object list because they have no SIMBAD entries.

After these checks, the list was run again through SIMBAD and further checked against DwarfArchives.org, against 
a list of published papers (on low-mass objects) likely not to have been ingested into SIMBAD by 2013 December, and 
against catalogs housed by VizieR. These checks (by GNM and AS) enabled us to eliminate other objects. Objects were 
kept in the list only when SIMBAD had no reported proper motion information and no links to published papers citing 
a proper motion value. Thus, a few objects with published spectral types are included in the list because this will 
be the first refereed publication in SIMBAD giving their motion information. Objects listed only in UCAC4 and other 
catalogs found in VizieR are noted as such in the final list.

Of the 26,850 motion candidates selected by our Quality Assurance criteria, we found that 18,862 objects (70\%) 
were already published and had motion information available in SIMBAD. Of the remainder, 3,583 objects 
(13\%) are new discoveries and 4,405 objects (16\%) were found to be non-motion sources after visual inspection of 
the DSS2 images.

The discovery list of 3,583 itself is comprised of two sublists. The first (Table 3) contains 3,525 objects whose motions 
could be confirmed using non-WISE data, and the second (Table 4) contains 58 unconfirmed
objects lacking counterparts in earlier surveys. For the list of confirmed objects, one of us (AS) consulted the 2MASS images 
to find the WISE source at the 2MASS epoch and then tabulated the 2MASS position and magnitudes from the 2MASS 
All-Sky Point Source Catalog. These 2MASS associations were then double checked (by JDK) by creating finder charts 
of DSS, 2MASS and WISE images centered at the position of the 2MASS source. Using these associations, a proper motion 
was measured for each source using the 2MASS position as the first epoch and the mean WISE All-Sky Source Catalog 
position as the second epoch.

Because the identification of motion candidates during Quality Assurance took place before selection of objects for 
the AllWISE Source Catalog (and before the SSST phenomenon was fully recognized), we had to retroactively replace a 
small number of our motion candidates that appeared only in the AllWISE Reject Table with their more appropriate 
entry from the AllWISE Source Catalog. In some cases, the entry in the AllWISE Source Catalog does not have a motion 
significant enough to have been selected by the Quality Assurance criteria. We retain such cases if our independent 
checks using the 2MASS-to-WISE time baseline 
confirm that the objects are truly moving. An example of this is WISEA J000205.60$-$322545.9, which has 
AllWISE motions of $0\pm74$ and $0\pm73$ mas yr$^{-1}$ in RA and Dec, respectively, but whose comparison with the 
earlier 2MASS position confirms a motion of $23.8\pm10.1$ and $42.0\pm9.8$ mas yr$^{-1}$ in RA and Dec. This motion 
is too small to have been legitimately uncovered by AllWISE, but we list it nonetheless as a serendipitous discovery.

The sky distribution of the confirmed motion objects is shown in Figure 14. The top panel shows the
ecliptic projection that is the natural system for the WISE scanning pattern. The
three-epoch coverage areas\footnote{See the depth-of-coverage sky map (Figure 5) in section IV.2 of the AllWISE Explanatory 
Supplement at http://wise2.ipac.caltech.edu/docs/release/allwise/expsup/
sec4\_2.html.} at ecliptic longitudes of $\sim$25$^\circ$-48$^\circ$ and $\sim$200$^\circ$-223$^\circ$ are, as expected,
seen as overdensities. An underdensity in the number of motion 
objects is seen toward the Galactic Center and along the Galactic mid-plane (bottom panel of Figure 14), as expected 
in these regions of source confusion and high backgrounds. 

\begin{figure}
\figurenum{14}
\includegraphics[scale=0.85,angle=0]{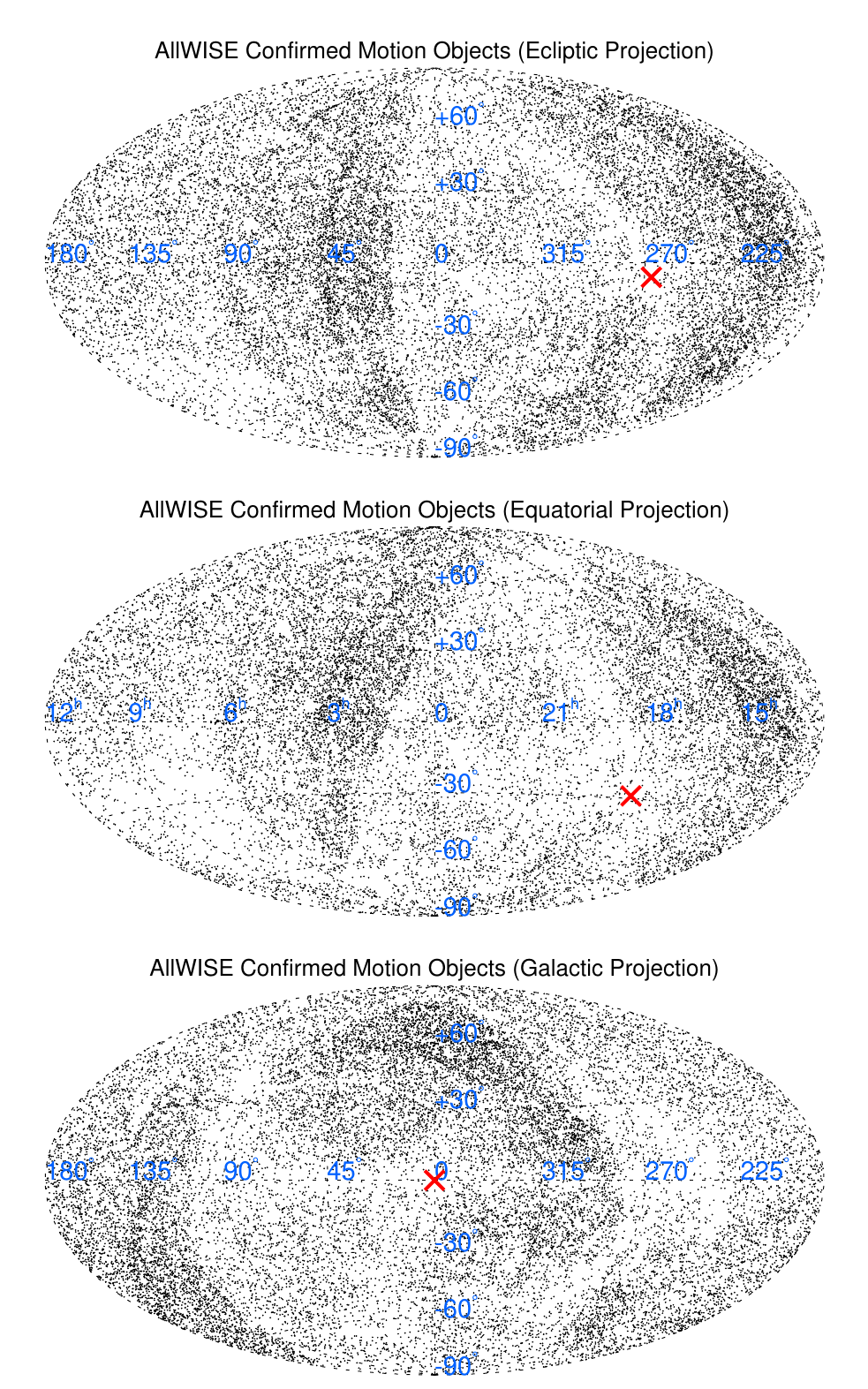}
\caption{
All-sky maps showing the locations of the 22,387 confirmed motion objects
identified in AllWISE. Of these, 18,862 are rediscoveries whose motions are given
in SIMBAD and the other 3,525 are new discoveries from Table 3. The projections shown
are (top) ecliptic, (middle) equatorial, and (bottom) galactic. A red ``X'' marks the 
location of the Galactic Center.
\label{skymap_all}}
\end{figure}

The sky distribution of the confirmed objects is compared in Figure 15 to the New Luyten Two Tenths Catalog 
(NLTT; \citealt{luyten1979nltt}), a large compilation of motion objects covering the entire sky.
As expected, the majority of new discoveries (middle panel of Figure 15) falls primarily in the southern hemisphere 
because the NLTT is much less complete in those regions (top panel of
Figure 15). The clumpiness of these discoveries in the southern hemisphere (e.g., at 
$11^h <$ RA $< 22^h$ and Dec $< -0^\circ$) is largely a consequence of the WISE three-epoch coverage
(middle panel of Figure 14) overlapping areas poorly covered by the NLTT. 

\begin{figure}
\figurenum{15}
\includegraphics[scale=0.85,angle=0]{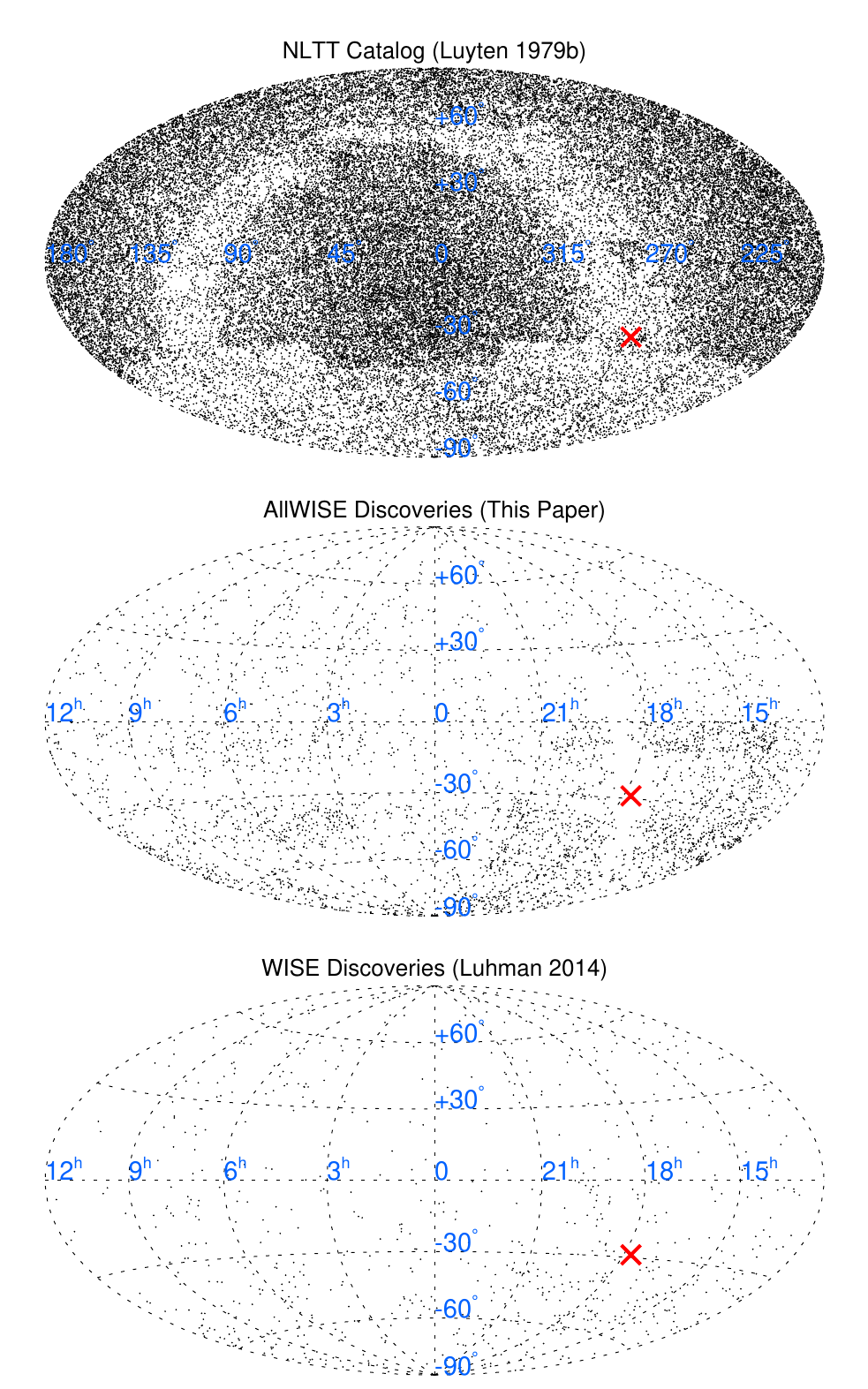}
\caption{
All-sky maps showing the locations of (top) the 58,845
motion objects from the New Luyten Two Tenths Catalog (\citealt{luyten1979nltt}),
(middle) the 3,525 motion objects from
Table 3 of this paper, and (bottom) the 762 motion objects from \cite{luhman2014}.
These are equatorial coordinate projections with the vernal equinox at the center and 
RA increasing to the left. A red ``X'' marks the location of the Galactic Center.
\label{skymap_various_catalogs}}
\end{figure}

\subsection{The Motion Catalog}

The list of all 3,525 confirmed AllWISE motion objects is given in Table 3. %~\ref{discoveries}. 
Column 1 gives the source 
designation from the AllWISE Source Catalog. Columns 2-4 give the 2MASS $J$, $H$, and $K_s$ magnitudes and 
associated errors of the source, and columns 5-6 give the AllWISE W1 and W2 profile-fit magnitudes and errors. 
Columns 7-8 give the AllWISE-measured motions in RA and Dec whereas columns 9-10 give the proper motions computed 
using the 2MASS-to-WISE time baseline. Column 11 is a flag column indicating whether the source has an entry in 
another motion catalog, even though the source itself has no publication history, and column 12 indicates whether 
additional information is available in the footnotes.

\begin{turnpage}
%\begin{center}
\begin{deluxetable*}{lccccccccccc}
\tabletypesize{\tiny}
%\rotate
\tablewidth{8.0in}
\tablenum{3}
\tablecaption{AllWISE Motion Discoveries\label{discoveries}}
\tablehead{
\colhead{WISEA Designation} &                          
\colhead{2MASS $J$} &  
\colhead{2MASS $H$} &     
\colhead{2MASS $K_s$} &
\colhead{W1} &
\colhead{W2} &
\colhead{AllWISE} &
\colhead{AllWISE} &
\colhead{Computed} &
\colhead{Computed} &
\colhead{Flag\tablenotemark{b}} &
\colhead{Note\tablenotemark{c}} \\
\colhead{} &                          
\colhead{(mag)} &  
\colhead{(mag)} &     
\colhead{(mag)} &
\colhead{(mag)} &
\colhead{(mag)} &
\colhead{RA Motion} &
\colhead{Dec Motion} &
\colhead{$\mu_\alpha$\tablenotemark{a}} &
\colhead{$\mu_\delta$\tablenotemark{a}} &
\colhead{} &
\colhead{} \\
\colhead{} &                          
\colhead{} &  
\colhead{} &     
\colhead{} &
\colhead{} &
\colhead{} &
\colhead{(mas/yr)} &
\colhead{(mas/yr)} &
\colhead{(mas/yr)} &
\colhead{(mas/yr)} &
\colhead{} &
\colhead{} \\
\colhead{(1)} &                          
\colhead{(2)} &  
\colhead{(3)} &     
\colhead{(4)} &
\colhead{(5)} &
\colhead{(6)} &
\colhead{(7)} &
\colhead{(8)} &
\colhead{(9)} &
\colhead{(10)} &
\colhead{(11)} &
\colhead{(12)}
}
\startdata
J000136.86$-$010146.9& 12.363$\pm$0.021& 11.831$\pm$0.020& 11.565$\pm$0.021& 11.374$\pm$0.022& 11.202$\pm$0.021&  -166$\pm$49&  -378$\pm$48&     -38.5$\pm$11.4& -233.1$\pm$10.5&  0& 0 \\
J000138.90$-$761350.1&  9.477$\pm$0.021&  8.893$\pm$0.024&  8.603$\pm$0.023&  8.460$\pm$0.023&  8.374$\pm$0.020&   284$\pm$30&  -170$\pm$32&     241.2$\pm$17.1&  -16.7$\pm$7.7&   0& 0 \\
J000205.60$-$322545.9& 14.079$\pm$0.026& 13.472$\pm$0.036& 13.216$\pm$0.037& 13.085$\pm$0.024& 13.023$\pm$0.026&     0$\pm$74&     0$\pm$73&      23.8$\pm$10.1&   42.0$\pm$9.8&   0& 0 \\
J000234.41+470030.9  & 10.288$\pm$0.019&  9.714$\pm$0.030&  9.468$\pm$0.021&  9.311$\pm$0.023&  9.276$\pm$0.021&   -69$\pm$31&  -202$\pm$30&     -93.4$\pm$9.7&  -107.1$\pm$9.5&   1& 0 \\
J000239.96+612015.0  & 12.709$\pm$0.028& 12.059$\pm$0.031& 11.728$\pm$0.021& 11.558$\pm$0.023& 11.390$\pm$0.021&   201$\pm$28&    92$\pm$27&     222.8$\pm$9.2&    68.5$\pm$9.1&   0& 0 \\
J000416.30$-$605925.3& 11.359$\pm$0.021& 10.842$\pm$0.022& 10.636$\pm$0.021& 10.515$\pm$0.022& 10.416$\pm$0.020&  -258$\pm$36&  -218$\pm$35&    -228.3$\pm$9.8&  -213.8$\pm$9.8&   1& 0 \\
J000533.57+280705.8  & 14.073$\pm$0.022& 13.489$\pm$0.033& 13.186$\pm$0.031& 12.975$\pm$0.024& 12.702$\pm$0.026&  -355$\pm$65&  -346$\pm$66&    -205.1$\pm$11.9& -264.8$\pm$10.0&  1& 1 \\
J000622.67$-$131955.6& 16.674$\pm$0.125& 15.548$\pm$0.105& 15.115$\pm$0.126& 14.239$\pm$0.027& 13.754$\pm$0.042&  -416$\pm$123& -852$\pm$128&   -239.2$\pm$17.3& -413.2$\pm$16.3&  2& 0 \\
J000915.71$-$285019.7& 14.744$\pm$0.035& 14.143$\pm$0.042& 13.800$\pm$0.045& 13.624$\pm$0.026& 13.435$\pm$0.033&    44$\pm$101& -229$\pm$103&     86.2$\pm$11.2&   32.0$\pm$10.9&  0& 0 \\
J001102.05$-$421417.7& 10.903$\pm$0.021&  10.37$\pm$0.026& 10.094$\pm$0.021&  9.912$\pm$0.022&  9.774$\pm$0.019&  -399$\pm$37&  -190$\pm$36&    -248.9$\pm$11.1&  -40.6$\pm$9.4&   1& 0 \\
\enddata
\tablecomments{Only a portion of this table is shown here to demonstrate its form and content. A machine-readable 
version of the full table is available.}
\tablenotetext{a}{This is the motion measured between the 2MASS and WISE epochs.}
\tablenotetext{b}{If the source is a motion discovery unique to AllWISE, Flag=0. If the only prior literature is an 
entry with similar motion in the UCAC4 Catalog or other VizieR catalog holding not incorporated into SIMBAD, Flag=1. 
If the object appears only in the \cite{luhman2014} list, Flag=2. If the object appears both in the \cite{luhman2014} 
list {\it and} in a prior catalog not incorporated into SIMBAD, Flag=3.}
\tablenotetext{c}{If there is an additional note about this source at the end of the table, Note=1.}
\end{deluxetable*}
%\end{center}
%\clearpage
\end{turnpage}

The 58 AllWISE motion candidates lacking counterparts in 2MASS and other earlier surveys are listed in 
Table 4. Sixty percent of the objects in this table (35 out of 58) are believed to be
flux transients based on their blue W1$-$W2 colors and appearance on the WISE All-Sky Release and AllWISE Release 
Atlas Images. These objects are slightly fainter versions of objects clearly seen in the W1 individual frames at one, 
and only one, WISE epoch, but because of their faintness, the variability cannot be confirmed by eye using the
individual frames. Most of the remaining forty percent of the objects may or may not be real motion objects. One of
these, WISEA J085510.74$-$071442.5, was also found by \cite{luhman2014}, and the AllWISE motion values of  $-4188\pm267$ 
and $226\pm283$ mas yr$^{-1}$ in RA and Dec, respectively, match well with his measures of
$-4800\pm300$ and $500\pm300$ mas yr$^{-1}$ as expected, since these motions are derived from the same data set.
Visual inspection of the individual frames also strongly suggests a large motion between the two WISE epochs, but we 
nonetheless list this source
as unconfirmed pending a third-epoch verification. 

%\begin{center}
\begin{deluxetable*}{lcccccc}
\tabletypesize{\tiny}
%\rotate
\tablewidth{6.0in}
\tablenum{4}
\tablecaption{AllWISE Motion Candidates Lacking 2MASS Counterparts}\tablenotemark{a}\label{unconfirmed_candidates}
\tablehead{
\colhead{WISEA Designation} &                          
\colhead{W1} &
\colhead{W2} &
\colhead{AllWISE} &
\colhead{AllWISE} &
\colhead{Flag\tablenotemark{b}} &
\colhead{Note\tablenotemark{c}} \\
\colhead{} &                          
\colhead{(mag)} &  
\colhead{(mag)} &     
\colhead{RA Motion} &
\colhead{Dec Motion} &
\colhead{} &                          
\colhead{} \\
\colhead{} &                          
\colhead{} &  
\colhead{} &     
\colhead{(mas/yr)} &
\colhead{(mas/yr)} &
\colhead{} &
\colhead{} \\
\colhead{(1)} &                          
\colhead{(2)} &  
\colhead{(3)} &     
\colhead{(4)} &
\colhead{(5)} &
\colhead{(6)} &
\colhead{(7)} 
}
\startdata
J001152.64-202826.4&      16.443$\pm$0.072&     15.261$\pm$0.093&        -8857$\pm$686&       16150$\pm$720& 0& 0\\
J001449.42-055522.7&      16.313$\pm$0.073&     16.217$\pm$0.233&         5931$\pm$929&        8629$\pm$936& 1& 0\\
J003314.92-465936.0&      16.094$\pm$0.051&     15.612$\pm$0.095&        -8412$\pm$436&        4870$\pm$442& 1& 0\\
J003647.68-122155.3&      15.863$\pm$0.054&     15.522$\pm$0.124&        -6302$\pm$619&       -1310$\pm$572& 1& 0\\
J005429.43-595619.9&      16.237$\pm$0.057&     15.772$\pm$0.118&        -2302$\pm$570&        4102$\pm$547& 0& 0\\
J005519.24-491935.3&      16.196$\pm$0.057&     16.069$\pm$0.153&         5645$\pm$552&        2333$\pm$557& 1& 0\\
J005853.15-555722.4&      15.540$\pm$0.038&     15.628$\pm$0.099&         5248$\pm$349&         826$\pm$327& 1& 0\\
J010133.97-551653.9&      17.206$\pm$0.122&     15.587$\pm$0.099&        -8224$\pm$861&        9173$\pm$886& 0& 0\\
J010213.98-344757.7&      16.059$\pm$0.050&     15.145$\pm$0.069&         1864$\pm$454&        5155$\pm$486& 0& 0\\
J010309.96-151101.0&      15.484$\pm$0.042&     15.500$\pm$0.109&        -4098$\pm$363&        1027$\pm$384& 1& 0\\
J011133.57-412722.5&      16.359$\pm$0.058&     16.203$\pm$0.148&       -19976$\pm$722&       13761$\pm$761& 1& 0\\
J011759.72-281749.6&      15.494$\pm$0.039&     15.399$\pm$0.091&        -1752$\pm$381&        3048$\pm$369& 1& 0\\
J012122.26-110214.0&      16.910$\pm$0.115&     15.257$\pm$0.095&        -2373$\pm$735&        5624$\pm$805& 0& 0\\
J012255.66-223007.0&      15.197$\pm$0.035&     14.886$\pm$0.067&        -1465$\pm$279&         226$\pm$281& 1& 0\\
J014224.00-361822.7&      16.046$\pm$0.050&     15.584$\pm$0.098&         -447$\pm$457&       -5412$\pm$442& 1& 0\\
J014446.74-203855.3&      16.205$\pm$0.063&     15.607$\pm$0.128&         6287$\pm$626&        2423$\pm$661& 1& 0\\
J014507.92-355631.8&      16.139$\pm$0.053&     15.915$\pm$0.123&         3276$\pm$485&        -633$\pm$492& 1& 0\\
J014804.04-712828.3&      16.360$\pm$0.051&     15.072$\pm$0.053&        -1697$\pm$428&       -3448$\pm$394& 0& 0\\
J020801.13-161630.1&      16.020$\pm$0.055&     15.991$\pm$0.152&        -3994$\pm$489&       11866$\pm$588& 1& 0\\
J020812.38-275731.9&      16.617$\pm$0.072&     16.350$\pm$0.191&        -9498$\pm$602&        -185$\pm$616& 1& 0\\
J024519.98+134918.6&      16.130$\pm$0.060&     15.977$\pm$0.154&         3028$\pm$307&        1042$\pm$333& 1& 0\\
J025246.35-262919.1&      16.385$\pm$0.059&     16.346$\pm$0.161&          -52$\pm$329&       -4234$\pm$340& 1& 0\\
J031159.56-195310.6&      15.970$\pm$0.043&     15.421$\pm$0.082&         -919$\pm$229&        5174$\pm$229& 0& 0\\
J031303.91-453358.2&      15.359$\pm$0.034&     15.389$\pm$0.071&        -1661$\pm$152&         510$\pm$155& 1& 0\\
J032307.25-364432.4&      15.885$\pm$0.042&     15.661$\pm$0.094&         1023$\pm$196&       -3686$\pm$194& 1& 0\\
J033231.23-245841.9&      16.376$\pm$0.054&     16.078$\pm$0.133&        -3320$\pm$288&       -2582$\pm$307& 1& 0\\
J034803.93-354036.0&      16.058$\pm$0.043&     15.580$\pm$0.075&        -1593$\pm$196&       -4013$\pm$197& 0& 0\\
J053133.00-304416.7&      15.447$\pm$0.038&     15.710$\pm$0.100&        -1246$\pm$335&       -2708$\pm$340& 1& 0\\
J053509.97-313855.9&      16.344$\pm$0.061&     16.244$\pm$0.181&         2027$\pm$749&       -4774$\pm$788& 1& 0\\
J054631.32-311749.1&      15.921$\pm$0.048&     15.706$\pm$0.108&        -1959$\pm$556&        3282$\pm$619& 1& 0\\
J055034.35-363805.8&      15.806$\pm$0.044&     15.291$\pm$0.076&        -3399$\pm$394&       -2114$\pm$430& 0& 0\\
J055202.16-534725.1&      16.253$\pm$0.043&     16.068$\pm$0.095&         3637$\pm$433&        1876$\pm$467& 1& 0\\
J055537.68-175339.9&      16.019$\pm$0.053&     15.611$\pm$0.107&         3813$\pm$545&        2242$\pm$596& 1& 0\\
J055658.63-140702.8&      15.367$\pm$0.040&     15.230$\pm$0.078&         -914$\pm$344&       -2203$\pm$364& 1& 0\\
J065959.23+780829.9&      15.600$\pm$0.039&     15.223$\pm$0.068&         -782$\pm$330&       -2872$\pm$367& 0& 0\\
J085510.74-071442.5&      16.231$\pm$0.064&     13.704$\pm$0.033&        -4188$\pm$267&         226$\pm$283& 2& 1\\
J105042.60+140202.3&      17.124$\pm$0.138&     15.568$\pm$0.124&        -1521$\pm$800&      -13560$\pm$912& 0& 0\\
J111202.32+324123.7&      17.128$\pm$0.112&     16.658$\pm$0.268&         5648$\pm$591&        5618$\pm$630& 0& 0\\
J113654.49-201658.7&      15.766$\pm$0.046&     15.718$\pm$0.159&         3239$\pm$407&       -4756$\pm$441& 1& 0\\
J121607.98+191003.1&      15.926$\pm$0.055&     15.313$\pm$0.103&         6708$\pm$519&       -4289$\pm$562& 1& 0\\
J121903.27-084702.2&      16.200$\pm$0.067&     16.594$\pm$0.283&        -3987$\pm$637&       -1822$\pm$703& 1& 0\\
J125728.42+464457.6&      16.574$\pm$0.075&     15.911$\pm$0.135&         6348$\pm$519&       -2077$\pm$543& 1& 0\\
J130825.50+331256.1&      15.736$\pm$0.044&     15.265$\pm$0.074&        -6619$\pm$288&       -2636$\pm$321& 1& 0\\
J131302.32+030607.4&      17.254$\pm$0.144&     $>$16.918       &        -1975$\pm$537&       -6980$\pm$618& 0& 0\\
J134115.18+280218.5&      17.410$\pm$0.140&     $>$16.837       &        -3137$\pm$676&       10163$\pm$776& 0& 0\\
J144817.44+490631.2&      16.053$\pm$0.046&     15.364$\pm$0.073&         5177$\pm$330&       -1900$\pm$357& 0& 0\\
J151813.45+311515.8&      17.376$\pm$0.103&     16.311$\pm$0.146&         7908$\pm$378&        -814$\pm$413& 0& 1\\
J163139.17-244942.7&      13.245$\pm$0.026&     11.004$\pm$0.021&         -235$\pm$64 &        -548$\pm$69 & 0& 1\\
J174254.29+761729.7&      15.997$\pm$0.036&     15.280$\pm$0.047&        -2100$\pm$328&       -2899$\pm$342& 0& 0\\
J180512.73+321459.3&      16.275$\pm$0.057&     14.869$\pm$0.051&        -2713$\pm$507&        3013$\pm$566& 0& 0\\
J185818.21+804757.5&      15.806$\pm$0.036&     15.556$\pm$0.066&        -1552$\pm$304&       -2236$\pm$328& 0& 0\\
J191837.08+833036.5&      15.891$\pm$0.038&     15.773$\pm$0.078&         3143$\pm$359&         817$\pm$380& 1& 0\\
J203712.92-071456.2&      16.198$\pm$0.064&     14.563$\pm$0.056&        -1541$\pm$297&        2346$\pm$289& 0& 0\\
J205029.36-344817.4&      16.044$\pm$0.056&     14.761$\pm$0.063&         2486$\pm$457&        1839$\pm$470& 0& 0\\
J225750.96-440429.5&      16.398$\pm$0.055&     16.842$\pm$0.257&       -11280$\pm$550&       11280$\pm$596& 1& 0\\
J230744.88-200218.5&      15.295$\pm$0.040&     15.455$\pm$0.125&        -2117$\pm$365&        6388$\pm$375& 1& 0\\
J232558.27-135406.6&      16.645$\pm$0.089&     16.629$\pm$0.307&        -2993$\pm$961&       17804$\pm$962& 1& 0\\
J232822.75-385208.8&      16.079$\pm$0.057&     15.582$\pm$0.111&        -1229$\pm$524&       -3938$\pm$527& 1& 0\\
\enddata
\tablenotetext{a}{Most of the objects in this list are believed not to be actual motion sources. See text for details.}
\tablenotetext{b}{If the source is a motion discovery unique to AllWISE, Flag=0. If this source is believed to be a flux 
transient and not a motion object because of its blue W1-W2 color, Note=1. If the object appears in 
\cite{luhman2014}, Flag=2.}
\tablenotetext{c}{If there is an additional note about this source at the end of the table, Note=1.}
\tablecomments{WISEA J085510.74-071442.5: This object is 28\arcsec from the radio source 3C 209, which has the 
identifier WISEA J085509.46-071502.9. Although the WISE imaging data show a clear photocentric shift between the two
epochs of data, additional imaging is needed to confirm the motion measure. WISEA J151813.45+311515.8: Brightest in W3 
band (11.073$\pm$0.097 mag) but also faintly visible at W1 and W2. WISEA J163139.17-244942.7: May correspond to the embedded, 
sub-mm source JCMTSE J163138.6-244950 in the $\rho$ Oph 
star formation complex.}
\end{deluxetable*}
%\clearpage
%\end{center}

\begin{figure*}
\figurenum{16}
\includegraphics[scale=0.7,angle=270]{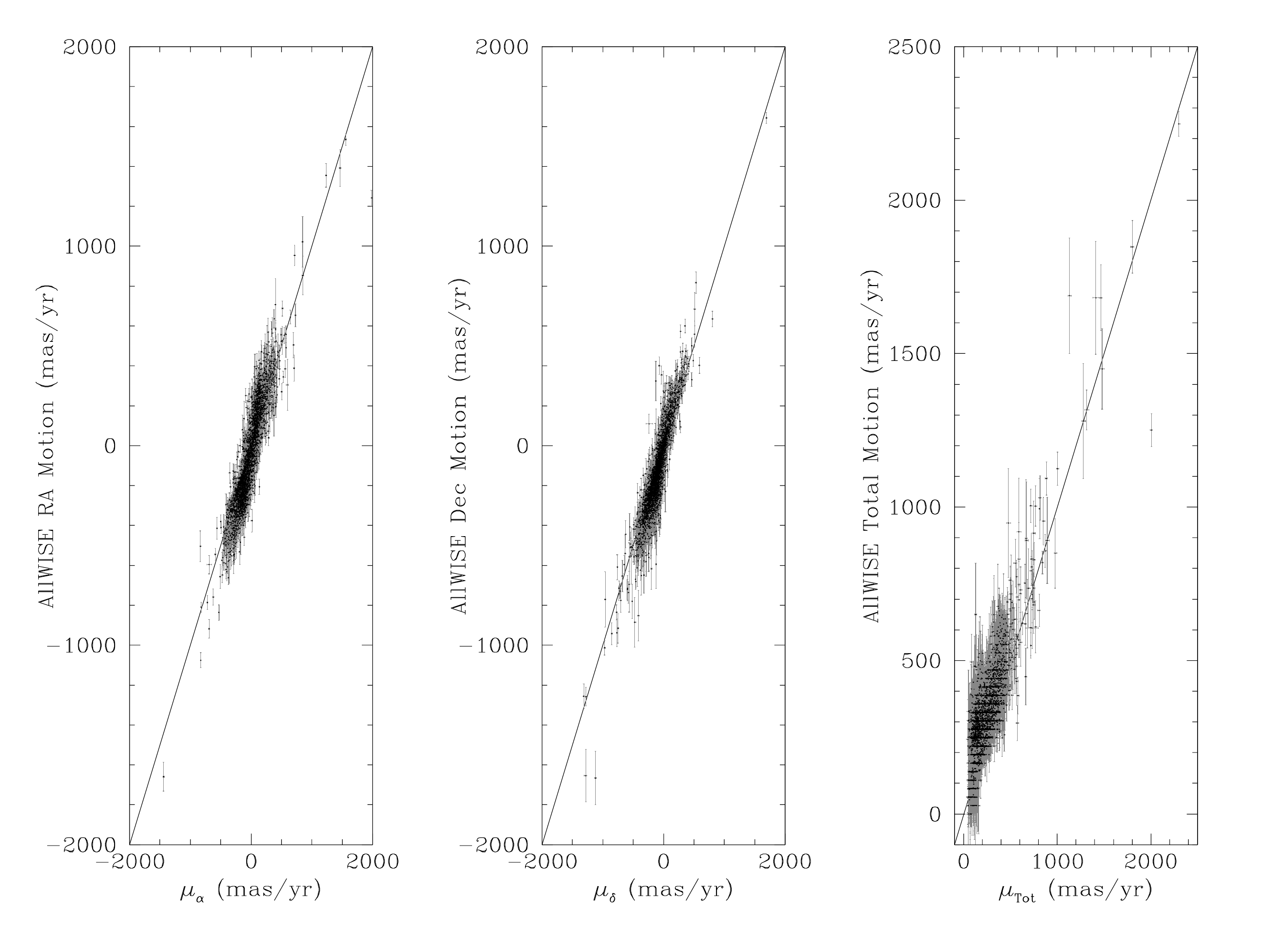}
\caption{The AllWISE-measured RA (left), Dec (middle), and total (right) motions plotted against proper motions 
measured using the 2MASS-to-WISE time baseline for objects in Table 3. %~\ref{discoveries}. 
The line of one-to-one correspondence is shown in each panel.
\label{truth_vs_newdiscoveries}}
\end{figure*}

The next set of plots aim to characterize objects in Table 3. Figure~\ref{truth_vs_newdiscoveries} shows the 
AllWISE-measured motions plotted against the 2MASS-to-WISE motions. Careful analysis of the left and middle 
panels of this plot shows that, unlike the truth-test cases in Figure~\ref{truth_vs_allwise}, the points do not 
fall along the line of one-to-one correspondence but rather would have a fitted slope much steeper than the 
one-to-one line. The reason for this is an inherent bias in our selection criteria. By demanding that the motion 
value be statistically significant compared to its measured errors means that at a given value of RA or Dec motion 
we preferentially select those that are measured on the high side of the average rather than on the low side. 
This effect is also seen in the right panel of Figure~\ref{truth_vs_newdiscoveries}. As expected given this bias, 
the AllWISE-measured motions tend to be higher than those measured independently.

\begin{figure}
\figurenum{17}
\includegraphics[scale=0.45,angle=0]{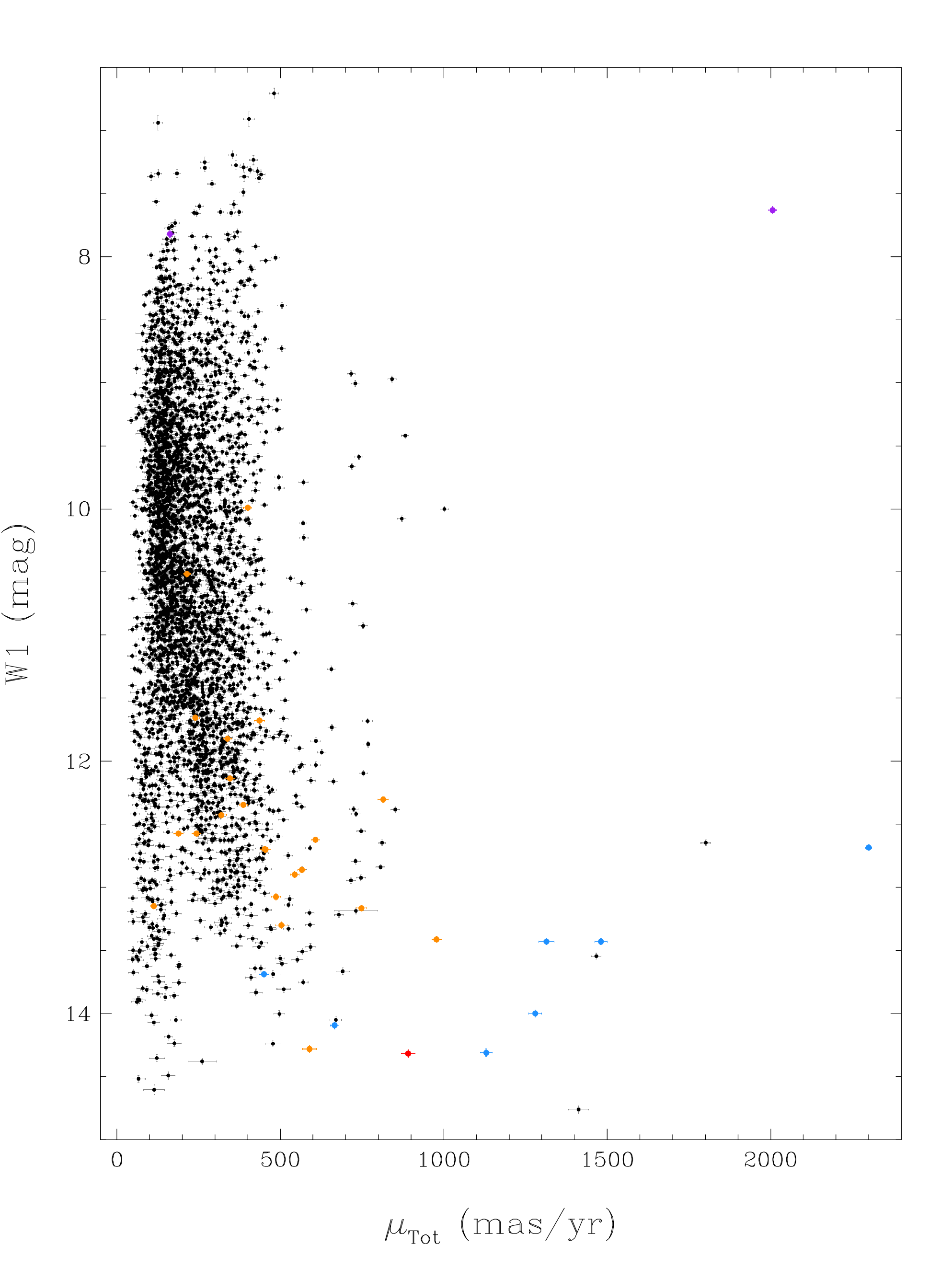}
\caption{AllWISE W1 profile-fit photometry plotted as a function of the total 2MASS-to-WISE proper motion for 
objects in Table 3. 
Objects with spectroscopic follow-up in Table 5 are color coded as follows: 
Light blue --early-L 
subdwarfs; red -- late-L subdwarfs; purple -- bright M dwarfs discussed in section 8; dark orange -- all others.
\label{w1_vs_totalmotion}}
\end{figure}

Figure~\ref{w1_vs_totalmotion} shows the AllWISE W1 profile-fit magnitude of each source plotted against the 
total motion computed from the 2MASS-to-WISE time baseline. As can be seen on the figure, ten of these objects 
have motions $\ge1000$ mas 
yr$^{-1}$, including the three highest movers WISEA J204027.30+695924.1 ($\mu=2300\pm10$ mas yr$^{-1}$; see 
section 7), WISEA J154045.67$-$510139.3 ($\mu=2006\pm12$ mas yr$^{-1}$; see 
section 8), 
and WISEA J070720.50+170532.7 ($\mu=1802\pm15$ mas yr$^{-1}$; \citealt{wright2013}).

\notetoeditor{I could not get the tag ``section~\ref{two_nearby_dwarfs}'' to work above, so I typed the 
section number above by hand.}

\begin{figure}
\figurenum{18}
\includegraphics[scale=0.45,angle=0]{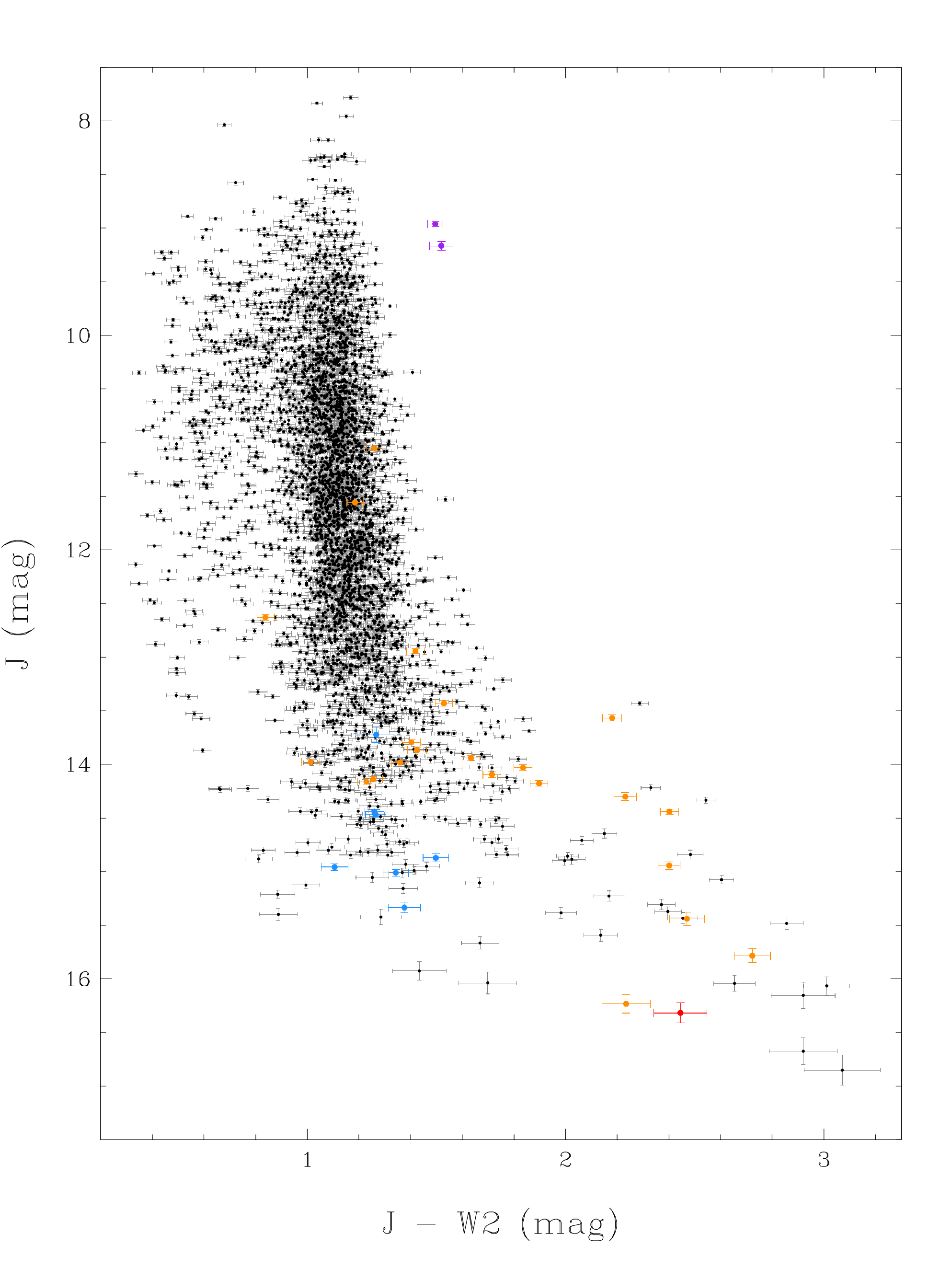}
\caption{2MASS $J$-band magnitude plotted against the $J$ $-$ W2 color for the motion objects in Table 3.
Color coding is the same as in Figure 17.
\label{J_vs_JW2}}
\end{figure}

\begin{figure}
\figurenum{19}
\includegraphics[scale=0.45,angle=0]{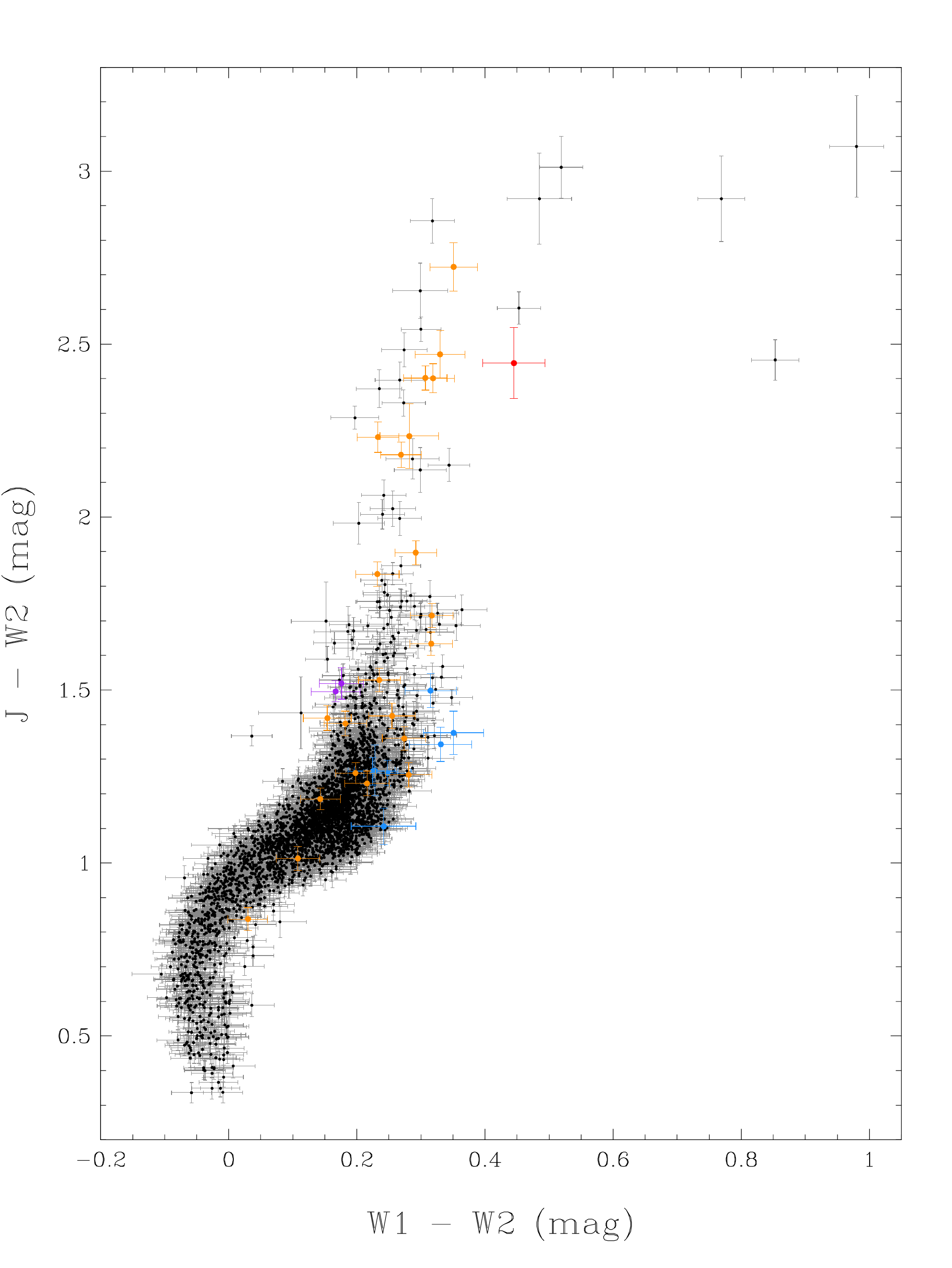}
\caption{$J$ $-$ W2 color versus W1 $-$ W2 color for the motion objects in Table 3. 
Color coding is the same as in Figure 17.
\label{W1W2_vs_JW2}}
\end{figure}

Figure~\ref{J_vs_JW2} shows the 2MASS $J$-band magnitude plotted as a function of the $J$ $-$ W2 color. As shown 
in Figure 7 of \cite{kirkpatrick2011}, L and T dwarfs generally have colors of $J$ $-$ W2 $>$ 2.0 mag, so the 
list of AllWISE motion objects contains approximately 30 new L and T dwarfs. Figure~\ref{W1W2_vs_JW2} (which can 
be compared to Figure 9 in \citealt{kirkpatrick2011}) breaks the degeneracy between the L and T dwarf populations. 
Figure 1 of \cite{kirkpatrick2011} shows that the break between the L and T dwarf populations is near W1$-$W2 = 
0.6 mag, which means that of the $\sim$30 new L and T dwarfs, only three are red enough to be T dwarfs themselves. 
These objects are WISEA J011154.36$-$505343.4, WISEA J210529.11$-$623559.3, and WISEA J212100.87$-$623921.6, all 
of which lack previous publication, likely because their very southern declinations have made follow-up somewhat 
more difficult. 

None of these objects is nearly red enough to fall in the Y dwarf regime, which falls very roughly at W1$-$W2 $>$ 
3.5 mag. It is natural, then, to ask why the AllWISE motion survey wasn't more efficient at picking up T and Y 
dwarfs. There are two reasons. First, the apparent magnitudes of the nearest T and Y dwarfs are sufficiently faint 
that AllWISE would only have been able to measure significant motions for the very fastest movers (see 
Figure~\ref{smallest_motions}). Second, several bright, very high motion T and Y dwarfs {\it were} picked up by 
our criteria, but these are not included in Table 3 %~\ref{discoveries} 
because they were previously known. In fact, 
it is hard to imagine a bright, nearby T or Y dwarf that would not have been picked up via color selection because, 
after all, WISE was designed to be highly sensitive to the methane absorption that defines these spectral types. 
With longer time baselines, WISE-like data would be able to probe deeper magnitude limits and smaller motions, 
enabling the discovery of fainter and/or colder objects than color selection itself can provide. Although it is 
possible that AllWISE has identified a moving W2-only source with a W1 $-$ W2 color limit sufficiently marginal 
that color-based criteria would not have selected it, our current motion-based criteria did not extract it. Modifying 
the selection criteria will be further discussed in section 5.

\begin{figure}
\figurenum{20}
\includegraphics[scale=0.45,angle=0]{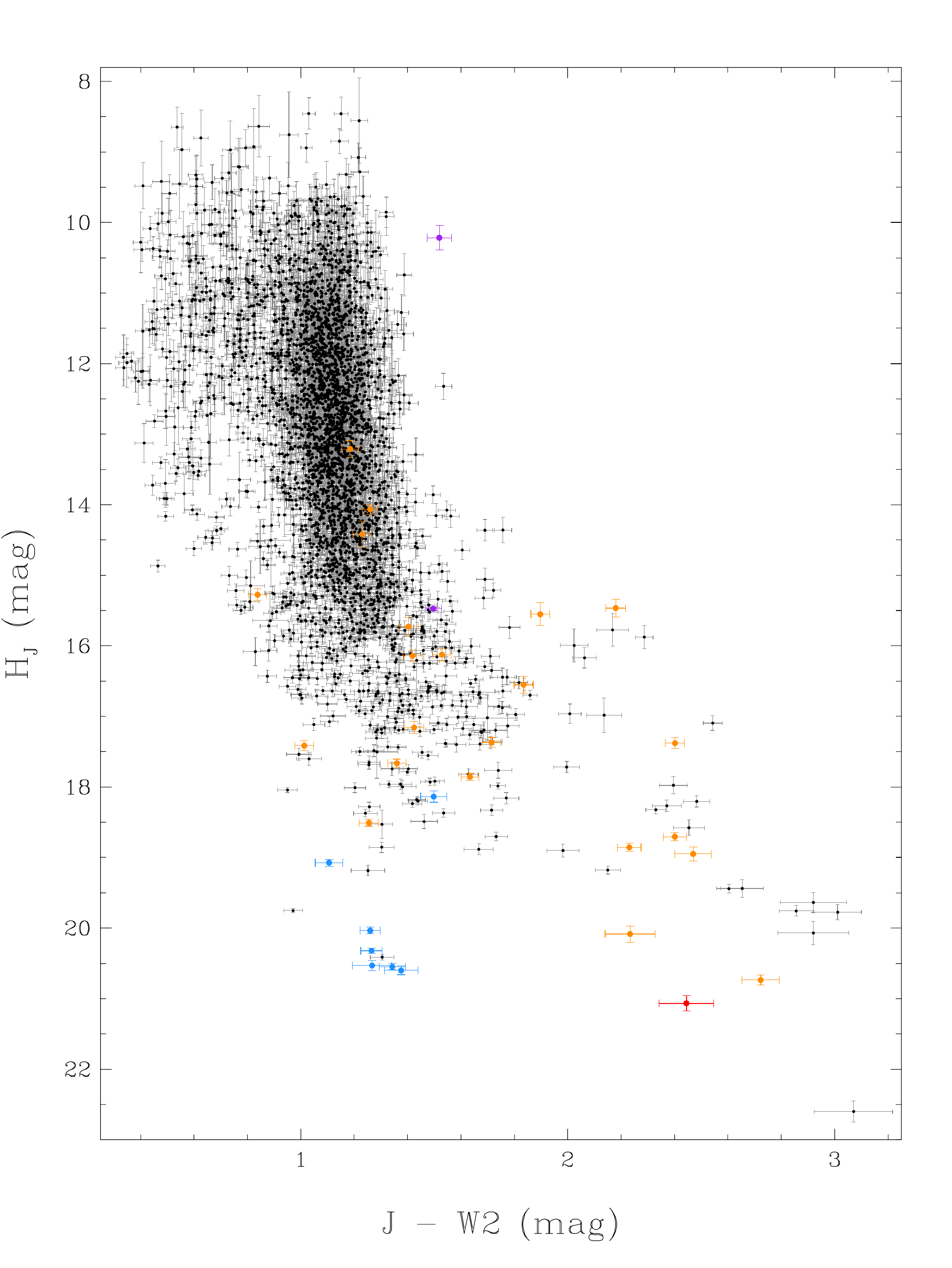}
\caption{Reduced proper motion at $J$-band plotted against the $J$ $-$ W2 color for the motion objects in Table 3.
Color coding is the same as in Figure 17.
\label{reducedPMJ_vs_JW2}}
\end{figure}

Finally, we present two plots that help to isolate potentially low-metallicity objects. 
Figure~\ref{reducedPMJ_vs_JW2} shows the reduced proper motion at $J$-band plotted against the $J$ $-$ W2 color. 
The reduced proper motion can be thought of as a poor man's substitute for an absolute magnitude measurement when 
a parallax is lacking. In the distance modulus equation, the total proper motion is used in place of the parallax 
(this is done because, to first order, the higher an object's proper motion, the larger its parallax is likely to 
be) and the ``absolute magnitude'' that results is called the reduced proper motion. At $J$-band this would be 
written as $H_J = J + 5\log(\mu) + 5$, where $\mu$ is expressed in arcsec yr$^{-1}$ and the reduced proper motion is in 
magnitudes. Objects with abnormally high space velocities confound this logic, however -- they have high motions 
not because they are close but because they are old. As a result, these objects have reduced motion values that 
are significantly fainter than objects of average kinematics. 
See \cite{reid1997} for a more extensive discussion.
Fortunately, this enables such objects (which tend 
to be low-metallicity because they are very old) to be easily selected from the bulk of objects (of near solar 
metallicity) in the Galactic disk. Several objects with reduced proper motion at $J$-band $>$ 19 mag and $0.8 < 
J -$ W2 $< 1.5$ mag can be seen in Figure~\ref{reducedPMJ_vs_JW2} and are prime low-metallicity candidates.

\begin{figure}
\figurenum{21}
\includegraphics[scale=0.35,angle=270]{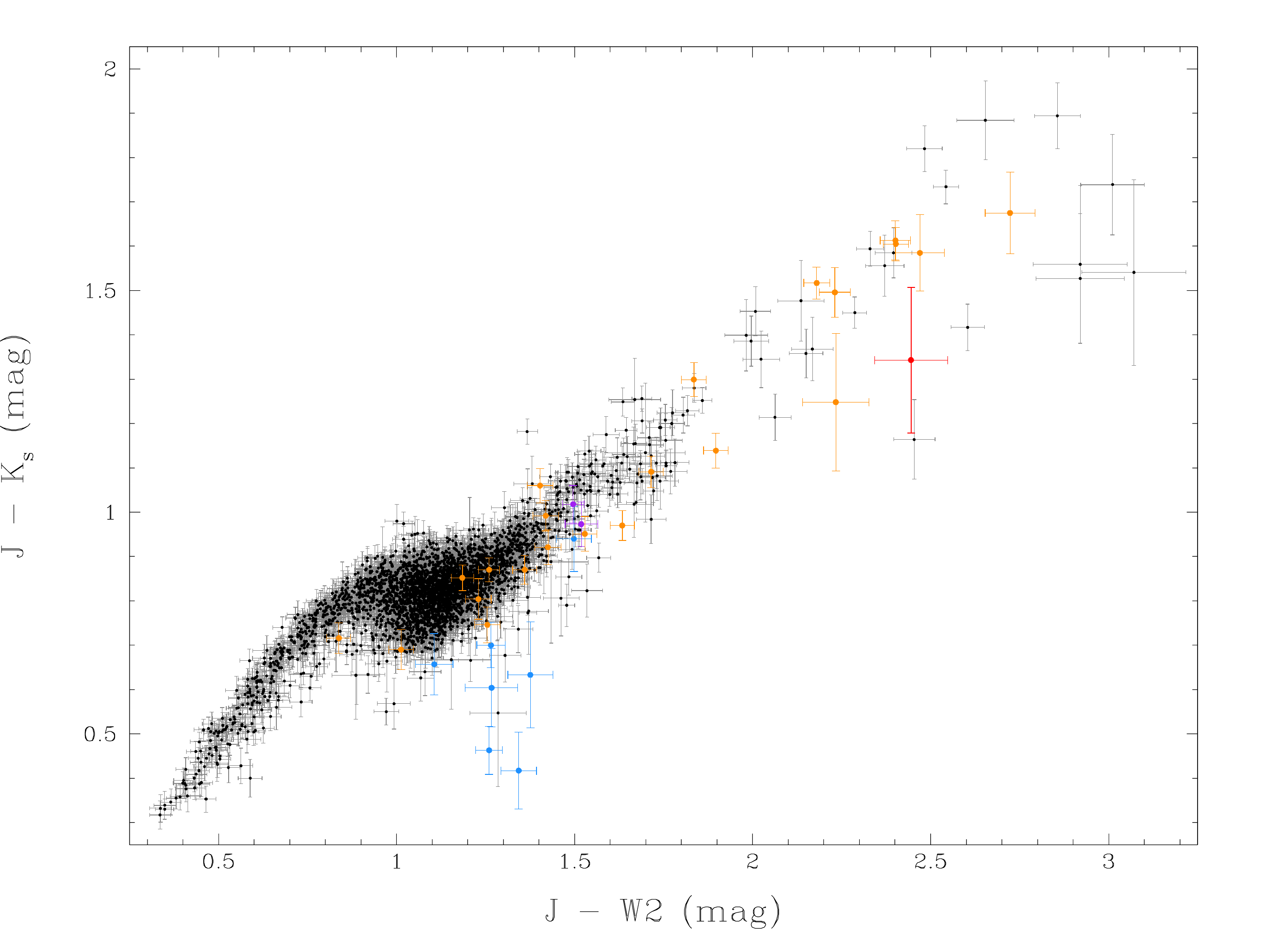}
\caption{$J-K_s$ color plotted against the $J$ $-$ W2 color for the motion objects in Table 3.
Color coding is the same as in Figure 17.
\label{JKs_vs_JW2}}
\end{figure}

Figure~\ref{JKs_vs_JW2} shows the $J-K_s$ color plotted against the $J$ $-$ W2 color. The locus at lower left near 
(0.3, 0.3) corresponds roughly to late-F or early-G spectral types (\citealt{ali2014}) and marks the earliest 
spectral types for which we have new motions. The bulk of objects clustered near (1.1, 0.8) are mid-M dwarfs, 
which by far dominate the Solar Neighborhood (Figure 11 of \citealt{kirkpatrick2012}). The trail of objects into 
the upper right corner of the figure is the $\sim$30 L and T dwarfs. This diagram also shows a collection of objects 
that fall below the standard main sequence. Qualitatively, if we take the long-baseline color, $J$ $-$ W2, to be 
indicative of temperature or spectral type, and the short-baseline color, $J - K_s$ to sample more specific 
atmospheric physics, we would expect objects at the same $J$ $-$ W2 value as a standard main sequence star but 
with a bluer $J - Ks$ value to be low-metallicity because of the increased contribution of collision-induced 
absorption by H$_2$ at $K$ band. Hence, this diagram provides another handy way -- using colors instead of motions, 
as in Figure~\ref{reducedPMJ_vs_JW2} -- to select low-metallicity candidates. 

\begin{turnpage}
\begin{center}
\begin{deluxetable*}{lcccccl}
\tabletypesize{\small}
\tablenum{5}
\tablecaption{Spectroscopic Follow-up of Motion Candidates\label{spectra}}
\tablehead{
\colhead{Object Name} &                          
\colhead{Opt} &  
\colhead{NIR} &  
\colhead{Telescope/} &     
\colhead{Obs.\ Date} &
\colhead{Exp.\ Time\tablenotemark{a}} &
\colhead{Corrector} \\
\colhead{} &                          
\colhead{Sp.\ Type} &  
\colhead{Sp.\ Type} &  
\colhead{Instrument} &     
\colhead{(UT)} &
\colhead{(s)} &
\colhead{Star} \\
\colhead{(1)} &                          
\colhead{(2)} &  
\colhead{(3)} &  
\colhead{(4)} &     
\colhead{(5)} &
\colhead{(6)} &
\colhead{(7)}
}
\startdata
WISEA J001450.17$-$083823.4    & sdL0  & ---         & Keck/DEIMOS   &    11 Dec 2013 & 600        & --- \\
WISEA J004326.26+222124.0      & ---   & sdL1        & IRTF/SpeX     &    24 Oct 2013 & 960        & HD 7215 \\
WISEA J004713.81$-$371033.7    & L4:   & ---         & Keck/DEIMOS   &    11 Dec 2013 & 600        & --- \\
WISEA J005757.64+201304.0      & sdL7  & ---         & Keck/LRIS     &    05 Oct 2013 & 1167/1130  & --- \\
...                            & ---   & sdL7        & IRTF/SpeX     &    24 Nov 2013 & 1800       & HD 16811 \\
WISEA J020201.25$-$313645.2    & sdL0  & ---         & Keck/LRIS     &    05 Oct 2013 & 600/600    & --- \\
WISEA J023421.83+601227.3      & M9.5e & ---         & Palomar/DSpec &    07 Jul 2013 & 2900       & --- \\
WISEA J030601.66$-$033059.0    & sdL0  & ---         & Keck/DEIMOS   &    11 Dec 2013 & 300        & --- \\  
...                            & ---   & sdL0        & IRTF/SpeX     &    22 Nov 2013 & 720        & HD 13936 \\
WISEA J040418.01+412735.5      & ---   & L3 pec (red)& IRTF/SpeX     &    24 Oct 2013 & 720        & HD 21038 \\
WISEA J043535.82+211508.9      & sdL0  & ---         & Keck/LRIS     &    04 Oct 2013 & 300/300    & --- \\
...                            & ---   & sdL0        & IRTF/SpeX     &    24 Oct 2013 & 960        & HD 27761 \\
...                            & ---   & sdL0        & Keck/NIRSPEC  &    14 Dec 2013 & 2400       & HD 35036 \\
WISEA J045921.21+154059.2      & ---   & sdL0        & IRTF/SpeX     &    22 Nov 2013 & 360        & HD 35036 \\
WISEA J053257.29+041842.5      & ---   & L3          & Keck/NIRSPEC  &    20 Nov 2013 & 720        & HD 39953 \\
WISEA J054318.95+642250.2      & ---   & L2          & Keck/NIRSPEC  &    14 Dec 2013 & 1440       & HD 33654 \\
WISEA J055007.94+161051.9      & ---   & L2          & IRTF/SpeX     &    24 Nov 2013 & 720        & HD 43583 \\
WISEA J060742.13+455037.0      & ---   & L2.5        & Keck/NIRSPEC  &    14 Dec 2013 & 1800       & HD 45105 \\
...                            & ---   & L2.5        & IRTF/SpeX     &    15 Dec 2013 & 960        & HD 45105 \\
WISEA J063957.71$-$034607.2    & sdM0.5& ---         & Keck/LRIS     &    04 Oct 2013 & 300/300    & --- \\
WISEA J065958.55+171710.9      & ---   & L2          & Keck/NIRSPEC  &    20 Nov 2013 & 960        & HD 39953 \\
WISEA J071552.38$-$114532.9    & ---   & L5 pec?     & Keck/NIRSPEC  &    20 Nov 2013 & 960        & HD 43607 \\
...                            & ---   & L4 pec (blue)& IRTF/SpeX    &    14 Dec 2013 & 960        & HD 56525 \\
WISEA J072003.20$-$084651.3    & ---   & M9          & IRTF/SpeX     &    14 Dec 2013 & 960        & HD 56525 \\
WISEA J074211.69$-$121151.6    & ---   & $<$M0 (wd?) & IRTF/SpeX     &    21 Nov 2013 & 1800       & HD 67725 \\
WISEA J082640.45$-$164031.8    & ---   & L9          & Keck/NIRSPEC  &    14 Dec 2013 & 2400       & HD 72282 \\
WISEA J085224.36+513925.5      & ---   & M7          & IRTF/SpeX     &    15 Dec 2013 & 960        & HD 45105 \\
WISEA J091657.18$-$112104.7    & ---   & M9          & IRTF/SpeX     &    15 Dec 2013 & 960        & HD 45105 \\
WISEA J102304.04+155616.4      & ---   & M8 pec      & IRTF/SpeX     &    14 Dec 2013 & 960        & HD 89239 \\
WISEA J154045.67$-$510139.3    & M6    & ---         & Magellan/IMACS&    27 Jan 2014 & 100        & --- \\
WISEA J162702.28$-$694411.8    & M4+M4 & ---         & Magellan/IMACS&    28 Jan 2014 & 300        & --- \\
WISEA J163605.71$-$044013.8    & ---   & M4.5        & IRTF/SpeX     &    14 Aug 2013 & 120        & HD 159008 \\
WISEA J182121.91$-$070008.6    & M7 pec& ---         & Keck/LRIS     &    04 Oct 2013 & 400/350    & ---\\
WISEA J204027.30+695924.1      & sdL0  & ---         & Keck/LRIS     &    05 Oct 2013 & 300/300    & --- \\
WISEA J204218.13$-$082137.8    & ---   & M7          & IRTF/SpeX     &    24 Oct 2013 & 960        & HD 193689 \\
WISEA J211543.59$-$322540.4    & ---   & M6:         & IRTF/SpeX     &    14 Aug 2013 & 960        & HD 199090 \\
WISEA J222013.75$-$361709.5    & ---   & M8 pec      & IRTF/SpeX     &    24 Oct 2013 & 960        & HD 194272 \\
WISEA J224128.33+043459.3      & ---   & M7.5        & IRTF/SpeX     &    24 Oct 2013 & 960        & HD 210501 \\
WISEA J232036.88+315739.5      & ---   & M4.5        & IRTF/SpeX     &    24 Oct 2013 & 960        & HD 210290 \\
WISEA J235459.79$-$185222.4    & ---   & L2          & Keck/NIRSPEC  &    14 Dec 2013 & 1800       & HD 219833 \\
\enddata
\tablenotetext{a}{For Keck/LRIS, the two exposure times refer to the blue-side spectrograph (left) and the red-side 
spectrograph (right).}
\end{deluxetable*}
\end{center}
\end{turnpage}

\section{Comparison to the Luhman (2014) Motion List\label{improved_criteria}}

While this paper was being finalized, \cite{luhman2014} published an independently selected set of motion objects 
uncovered in the same WISE data set. Rather than using AllWISE results, his methodology started with the publicly 
available single exposure catalogs from each phase of the mission. The ecliptic polar regions were ignored (to save 
on processing time) and the rest of the WISE data divided into discrete epochs encompassing the twelve or more 
revisits of the same patch of sky with an overall time difference generally less than one day. The mean coordinates 
of the group at a single epoch were computed, and these were compared to other groups at other epochs as long as 
two groups were within $1\farcs5$ of one another. Motions were thus computed for these groups paired across epochs, 
and groups for which no pairings were found were examined separately in case they corresponded to very high-motion 
objects. The \cite{luhman2014} list contains 762 objects, and its sky distribution (lower panel of
Figure 15) is similar to ours, although the source density is considerably lower. Of those 762 objects,
321 are also included in our 
Table 3, one (the unconfirmed object WISEA J085510.74$-$071442.5) is included in Table 4, and another 
(WISEA J104915.52$-$531906.1) was eliminated from consideration because it was
published earlier by \cite{luhman2013}.  

The remaining 439 objects enable us to determine which of our selection criteria were set 
too tightly, so that future researchers can perform their own, refined searches.
Some sources fail more than one of the criteria discussed in section~\ref{selection_criteria}. Specifically, 
a total of 80\% 
(351/439) of the objects were missed because the $rchi2/rchi2\_pm$ ratio failed to exceed a threshold of 
1.03. That is, the reduced $\chi^2$ from the motion fit was not markedly better than that from the stationary fit. 
In most cases, the reduced $\chi^2$ was improved, but this improvement was less than 3\%. 

The next most problematic criteria were the restrictions that $w?rchi2$ be less than 2.0 to ensure a point-like 
source. These failed in 39\% (172/439) of the cases. In a similar vein, 13\% (56/439)
of the sources failed the $nb = 1$ 
criterion to assure that the source was not blended. In most of these cases, a check of the WISE images confirms 
that the motion source from \cite{luhman2014} is blended with another source and/or appears slightly elongated, 
but AllWISE processing was nonetheless able to handle these cases. (It should be expected, however, that there 
are plenty of other cases of blending and elongation that are not so elegantly handled.) 
Finally, 15\% (64/439) of the sources failed the $Q$ criterion, 2.5\% (11/439) failed the $cc\_flags$ criterion, 
1.4\% (6/439) failed the $pmcode$ criterion, 1.4\% (6/439) were missed during by-eye Quality Assurance reviewing 
(one of these was caused by a poor stretch on the images used
to scrutinize candidates),
0.7\% (3/439) were lost due to transcription errors, and 0.2\% (1/439) is a previously published source
that was removed from our list because it was not a new discovery.

Figure 22 compares the \cite{luhman2014} motion objects (760 total after removing WISEA J085510.74$-$071442.5 
and WISEA J104915.52$-$531906.1) to our list of 3,525 objects with confirmed motions (Table 3). The plot shows histograms
in W1 magnitude and in total proper motion computed from the 2MASS-to-WISE time baseline. Our
survey tends to sample brighter objects than the \cite{luhman2014} catalog and,
as a result, also probes to considerably smaller motion. Both surveys are able to select high motion objects,
although there are somewhat fewer objects with $\mu > 750$ mas yr$^{_1}$ in the \cite{luhman2014} list than in
ours.

\begin{figure*}
\figurenum{22}
\includegraphics[scale=0.65,angle=270]{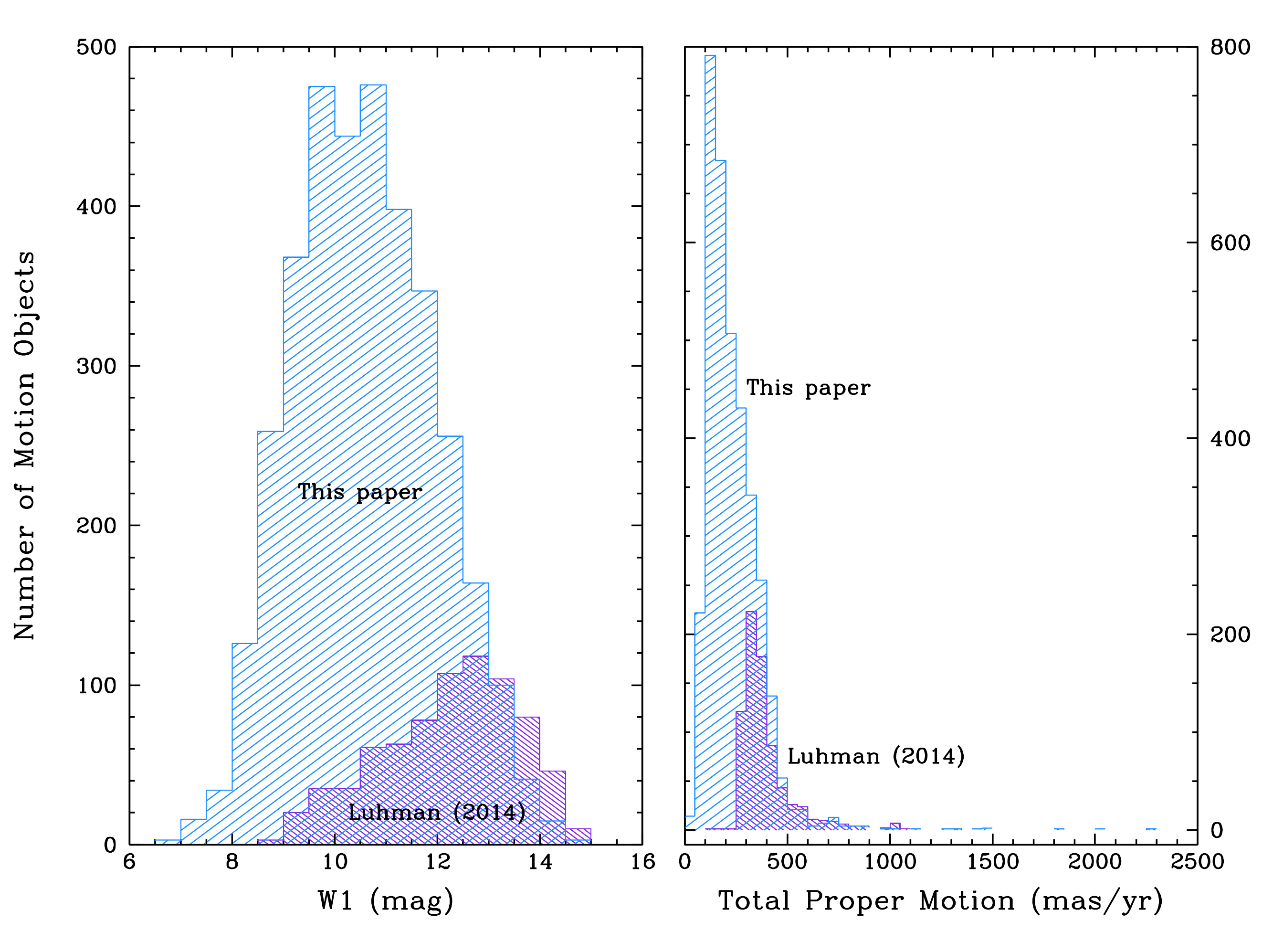}
\caption{Histograms of the W1 mag (left panel) and total 2MASS-to-WISE proper motion (right panel) for our 
survey (light blue; 3,525 objects from Table 3) compared to that of \cite{luhman2014} (violet; 760 objects total).
\label{catalog_histograms}}
\end{figure*}

Our criteria missed objects found by \cite{luhman2014}, but \cite{luhman2014} also missed objects found by us. 
Our second-highest motion object, WISEA J154045.67$-$510139.3, is one notable example of the latter case. Clearly, the AllWISE data 
products hold many other potential discoveries. Moreover, with the WISE satellite having been reactivated in 2013 
December for an additional three years of data taking, extending this all-sky motion survey to even smaller motion 
limits via the extended time baseline will also be possible. Such a survey would allow researchers to expand the 
volume and spectral type range over which WISE motion discoveries can be made.

\section{Spectroscopic Follow-up of Selected Discoveries}

We have selected objects for spectroscopic observation as follows. Figure 17 was used primarily to select objects of
highest motion, since these tend to be either close or low-metallicity. Figure 18 was used to look for objects that
were bright in $J$-band magnitude relative to their peers in $J -$ W2 color, since such objects are generally the
closest objects of their type. Figure 19 was used to select late-type objects, although all of the T dwarf candidates
were at unfavorable sky positions for the follow-up obtained for this paper. Figures 20 and 21 were used to select
low-metallicity objects since, as explained earlier, such objects tend to fall in the lower parts of these diagrams,
thereby distinguishing themselves from the bulk of objects, which have solar metallicities. Objects for which we have obtained
spectroscopic follow-up are color coded in Figures 17 through 21, as described in the legend to Figure 17. These
follow-up observations are detailed below and summarized in Table 5.

\subsection{Optical Spectroscopy}

One source was observed (by SFA) on the night of 2013 July 07 UT with the Double Spectrograph on the Hale 5m 
telescope on Palomar Mountain, California. In the red-side spectrograph, a 316 lines mm$^{-1}$ grating blazed at 
7500 \AA\ was used to produce a spectrum covering the range 3800 to 10500 \AA. Standard reduction procedures for 
CCD data were employed. This night had cirrus clouds.

Six sources were observed (by DS, MB, and GBL) on the nights of 2013 October 04-05 UT with the Low Resolution 
Imaging Spectrometer (LRIS, \citealt{oke1995}) at the 10m W.\ M.\ Keck Observatory on Mauna Kea, Hawai'i. The blue 
side was used with a 600 lines mm$^{-1}$ grating blazed at 4000 \AA, and the red side was used with a 400 lines 
mm$^{-1}$ grating blazed at 8500 \AA. Objects had very little flux in the blue side so the blue-side spectra are 
not considered further. 
The red side produced a spectrum covering the range from 5500 to 10000 \AA. Standard reduction procedures for CCD 
data were employed. The first night had clouds and the second night was clear.

Three sources were observed (by DS) on the clear night of 2013 December 11 UT with the Deep Imaging 
Multi-object Spectrograph (DEIMOS; \citealt{faber2003}) at the 10m W.\ M.\ Keck Observatory on Mauna Kea, Hawai'i. 
The instrument was used in single-object mode utilizing the 1200 line/mm grating blazed at 7500 \AA\ to provide 
continuous wavelength coverage from 6650 to 9300 \AA. Standard reduction procedures for CCD data were employed.

Two objects were observed (by JAR) on the clear nights of 2014 January 27 and 28 UT with the Inamori-Magellan Areal
Camera and Spectrgraph (IMACS; \citealt{dressler2011}) on the 6.5m Walter Baade Telescope at Las Campanas Observatory, Chile.
The f/4 camera was used with a Bessel $V$-band or a CTIO $I$-band filter for imaging observations and with a 600 lines 
mm$^{-1}$ grating for the spectroscopic observations. The grating angle was set for a wavelength coverage of 7100 to 
104000 \AA. Standard reduction procedures 
for CCD data were again employed.

\subsection{Near-infrared Spectroscopy}

Several sources were observed with SpeX (\citealt{rayner2003}) on the NASA 3m Infrared Telescope Facility (IRTF) 
on Mauna Kea, Hawai'i. The UT dates of observation were 2013 August 14 (by AJB; cirrus and patchy clouds), 2013 
October 24 (by AS; clear skies), 21-22 and 24 November 2013 (by JKF and NS; thick cirrus on the first night but 
then clear), and 14-15 December 2013 (by AS; clear). For all but the October night, SpeX was used in prism mode 
with a 0$\farcs$5 or 0$\farcs$8 wide slit to achieve a resolving power of $R\equiv\lambda / \Delta \lambda 
\approx 100-150$ over 
the range 0.8-2.5 $\mu$m. For the October run, SpeX was used in cross-dispersed mode to produce spectra over the 
range 0.9-2.4 $\mu$m with a resolving power of $R\equiv\lambda / \Delta \lambda \approx 1200$. All data were 
reduced using Spextool (\citealt{cushing2004}); A0 stars were used for the telluric correction and flux calibration 
steps following the technique described in \cite{vacca2003}. 

Several sources were also observed on the UT nights of 2013 November 20 and 2013 December 14 (by GNM and SEL) 
with the Near-Infrared Spectrometer (NIRSPEC, \citealt{mclean1998,mclean2000}) at the 10m W.\ M.\ Keck Observatory 
on Mauna Kea, Hawai'i. There was cirrus on the November night and heavier clouds on the December one. In 
low-resolution mode, use of the 42\arcsec$\times$0${\farcs}$38 slit results in a resolving power of 
R~$\equiv~{\lambda}/{\Delta}{\lambda}~{\approx}~2500$. Our brown dwarf candidates were observed in the N3 
configuration (see \citealt{mclean2003}) that covers part of the $J$-band window from 1.15 to 1.35 $\mu$m. Data
 were reduced using the REDSPEC package, as described in \cite{mclean2003}.  

\subsection{Spectral Classification}

We have typed the optical spectra as follows. Each spectrum was normalized 
at 7500 \AA\ or 8250 \AA\ and overplotted on a suite of like-normalized LRIS spectra of primary M and L optical spectral standards 
from \cite{kirkpatrick1991} and \cite{kirkpatrick1999}. These plots were examined by eye to determine the best match 
and to look for any peculiarities with respect to the standard sequence. Objects falling midway between integral 
classes (such as M9 and L0) were assigned the half class in between (in this case, M9.5). Objects showing notable 
peculiarities were given a suffix of ``pec'' unless the peculiarities were determined to be caused by low-metallicity, 
in which case these were reclassified against published optical spectra typed as late-M and early-L subdwarfs and 
given a prefix of ``sd''.

\begin{figure}
\figurenum{23}
\includegraphics[scale=0.45,angle=0]{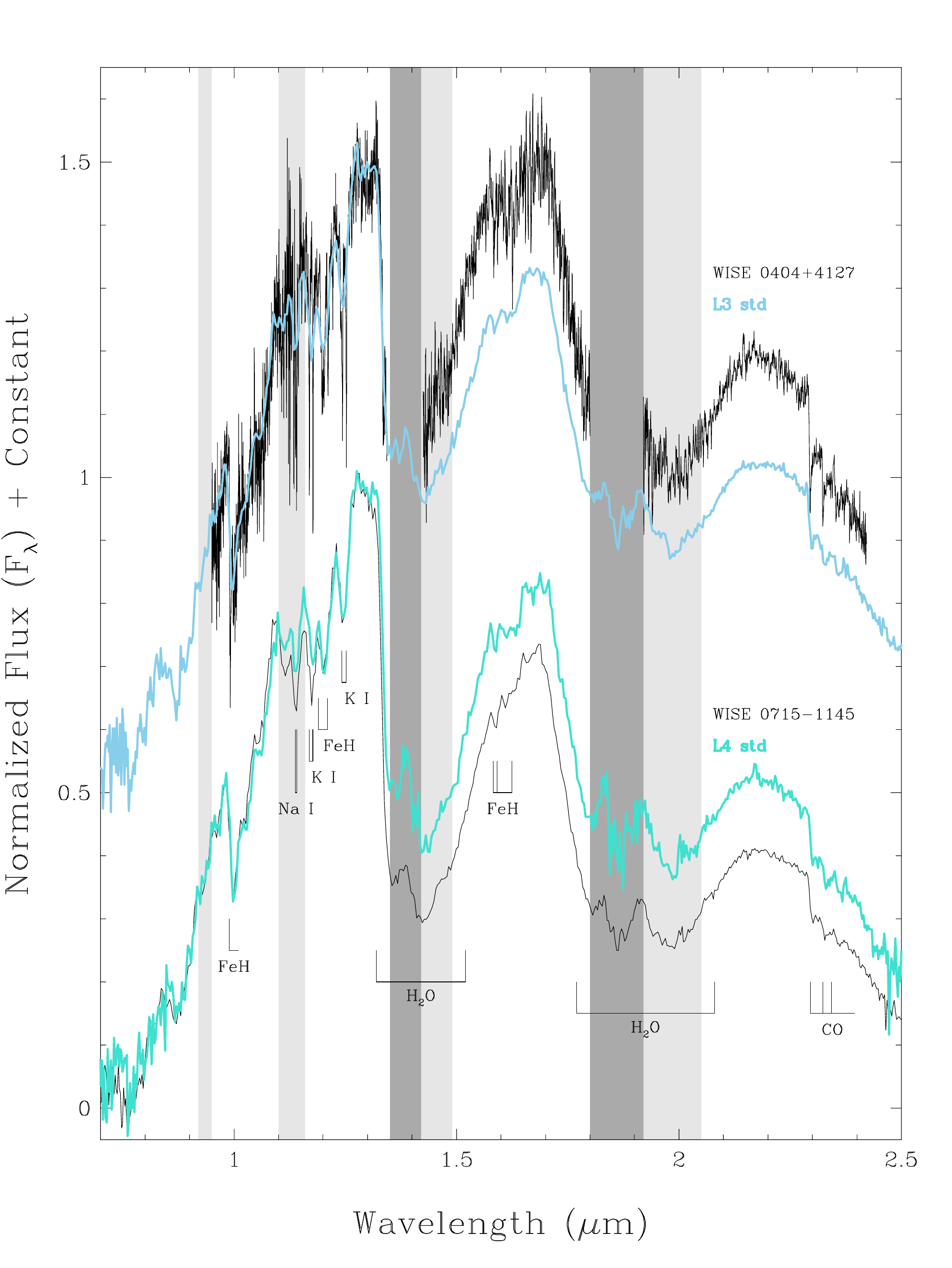}
\caption{Near-infrared spectra of WISEA J040418.01+412735.5 (discovered earlier by \citealt{castro2013} and hence 
removed from our discovery list in Table 3) %~\ref{discoveries}) 
and WISEA J071552.38$-$114532.9 overplotted with the 
near-infrared spectral standard (\citealt{kirkpatrick2010}) providing the best match throughout $J$-band. Note that 
the $H$- and $K$-band spectra of WISEA J040418.01+412735.5 are much redder than the standard, and that of WISEA 
J071552.38$-$114532.9 is much bluer than the standard. Prominent spectral 
features are marked. Per \cite{rayner2009}, regions of telluric absorption are 
marked by the dark grey (atmospheric transmission $<$20\%) and light grey (20\% $<$ atmospheric transmission 
$<$ 80\%) zones in wavelength. Spectra have been normalized at 1.28 $\mu$m and a constant offset added to 
the flux to separate the spectra vertically except where overplotting was intended.
\label{allwise_redblue_Ls}}
\end{figure}

We have typed the near-infrared spectra as follows. Each target spectrum was normalized 
to one at 1.28 $\mu$m and compared to the near-infrared spectral standards from \cite{kirkpatrick2010} normalized 
the same way. Using the methodology outlined in that paper, the core near-infrared type was determined {\it only} 
from the 0.9-1.4 $\mu$m portion, and then the corresponding goodness of fit to the same spectral standard from 
1.4-2.5 $\mu$m was judged. In most cases the same spectral standard also provided the best fit in this 
longer-wavelength region. In other cases, the target spectrum was notably much bluer or redder so the fit across 
the $H$ and $K$ windows was very poor despite the excellent fit in the $J$ window. These peculiar objects were 
given suffixes of ``pec (blue)'' or ``pec (red)'' to denote the slope of the spectrum relative to the standard 
(see Figure~\ref{allwise_redblue_Ls}). Some of the peculiar spectra were deemed to be low-metallicity (prefix of 
``sd'') and were typed against near-infrared spectra of similar subdwarfs from the published literature. 

Optical and/or near-infrared spectral types are listed in Table 5. 
Subdwarfs are discussed in section 7, and two bright M dwarf systems are discussed in section 8.

\section{A Trove of L Subdwarf Discoveries\label{L_subdwarfs}}

A number of discoveries have the photometric characteristics of subdwarfs (i.e., low-metallicity objects) and were 
confirmed as such by our follow-up spectroscopy. Two of these are early-M subdwarfs -- the sdM0.5 WISEA 
J063957.71$-$034607.2 and the usdM3 discovery by \cite{wright2013} WISEA J070720.50+170532.7. Six others fall near 
late-M/early-L and another, WISEA J005757.64+201304.0, falls at late-L.

\begin{figure}
\figurenum{24}
\includegraphics[scale=0.45,angle=0]{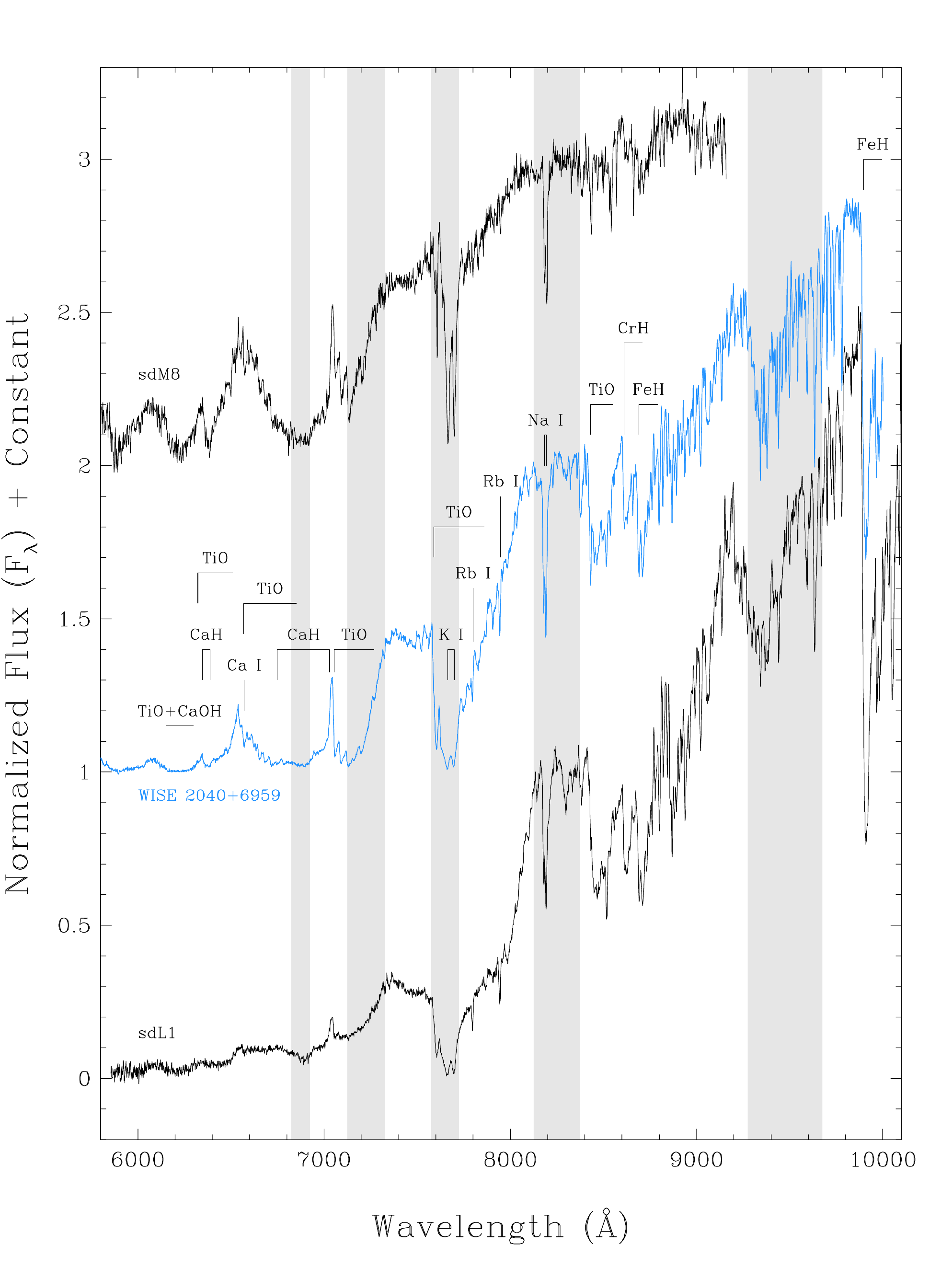}
\caption{Optical spectra of the sdM8 LSPM J1425+7102 (\citealt{lepine2007}) and the sdL1 2MASS 17561080+2815238 
(\citealt{kirkpatrick2010}) compared to the optical spectrum of WISEA J204027.30+695924.1. Prominent spectral 
features are marked. Spectra have been normalized at 8250 \AA\ and a constant 
offset added to the flux to separate the spectra vertically. Regions of telluric absorption are marked by the light 
grey bands.
\label{allwise_optical_sdL_features}}
\end{figure}

\begin{figure}
\figurenum{25}
\includegraphics[scale=0.45,angle=0]{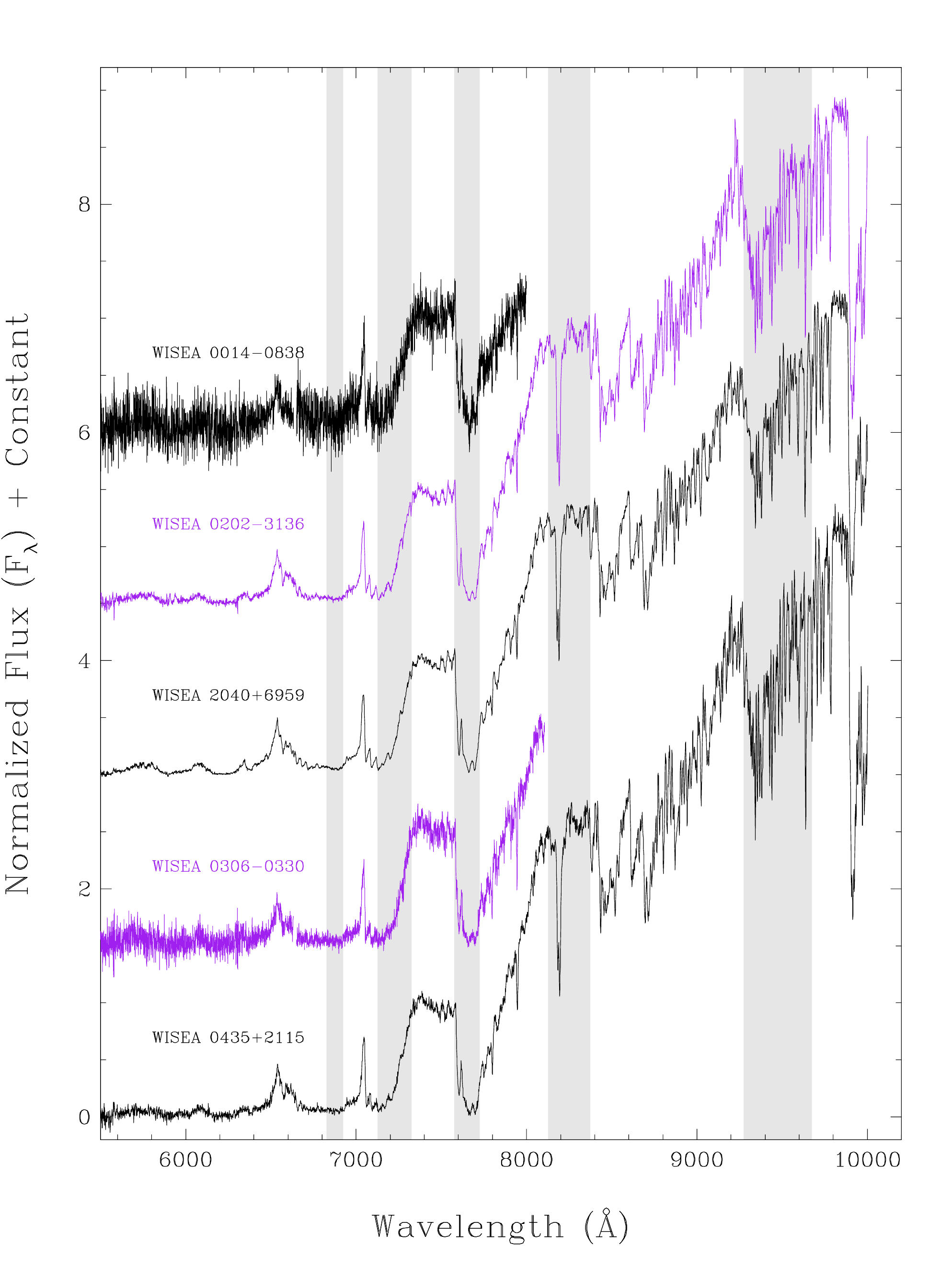}
\caption{Optical spectra of five early-L subdwarf discoveries from AllWISE: WISEA J001450.17-083823.4, WISEA 
J020201.25-313645.2, WISEA J204027.30+695924.1, WISEA J030601.66-033059.0, and WISEA J043535.82+211508.9. Spectra 
have been normalized at 7500 \AA\ and a constant offset added to the flux to separate the spectra vertically. 
Regions of telluric absorption are marked by the light grey bands.
\label{allwise_new_sdLs_optical}}
\end{figure}

One of the subdwarfs near the M/L transition is our highest motion discovery, WISEA J204027.30+695924.1. As 
Figure~\ref{allwise_optical_sdL_features} shows, the optical spectrum of this object is intermediate in type 
between an sdM8 and an sdL1. One of the most distinctive features in the optical spectra of subdwarfs at the M/L 
boundary is the emission-like feature near 7050 \AA\ that is, in fact, a narrow wavelength region relatively free 
of opacity between the strong CaH band to the blue and the strong TiO band to the red. Optical spectra of four 
other AllWISE discoveries (Figure~\ref{allwise_new_sdLs_optical}) show this same telltale signature and have 
overall spectral morphologies very similar to WISEA J204027.30+695924.1. 

\begin{figure}
\figurenum{26}
\includegraphics[scale=0.45,angle=0]{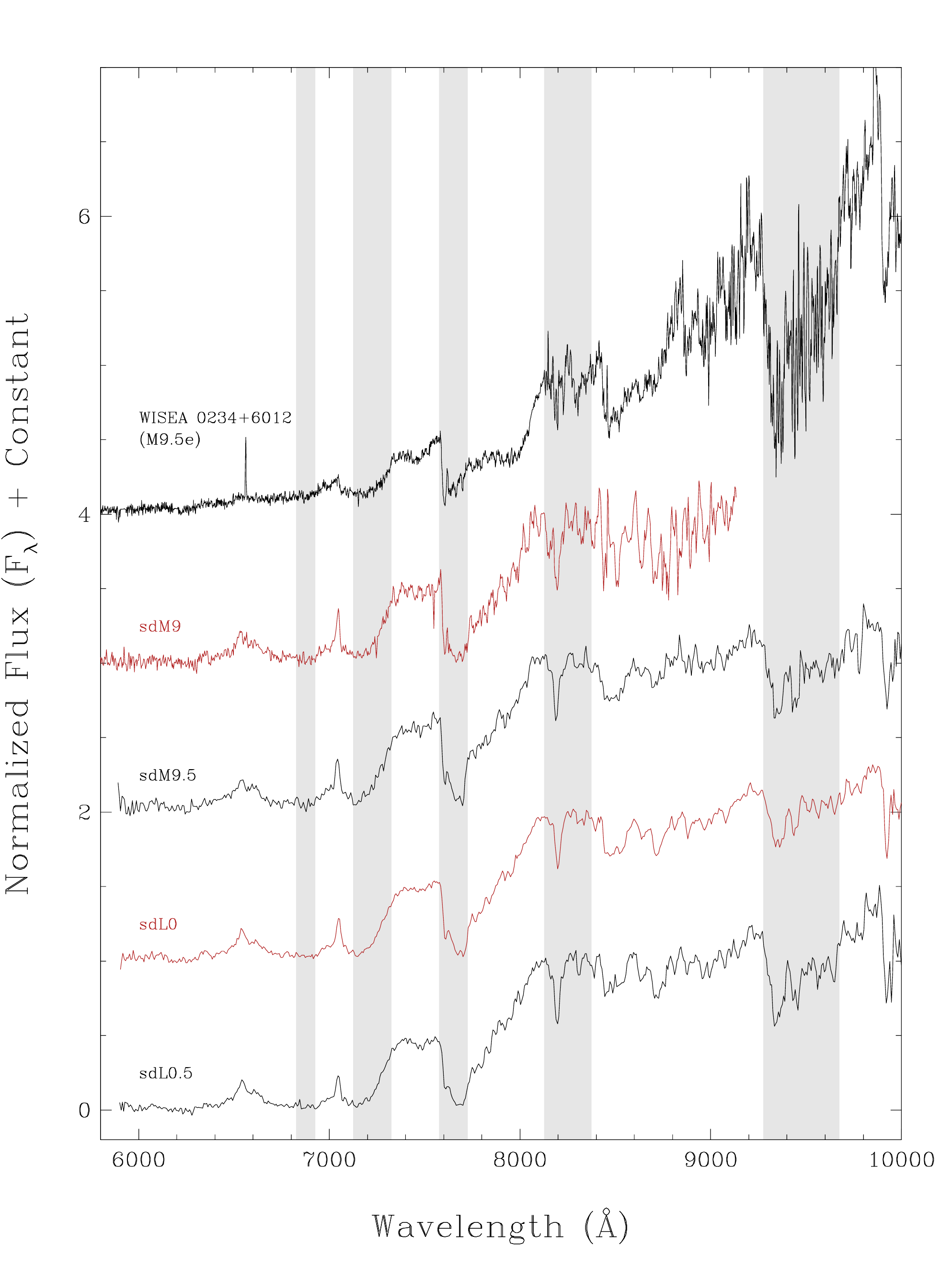}
\caption{Optical spectrum of a normal M9.5e dwarf (the motion star WISEA J023421.83+601227.3) compared to optical 
spectra of very late-M and early-L subdwarfs from the literature: the sdM9 2MASS J16403197+1231068 from 
\cite{gizis2006}, the sdM9.5 ULAS J115826.62+044746.8 from \cite{lodieu2010}, and the sdL0 ULAS J033350.84+001406.1 
and the sdL0.5 ULAS J124425.90+102441.9 from \cite{lodieu2012}. Spectra have been normalized at 7500 \AA\ 
and a constant offset added to the flux to separate the spectra vertically. Regions of telluric absorption are 
marked by the light grey bands.
\label{allwise_optical_sdL_sequence}}
\end{figure}

Note that this spectral morphology is distinctly different from that of a normal late-M dwarf (top spectrum in 
Figure~\ref{allwise_optical_sdL_sequence}), due mainly to the relatively weaker hydride bands in the subdwarfs. 
Only a few subdwarfs are known in this spectral range, so their classification is still in its infancy. In fact, 
no spectroscopic standards yet exist at these types, but we can compare to other published discoveries and use 
their tentative types as guides. A sampling of these very late-M and early-L subdwarfs from the literature is 
shown by the bottom four spectra in Figure~\ref{allwise_optical_sdL_sequence}. There is a plateau in these spectra
 between 7300 and 7500 \AA\ that is bounded on the blue side by TiO absorption and on the red side by TiO and a 
strong \ion{K}{1} doublet. The slope at the top of this plateau slowly changes from slightly red to flat through 
the sdM9-to-sdL0.5 sequence. That is, the flux (in units of $f_\lambda$) at the left edge of the plateau near 
7300 \AA\ is lower or equal to the flux at the right edge of the plateau near 7500 \AA. A check of the sdM8 
spectrum in Figure~\ref{allwise_optical_sdL_features} shows that the plateau has a redward slope, whereas the 
sdL1 spectrum in the same figure has a plateau with a blueward slope. Because all of our spectra in 
Figure~\ref{allwise_new_sdLs_optical} have plateaus with flat or slightly blueward slopes, we choose to classify 
all of these as sdL spectra. None, however have a spectral morphology as extreme as the sdL1 in 
Figure~\ref{allwise_optical_sdL_features} and look more similar to the sdM9-to-sdL0.5 sequence in 
Figure~\ref{allwise_optical_sdL_sequence}. Thus, we tentatively classify all five of these spectra as sdL0.

\begin{figure}
\figurenum{27}
\includegraphics[scale=0.45,angle=0]{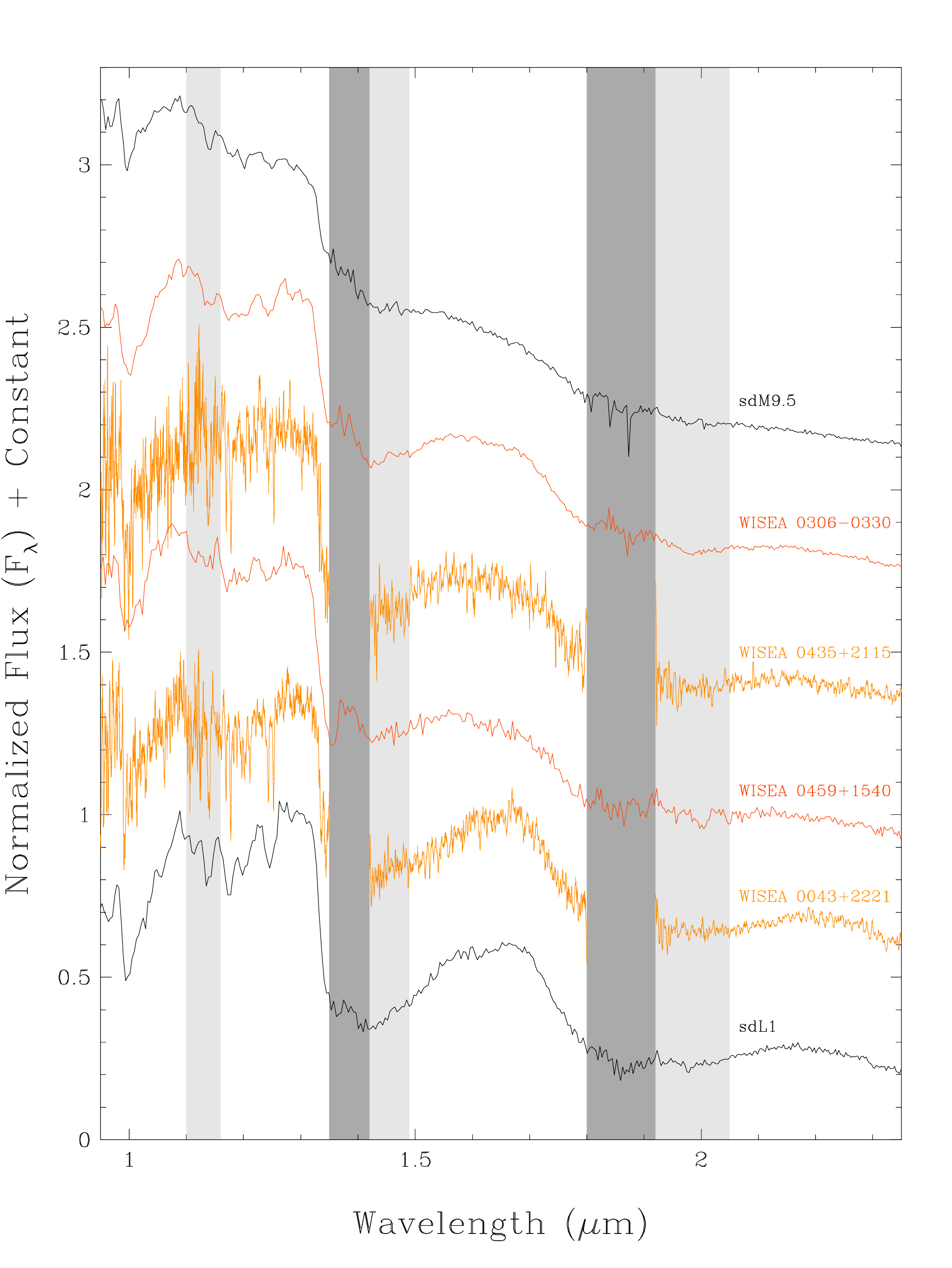}
\caption{Near-infrared spectra of the sdM9.5 J1013$-$1356 (\citealt{burgasser2007}) and the sdL1 2MASS 
J17561080+2815238 (\citealt{kirkpatrick2010}) compared to the near-infrared spectra of four discoveries from 
AllWISE: WISEA J030601.66$-$033059.0, WISEA J043535.82+211508.9, WISEA J045921.21+154059.2 and WISEA 
J004326.26+222124.0. Regions of telluric absorption are 
marked by the dark grey and light grey bands as explained in the caption to Figure~\ref{allwise_redblue_Ls}. Spectra 
have been normalized at 1.28 $\mu$m and a constant offset added to the flux to separate the spectra vertically.
\label{allwise_new_sdLs_IR}}
\end{figure}

Near-infrared spectra of four of the AllWISE motion discoveries are shown in Figure~\ref{allwise_new_sdLs_IR} 
and are compared with near-infrared spectra of an sdM9.5 and the optically-classified sdL1 from 
Figure~\ref{allwise_optical_sdL_features}. These spectra have 
a similar spectral morphology -- notably, strong FeH at 9896 \AA\ and suppressed fluxes at $H$ and $K$ bands 
relative to $J$. This morphology is similar to that of the 
sdM9.5 and the sdL1. Two of the AllWISE objects, WISEA J030601.66$-$033059.0 and  WISEA J043535.82+211508.9, 
also have optical spectra, shown in Figure~\ref{allwise_new_sdLs_optical}, that we classify as sdL0. 
Thus, we tentatively classify these of these AllWISE sources as sdL0 objects in the near-infrared, whereas
we classify WISEA J004326.26+222124.0 -- because its spectral morphology is more like 2MASS J17561080+2815238
-- as an sdL1.

\begin{figure}
\figurenum{28}
\includegraphics[scale=0.45,angle=0]{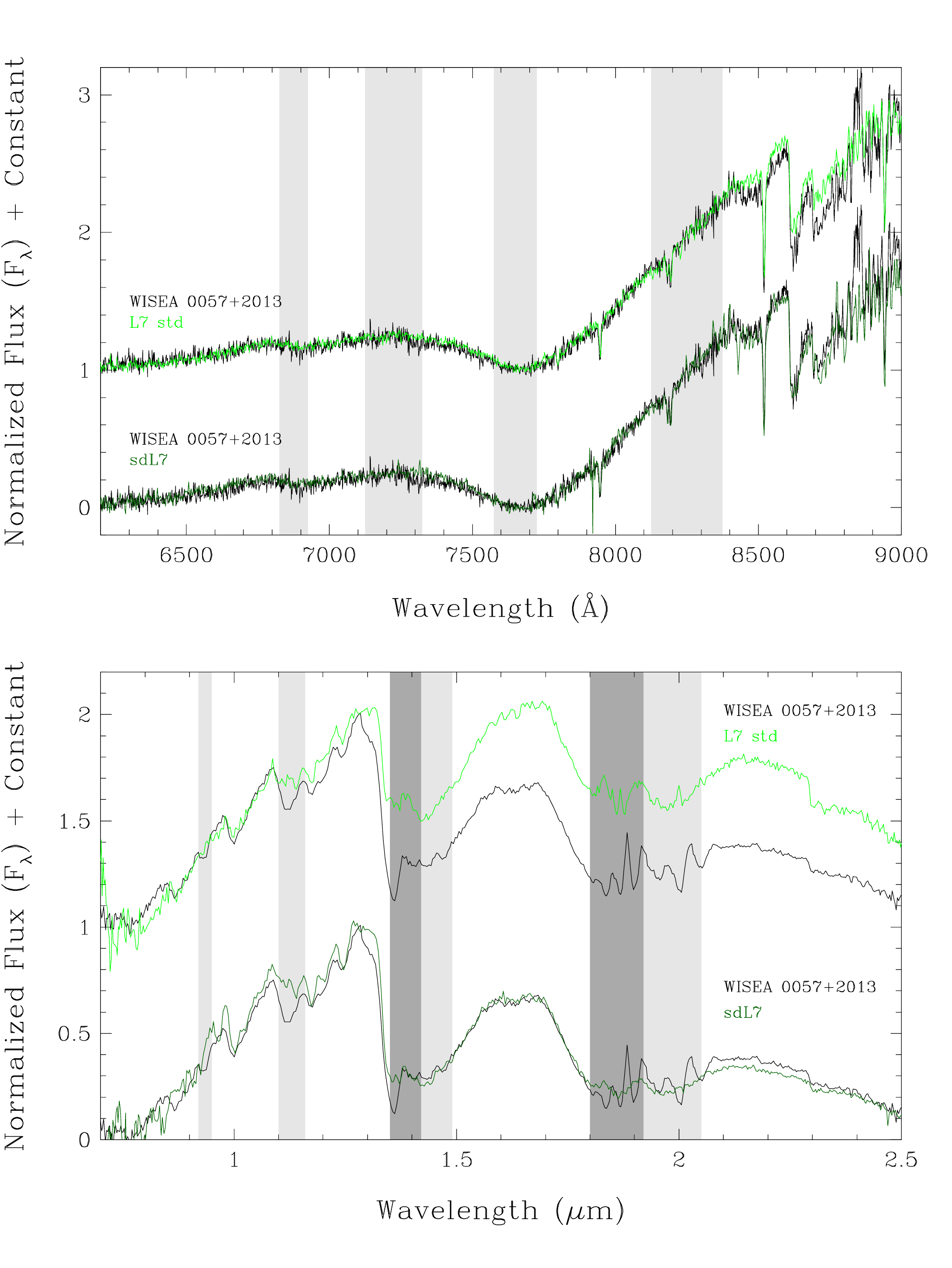}
\caption{Spectra of WISEA J005757.64+201304.0 compared to the L7 standard and an sdL7 from the literature. 
(Upper panel) The optical spectrum of the AllWISE object compared to the optical L7 standard DENIS-P J0205.4$-$1159 
(\citealt{kirkpatrick1999}) and the sdL7 2MASS J11582077+0435014 (\citealt{kirkpatrick2010}). (Lower panel) The 
near-infrared spectrum of the AllWISE object compared to the near-infrared L7 standard 2MASS J01033203+1935361 
and the sdL7 2MASS J11582077+0435014 (both from \citealt{kirkpatrick2010}). Normalizations, offsets, and 
telluric-zone shading are the same as in previous figures.
\label{WISE0057_optical_IR}}
\end{figure}

One final L subdwarf discovery from AllWISE, WISEA J005757.64+201304.0, falls at late-L. 
Figure~\ref{WISE0057_optical_IR} compares the optical and near-infrared spectra of this object with the L7 
standard and with a known sdL7 from the literature. In the optical, WISEA J005757.64+201304.0 fits the overall 
spectral shape of the normal L7 well except in the strength of the TiO band at 8432 \AA, the CrH band at 8611 \AA, 
and the FeH band at 8692 \AA, all of which are stronger in the AllWISE object than in the standard. The published 
sdL7 spectrum, while fitting the overall shape as well as the normal L7 standard, also nicely matches the strengths 
of the these three discrepant bands. In the near-infrared, the spectrum of the AllWISE object best matches the L7 
standard in the $J$-band -- although the match is rather poor -- but the standard is much redder in the $H$ and 
$K$ portions of the spectrum. The near-infrared spectrum of the sdL7, on the other hand, provides a better match 
at $J$-band, an excellent match at $H$, and a much improved match at $K$. For these reasons, we tentatively 
classify this object as an sdL7 in both the optical and the near-infrared.

%Is our brightest L subdwarf discovery, WISEA J204027.30+695924.1, close enough to the Sun that we can measure 
%its parallax via extant data? We have amassed data from various archival sources, as listed in 
%Table 6. %~\ref{additional_astrometry}. 
%Using our fitting code, which is explained in section 5.2 of 
%\cite{kirkpatrick2011}, we obtain the following astrometric solution with $\chi^2 = 1.27$ and seven degrees of 
%freedom: $\mu_{\alpha} = 1.555\pm0.007$ arcsec yr$^{-1}$, $\mu_{\delta} = 1.694\pm0.006$ arcsec yr$^{-1}$, and 
%$\pi = 0.036\pm0.030$ arcsec. In other words, the parallax is not large enough to measure with these data. 
%\cite{schilbach2009} provide a trigonometric parallax for the sdM9.5 SSSPM J1013$-$1356 of 20$\pm$2 mas; if we 
%assume that the AllWISE object and SSSPM J1013$-$1356 have similar absolute magnitudes given their similar 
%spectroscopic classifications, then this would place WISEA J204027.30+695924.1 at a distance of roughly 35-45 pc, 
%which is beyond our ability to measure using the astrometric data currently in hand.

Our follow-up from AllWISE has already added several new L subdwarfs to the number published to date. 
Table 6 
compiles all of the late-M and L subdwarfs currently recognized. Of these, twenty 
were known prior to AllWISE follow-up and eight are from this paper. Additional follow-up of other AllWISE motion 
sources is certain to reveal others. It is curious that the AllWISE subdwarf discoveries tend to cluster around 
a type of sdL0. Is this providing a hint regarding the number density of these objects in the Solar Neighborhood? 
Table 6 
lists the AllWISE motions for all of the known late-M and L subdwarfs detected by 
WISE. Although a few of the detected objects do not have significant AllWISE motions, most could be easily selected 
(and many of these were noted as having been found during Quality Assurance reviews). While AllWISE is capable of 
uncovering subdwarfs throughout the L subdwarf range, objects at mid-L types seem to be lacking. Of the twenty-eight 
objects in this table, only five fall in the broad range from sdL2 to sdL6. Seventeen objects alone are known within 
sdM9-sdL1 and another five are known within the sdL7-sdL8 range. (The ``schizophrenic'' object LSR J1610$-$0040 
is not included in this list since its spectrum defies classification.)

%\clearpage
%\LongTables
\begin{turnpage}
%\begin{center}
\begin{deluxetable*}{lcccccccccc}
\tabletypesize{\tiny}
%\rotate
%\tablewidth{8.5in}
\tablenum{6}
\tablecaption{A List of Known Late-M ($\ge$M9) and L Subdwarfs\label{known_subdwarfs}}
\tablehead{
\colhead{Designation} & 
\colhead{Opt.} &
\colhead{NIR} &
\colhead{Ref.} &
\colhead{2MASS $J$} &  
\colhead{2MASS $H$} &     
\colhead{2MASS $K_s$} &
\colhead{W1} &
\colhead{W2} &
\colhead{AllWISE} &
\colhead{AllWISE} \\
\colhead{} & 
\colhead{Sp.\ Type} &
\colhead{Sp.\ Type} &
\colhead{} &
\colhead{(mag)} &  
\colhead{(mag)} &     
\colhead{(mag)} &
\colhead{(mag)} &
\colhead{(mag)} &
\colhead{RA Motion} &
\colhead{Dec Motion} \\
\colhead{} & 
\colhead{} &
\colhead{} &
\colhead{} &
\colhead{} &  
\colhead{} &     
\colhead{} &
\colhead{} &
\colhead{} &
\colhead{(mas/yr)} &
\colhead{(mas/yr)} \\
\colhead{(1)} &
\colhead{(2)} &
\colhead{(3)} &
\colhead{(4)} &
\colhead{(5)} &
\colhead{(6)} &
\colhead{(7)} &
\colhead{(8)} &
\colhead{(9)} &
\colhead{(10)} &
\colhead{(11)} 
}
\startdata
WISEA J001450.17$-$083823.4& sdL0        & ---          & 1 &  14.469$\pm$0.026&     13.950$\pm$0.026&      13.769$\pm$0.044&    13.429$\pm$0.025&     13.204$\pm$0.030&       1392$\pm$92 &    -406$\pm$93 \\
2MASS J00412179+3547133   & ---          & sdL?         &19 &  15.935$\pm$0.081&     15.728$\pm$0.152&      15.166$\pm$0.121&    14.743$\pm$0.032&     14.454$\pm$0.049&         13$\pm$133&      54$\pm$136\\
...                       & ---          & sdM9         &20 &  ...             &     ...             &      ...             &    ...             &     ...             &         ...       &      ...       \\
WISEA J004326.26+222124.0 & ---          & sdL1         & 1 &  14.871$\pm$0.038&     14.226$\pm$0.039&      13.931$\pm$0.063&    13.688$\pm$0.025&     13.373$\pm$0.031&        270$\pm$59 &    -381$\pm$60 \\
WISEA J005757.65+201304.0 & sdL7         & sdL7         & 1 &  16.317$\pm$0.095&     15.449$\pm$0.088&      14.974$\pm$0.134&    14.317$\pm$0.029&     13.872$\pm$0.039&        853$\pm$98 &    -260$\pm$101\\
WISEA J020201.25$-$313645.2& sdL0        & ---          & 1 &  15.335$\pm$0.050&     14.937$\pm$0.098&      14.702$\pm$0.108&    14.310$\pm$0.028&     13.959$\pm$0.038&       -278$\pm$132&   -1666$\pm$134\\
WISEA J030601.66$-$033059.0& sdL0        & sdL0         & 1 &  14.441$\pm$0.026&     14.060$\pm$0.040&      13.978$\pm$0.048&    13.429$\pm$0.025&     13.181$\pm$0.027&        388$\pm$44 &   -1258$\pm$47 \\
ULAS J033350.84+001406.1  & sdL0         & ---          & 2 &  16.018$\pm$0.111&     15.698$\pm$0.178&      $>$16.630       &    15.077$\pm$0.038&     14.765$\pm$0.071&        663$\pm$284&     -70$\pm$300\\
WISEA J043535.82+211508.9 & sdL0         & sdL0         & 1 &  15.011$\pm$0.031&     14.682$\pm$0.053&      14.594$\pm$0.081&    13.999$\pm$0.029&     13.668$\pm$0.039&       1022$\pm$127&    -771$\pm$138\\
WISEA J045921.22+154059.2 & ---          & sdL0         & 1 &  14.957$\pm$0.031&     14.613$\pm$0.047&      14.300$\pm$0.062&    14.093$\pm$0.028&     13.851$\pm$0.042&        707$\pm$130&    -553$\pm$142\\
2MASS J05325346+8246465   & late sdL     & late sdL     &17 &  15.179$\pm$0.058&     14.904$\pm$0.091&      14.918$\pm$0.145&    13.824$\pm$0.025&     13.260$\pm$0.028&       2416$\pm$77 &   -1450$\pm$83\\
...                       & (e?)sdL7     & (e?)sdL7     & 4 &  ...             &     ...             &      ...             &    ...             &     ...             &         ...       &      ...       \\
...                       & (e)sdL7:$\alpha$& (e)sdL7:$\alpha$ & 18 &  ...             &     ...             &      ...             &    ...             &     ...             &         ...       &      ...       \\
...                       & sdL7         & ---          &12 &  ...             &     ...             &      ...             &    ...             &     ...             &         ...       &      ...       \\
...                       & esdL7        & esdL7        &24 &  ...             &     ...             &      ...             &    ...             &     ...             &         ...       &      ...       \\
2MASS J06164006$-$6407194 & sdL5         & sdL5         & 9 &  16.403$\pm$0.112&     16.275$\pm$0.228&      $>$16.381       &    15.646$\pm$0.030&     15.183$\pm$0.042&       1438$\pm$162&     -63$\pm$177\\ 
2MASS J06453153$-$6646120 & sdL8         & sdL8         & 4 &  15.602$\pm$0.067&     14.696$\pm$0.070&      14.372$\pm$0.084&    13.761$\pm$0.023&     13.308$\pm$0.023&       -851$\pm$45 &    1134$\pm$44\\
SSSPM J1013$-$1356        & sdM9.5       & sdM9.5       &12 &  14.621$\pm$0.029&     14.382$\pm$0.048&      14.398$\pm$0.077&    13.797$\pm$0.026&     13.604$\pm$0.032&         90$\pm$93 &   -1087$\pm$100\\
...                       & sdM9.5       & ---          &21 &  ...             &     ...             &      ...             &    ...             &     ...             &         ...       &      ...       \\
2MASS J11582077+0435014   & sdL7         & sdL7         & 4 &  15.611$\pm$0.054&     14.684$\pm$0.064&      14.439$\pm$0.063&    13.695$\pm$0.026&     13.361$\pm$0.033&        871$\pm$103&    -633$\pm$109\\
ULAS J115826.62+044746.8  & sdM9.5       & ---          & 2 &  16.536$\pm$0.130&     16.079$\pm$0.201&      15.725$\pm$0.186&    15.655$\pm$0.051&     15.408$\pm$0.121&        127$\pm$469&    -913$\pm$525\\
ULAS J124425.90+102441.9  & sdL0.5       & ---          & 2 &  16.198$\pm$0.128&     15.650$\pm$0.160&      $>$15.400       &    15.453$\pm$0.041&     15.138$\pm$0.088&        -84$\pm$310&    -174$\pm$337\\
...                       & sdL2?        & ---          &25 &  ...             &     ...             &      ...             &    ...             &     ...             &         ...       &      ...        \\
SDSS J125637.13$-$022452.4& sdL4?        & ---          & 8 &  16.099$\pm$0.105&     15.792$\pm$0.148&      $>$15.439       &    15.210$\pm$0.036&     15.005$\pm$0.080&       -604$\pm$183&     -50$\pm$195\\
...                       & sdL3.5       & sdL3.5       &10 &  ...             &     ...             &      ...             &    ...             &     ...             &         ...       &      ...       \\
...                       & ---          & esdL3.5      &24 &  ...             &     ...             &      ...             &    ...             &     ...             &         ...       &      ...       \\
HD 114762B\tablenotemark{a}&---          & d/sdM9       &11 &  ...             &     ...             &      ...             &    ...             &     ...             &         ...       &      ...       \\
SDSS J133348.24+273508.8  & sdL3         & ---          &25 &  ...             &     ...             &      ...             &    ...             &     ...             &         ...       &      ...       \\
ULAS J135058.86+081506.8  & sdL5$\pm$1   & esdL4        & 3 &  ...             &     ...             &      ...             &    ...             &     ...             &         ...       &      ...       \\
SDSS J141624.08+134826.7  & sdL7         & sdL7         & 4 &  13.148$\pm$0.021&     12.456$\pm$0.027&      12.114$\pm$0.021&    11.363$\pm$0.022&     11.026$\pm$0.020&          5$\pm$27 &      80$\pm$28\\
...                       & d/sdL7       & ---          & 5 &  ...             &     ...             &      ...             &    ...             &     ...             &         ...       &      ...       \\
...                       & L5           & L4$\pm$1.5 (blue)  & 6 &  ...             &     ...             &      ...             &    ...             &     ...             &         ...       &      ...       \\
...                       & L6$\pm$0.5   & L6$\pm$ pec  & 7 &  ...             &     ...             &      ...             &    ...             &     ...             &         ...       &      ...       \\
...                       & d/sdL7       & d/sdL7       &24 &  ...             &     ...             &      ...             &    ...             &     ...             &         ...       &      ...       \\
2MASS J14343616+2202463   & ---          & sdM9         &20 &  14.519$\pm$0.034&     13.833$\pm$0.040&      13.545$\pm$0.041&    13.248$\pm$0.024&     12.877$\pm$0.026&       -752$\pm$40 &     177$\pm$42\\
SSSPM J1444$-$2019        & d/sdM9       & ---          &12 &  12.546$\pm$0.023&     12.142$\pm$0.024&      11.933$\pm$0.024&    11.471$\pm$0.023&     11.211$\pm$0.022&      -3329$\pm$45 &   -1864$\pm$47\\
...                       & sdM9 or sdL  & ---          &14 &  ...             &     ...             &      ...             &    ...             &     ...             &         ...       &      ...       \\
LSR J1610$-$0040          & sdL          & ---          &16 &  12.911$\pm$0.018&     12.302$\pm$0.020&      12.019$\pm$0.026&    11.637$\pm$0.243&     11.537$\pm$0.023&       -906$\pm$51 &   -1220$\pm$51\\
...                       & sd?M6 pec    & ---          &22 &  ...             &     ...             &      ...             &    ...             &     ...             &         ...       &      ...       \\
...                       & ---          & d/sdM v. pec?&23 &  ...             &     ...             &      ...             &    ...             &     ...             &         ...       &      ...       \\
2MASS J16262034+3925190   & ---          & sdL          &15 &  14.435$\pm$0.026&     14.533$\pm$0.049&      14.466$\pm$0.074&    13.482$\pm$0.024&     13.136$\pm$0.026&      -1539$\pm$68 &     398$\pm$71\\
...                       & sdL4         & ---          &13 &  ...             &     ...             &      ...             &    ...             &     ...             &         ...       &      ...       \\
...                       & sdL4         & ---          &12 &  ...             &     ...             &      ...             &    ...             &     ...             &         ...       &      ...       \\
...                       & ---          & esdL4        &24 &  ...             &     ...             &      ...             &    ...             &     ...             &         ...       &      ...       \\
2MASS J16403197+1231068   & d/sdM9       & ---          &12 &  15.946$\pm$0.077&     15.605$\pm$0.112&      $>$15.520       &    15.033$\pm$0.036&     14.846$\pm$0.068&       -365$\pm$250&    -324$\pm$267\\
                          & sdM9/sdL?    & ---          &13 &  ...             &     ...             &      ...             &    ...             &     ...             &         ...       &      ...       \\
                          & ---          & sdM8?        &19 &  ...             &     ...             &      ...             &    ...             &     ...             &         ...       &      ...       \\
2MASS J17561080+2815238   & sdL1         & L1 pec (blue)& 4 &  14.712$\pm$0.031&     14.135$\pm$0.040&      13.813$\pm$0.040&    13.384$\pm$0.024&     13.068$\pm$0.026&       -748$\pm$67 &    -541$\pm$73 \\
WISEA J204027.30+695924.1 & sdL0         & ---          & 1 &  13.723$\pm$0.070&     13.313$\pm$0.066&      13.119$\pm$0.054&    12.684$\pm$0.023&     12.456$\pm$0.022&       1535$\pm$29 &    1643$\pm$28 \\
\enddata
\tablenotetext{a}{Also known as 2MASS J13121982+1731016B.}
\tablerefs{
(1) this paper,
(2) \cite{lodieu2012},
(3) \cite{lodieu2010},
(4) \cite{kirkpatrick2010},
(5) \cite{burningham2010},
(6) \cite{schmidt2010a},
(7) \cite{bowler2010},
(8) \cite{sivarani2009},
(9) \cite{cushing2009},
(10) \cite{burgasser2009},
(11) \cite{bowler2009},
(12) \cite{burgasser2007},
(13) \cite{gizis2006},
(14) \cite{scholz2004a},
(15) \cite{burgasser2004a},
(16) \cite{lepine2003} and \cite{dahn2008},
(17) \cite{burgasser2003},
(18) \cite{kirkpatrick2005},
(19) \cite{burgasser2004b},
(20) \cite{sheppard2009},
(21) \cite{scholz2004b},
(22) \cite{reiners2006},
(23) \cite{cushing2006},
(24) \cite{zhang2013},
(25) \cite{zhang2012}.
}
\end{deluxetable*}
%\end{center}
%\clearpage
\end{turnpage}

\begin{figure*}
\figurenum{29}
\includegraphics[scale=0.375,angle=0]{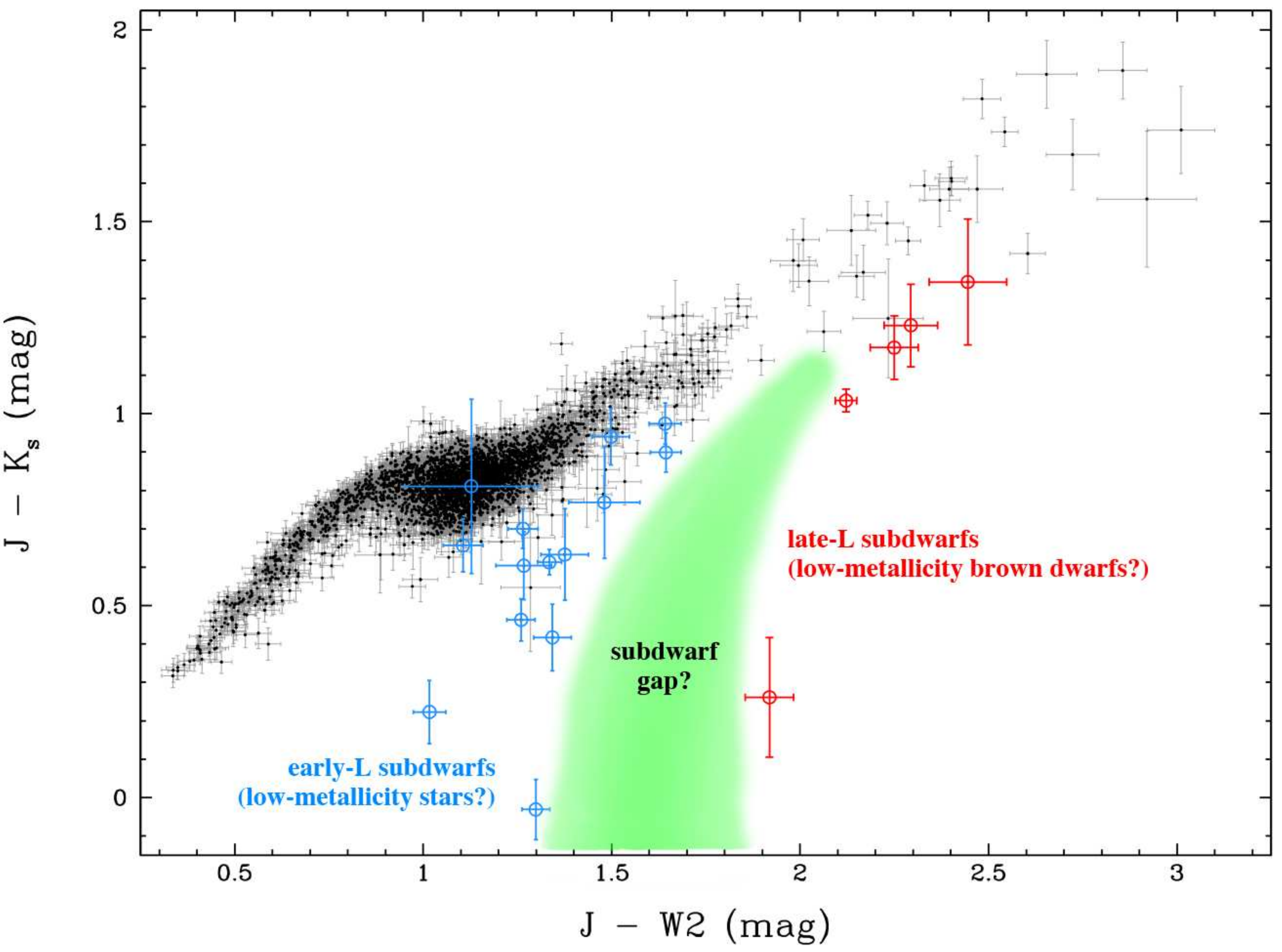}
\caption{AllWISE motion stars (solid black dots) from Figure~\ref{JKs_vs_JW2}, with the three possible T dwarfs 
omitted for clarity. The locations of known late-M and L subdwarfs from Table 6 are shown by 
open circles. Those with spectral types earlier than sdL5 are shown in blue and those with types later than sdL5 
are shown in red. The wedge (green zone) depicts an area of color space where L subdwarfs may rarely be found. 
See text for details.
\label{sd_gap}}
\end{figure*}

Figure~\ref{sd_gap} is a rehash of Figure~\ref{JKs_vs_JW2} with all known spectroscopically identified late-M
and L-type subdwarfs 
color coded. This diagram hints at the possibility that a lightly populated gap may exist between the early-L 
subdwarf and late-L subdwarf populations. Such a gap would be expected on theoretical grounds: Lower metallicities 
generally correspond to older ages. At these older ages, substellar objects have had enough time to cool that a gap 
in temperature will have opened up between the hottest members of their population and the coldest, lowest mass members of 
similarly aged hydrogen-burning stars. Such a temperature gap would be wider the older the population, and hence, 
the lower the metallicity. Figure~\ref{sd_gap} shows the first signs of what may be a sparsely populated locus of 
color space corresponding to this temperature gap. 

The late-L subdwarf at (1.92,0.26) on Figure~\ref{sd_gap} is the (e?)sdL7 2MASS J05325346+8246465 
(\citealt{burgasser2003}). That object has a trigonometric parallax measurement; after comparison to theoretical 
evolutionary models, \cite{burgasser2008} find that the resulting luminosity is consistent with the object being 
a very high-mass brown dwarf just below the hydrogen-burning limit. This would place the object on the brown dwarf 
side of the L subdwarf gap, as shown in the figure. The early-L subdwarf at (1.30,-0.03) is the sdL4 2MASS 
J16262034+3925190; \cite{burgasser2004a} finds that this object is near or below the hydrogen-burning minimum mass 
limit, consistent with our placement of it on the opposite side of the sparsely populated ``subdwarf gap.'' 

Despite these checks against 
limited published results, the zone of L subdwarf avoidance shown on Figure~\ref{sd_gap} should be regarded only as 
a cartoon depiction of expectations. With continued follow-up of motion discoveries from AllWISE, it is hoped that 
researchers will be able to more fully explore the boundaries of the gap, if it is real.

\section{Two New, Nearby M Dwarf Systems\label{two_nearby_dwarfs}}

\begin{figure*}
\figurenum{30}
\includegraphics[scale=0.475,angle=0]{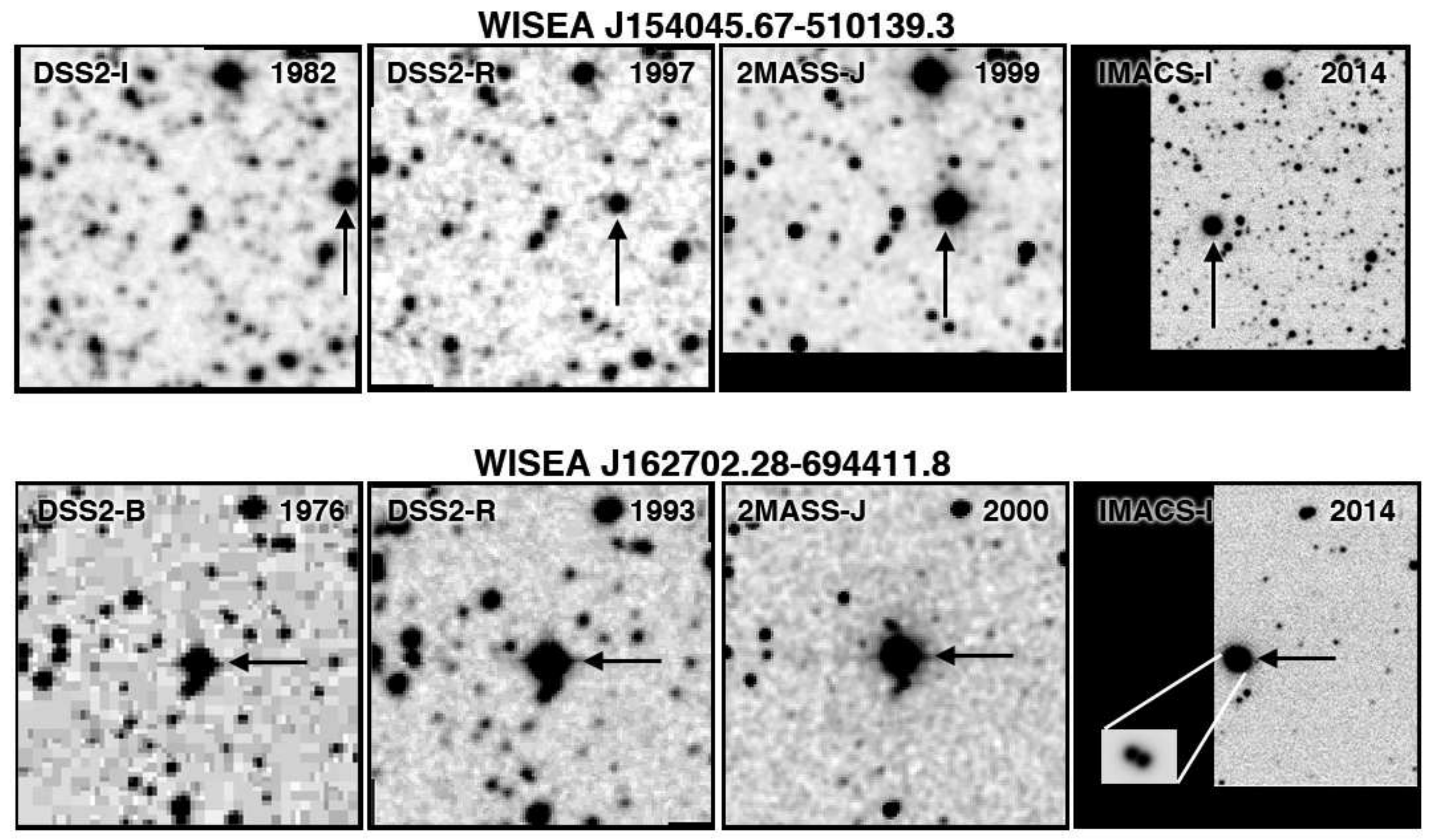}
\caption{Images from DSS2, 2MASS, and Magellan/IMACS for two nearby M dwarf systems. (Upper
panel) DSS2-$I$, DSS2-$R$, 2MASS-$J$, and IMACS-$I$ images for the motion object 
WISEA J154045.67$-$510139.3. 
(Lower panel) DSS2-$B$, DSS2-$R$, 2MASS-$J$, and IMACS-$I$ images for the motion object 
WISEA J162702.28$-$694411.8.
In both panels the epoch of each image is labeled and the position of the motion star in indicated by an 
arrow. Each subpanel is 2$\arcmin$ square with north up and east to the left. The inset on the IMACS-$I$
image for WISEA J162702.28$-$694411.8AB shows the individual components of the system.
\label{M_dwarf_finders}}
\end{figure*}

There is a single object, WISEA J154045.67$-$510139.3, occupying the upper right quadrant of 
Figure~\ref{w1_vs_totalmotion}. This object is very bright, $J = 8.96\pm0.02$ 
mag and W2 $= 7.47\pm0.02$ mag, and has a very large motion, $\mu=2006\pm12$ mas yr$^{-1}$, which is the 
second-highest motion of any of our discoveries. A finder chart is shown in 
the upper panel of Figure~\ref{M_dwarf_finders}. 
Our optical spectrum (Figure 31) shows this object to be a normal M6 dwarf.
The absolute $J$-band magnitude for a typical M6 dwarf is 10.12 mag 
(equation 6 of \citealt{cruz2003}), which would place this object at a distance of $\sim$5.9 pc if single or 
$\sim$8.3 pc if an equal-magnitude double.
We have gathered positional information on this source from a variety of online archives, as listed in 
Table 7.
Using our fitting code, which is explained in section 5.2 of 
\cite{kirkpatrick2011}, we obtain the following astrometric solution with $\chi^2 = 9.5$ and five degrees of 
freedom: $\mu_{\alpha} = 1\farcs951\pm0\farcs006$ yr$^{-1}$, $\mu_{\delta} = -0\farcs332\pm0\farcs006$ yr$^{-1}$, 
and $\pi = 0\farcs165\pm0\farcs041$ ($d \approx 6$ pc). Although the value of the parallax is very fragile given 
the size of its uncertainty, the available astrometry strongly suggests that this object is close to the Sun,
in agreement with our spectrophotometric distance estimate.

\begin{figure}
\figurenum{31}
\includegraphics[scale=0.325,angle=270]{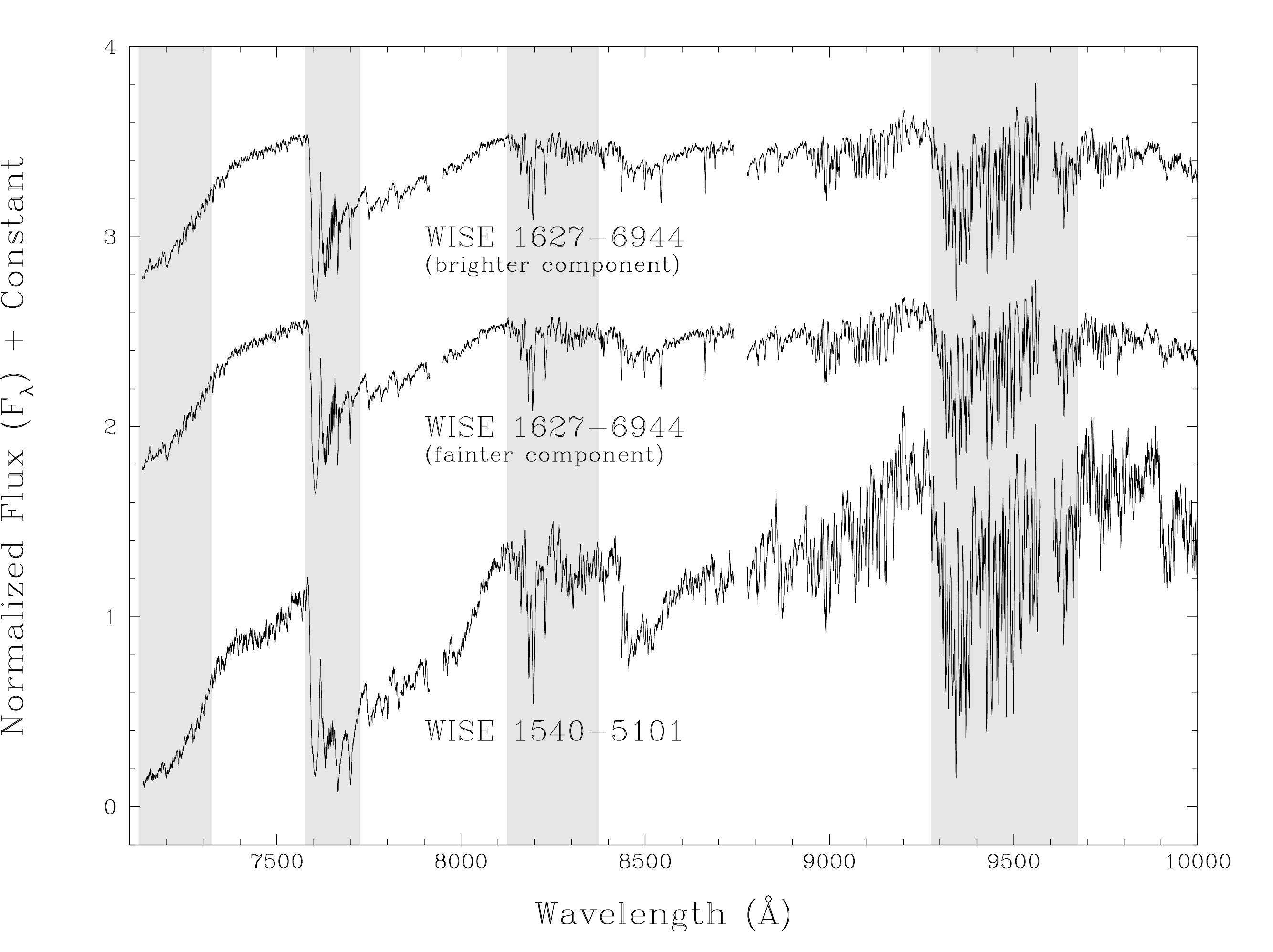}
\caption{
Optical spectra of the two components of the WISEA J162702.28$-$694411.8AB system (top spectra) and
WISEA J154045.67$-$510139.3 (bottom spectrum). Spectra 
have been normalized at 7500 \AA\ and a constant offset added to the flux to separate the spectra vertically.
Regions of telluric absorption are marked by the light grey bands.
\label{M_dwarf_spectra}}
\end{figure}

\begin{deluxetable*}{lllllll}
\tabletypesize{\tiny}
%\rotate
\tablewidth{6.0in}
\tablenum{7}
\tablecaption{Additional Astrometric Data for The Two Nearby M Dwarf Systems\label{additional_astrometry}}
\tablehead{
\colhead{MJD} & 
\colhead{Calendar} &   
\colhead{J2000 RA} &                     
\colhead{J2000 Dec} & 
\colhead{RA error} &                    
\colhead{Dec error} & 
\colhead{Notes} \\
\colhead{} & 
\colhead{Date} &   
\colhead{(deg)} &                     
\colhead{(deg)} & 
\colhead{($\arcsec$)} &                    
\colhead{($\arcsec$)} & 
\colhead{} \\
\colhead{(1)} &    
\colhead{(2)} &                     
\colhead{(3)} & 
\colhead{(4)} &                    
\colhead{(5)} & 
\colhead{(6)} &
\colhead{(7)}
}
\startdata
\\
%\multicolumn{7}{l}{WISEA J204027.30+695924.1}  \\            
%34626.710 & 1953 Sep 06& 310.04197  & 69.96312  & 1.00 & 1.00 & DSS1 $R$-band \\   
%50119.501 & 1996 Feb 06& 310.095939 & 69.983178 & 1.00 & 0.81 & USNO-B1 \\
%51449.147 & 1999 Sep 28& 310.100080 & 69.984940 & 0.07 & 0.06 & 2MASS \\
%55213.408 & 2010 Jan 17& 310.113102 & 69.989774 & 0.05 & 0.05 & WISE epoch 1 \\
%55401.011 & 2010 Jul 24& 310.113721 & 69.990033 & 0.05 & 0.05 & WISE epoch 2 \\
%55578.619 & 2011 Jan 17& 310.114362 & 69.990238 & 0.05 & 0.05 & WISE epoch 3 \\
%\\
\multicolumn{7}{l}{WISEA J154045.67$-$510139.3}  \\            
45078.690 & 1982 Apr 19& 235.166113 & -51.025275 & 0.333   & 0.333 & DSS2 $I$-band \\     
50554.606 & 1997 Apr 16& 235.179017 & -51.026369 & 0.333   & 0.333 & DSS2 $R$-band \\     
51364.022 & 1999 Jul 05& 235.180914 & -51.026588 & 0.06    & 0.06  & 2MASS \\             
55251.096 & 2010 Feb 24& 235.190210 & -51.027578 & 0.071   & 0.067 & WISE All-Sky \\      
55432.482 & 2010 Aug 24& 235.190481 & -51.027619 & 0.044   & 0.041 & WISE 3-band Cryo \\  
\\
\multicolumn{7}{l}{WISEA J162702.28$-$694411.8}  \\            
42869.726 & 1976 Apr 01& 246.75713  & -69.73792  & 0.5   &  0.5    &  DSS2 $B$-band \\    
48352.721 & 1991 Apr 06& 246.75809  & -69.737244 & 0.33  &  0.33   &  DSS2 $I$-band \\    
51652.288 & 2000 Apr 18& 246.758802 & -69.736984 & 0.06  &  0.06   &  2MASS \\            
55264.727 & 2010 Mar 09& 246.759567 & -69.736625 & 0.066 &  0.064  &  WISE All-Sky \\     
55444.838 & 2010 Sep 05& 246.759596 & -69.736595 & 0.041 &  0.039  &  WISE 3-band Cryo \\ 
\enddata
\end{deluxetable*}

A second, bright object is seen next to WISEA J154045.67$-$510139.3 on 
Figure~\ref{J_vs_JW2}. This object, WISEA J162702.28$-$694411.8, has magnitudes and colors of 
$J = 9.16\pm0.04$ mag, W2 $= 7.65\pm0.02$ mag, $J$ $-$ W2 $= 1.52\pm0.05$ mag, and 
$J - K_s = 0.97\pm0.05$ mag. A finder chart is shown in the lower panel of Figure~\ref{M_dwarf_finders}.
Our follow-up imaging shows it to be a near equal-magnitude binary separated by 2$\farcs$0. Magnitude differences
between the two components are found to be ${\Delta}V$ = 0.18$\pm$0.01 mag and ${\Delta}I$ = 0.13$\pm$0.01 mag.
Optical spectra of the two components (Figure 31) show that both are normal M4 dwarfs. 
The absolute $J$-band magnitude for a typical M4 dwarf is 7.89 mag 
(equation 6 of \citealt{cruz2003}), which would place this near-equal-magnitude binary at a distance 
of $\sim$25 pc.
Using additional positional information listed in Table 7, 
we obtain the following astrometric solution with $\chi^2 = 1.22$ and five degrees of freedom: $\mu_{\alpha} = 
0\farcs096\pm0\farcs007$ yr$^{-1}$, $\mu_{\delta} = 0\farcs133\pm0\farcs006$ yr$^{-1}$, and $\pi = 
0\farcs008\pm0\farcs039$. At face value, a nearby distance seems not to be supported by these data, but we note 
that the first four (of five) data points were all taken in March or April -- i.e., at nearly the same parallax 
factor. Only the final observation, taken in September, provides any leverage regarding the size of the parallax. 
Moreover, the binarity of the source may be complicating clean measurements of the photocenter. 
Additional astrometric monitoring is needed.

WISEA J154045.67$-$510139.3 and WISEA J162702.28$-$694411.8 have galactic coordinates of ($328\fdg0$,$+3\fdg4$) and 
($319\fdg7$,$-14\fdg2$), respectively, and fall in the lower right quadrant of Figure 15 (middle panel).
One of the objects in the \cite{luhman2014} list, WISE J163348.95-680851.6 ($J$ = 11.19 mag, $J - K_s$ = 1.12 mag,
$J$ - W2 = 2.20 mag), is located at galactic coordinates of ($321\fdg0$,$-13\fdg4$), and may be another close
object, likely a late-M or early-L dwarf.
The density of WISE motion discoveries in this quadrant demonstrates that a large area around the Galactic Center
has been poorly surveyed by earlier searches for nearby stars. 

\section{Conclusions}

We have characterized the motion measurements contained within the AllWISE Data Release. We have presented a list 
of 3,525 motion objects lacking previous literature in SIMBAD and find two very interesting results. The first is 
that AllWISE is revealing many more examples of low-metallicity objects near the stellar/substellar break, allowing 
researchers to probe the region near the hydrogen-burning minimum mass for objects formed early in the history of 
the Milky Way. The second is that AllWISE is continuing the WISE legacy of uncovering previously overlooked, very 
nearby objects. Prior to AllWISE processing, WISE had already revealed several nearby systems never before 
recognized --e.g., the L+T binary WISE J104915.57-531906.1 at 2.0 pc (\citealt{luhman2013}) and the T6 dwarf WISEPC 
J150649.97+702736.0 (\citealt{kirkpatrick2011}) possibly as close as 3.4 pc (\citealt{marsh2013}); 
AllWISE has 
already contributed another overlooked object -- WISEA J154045.67$-$510139.3 ($\sim$5.9 pc)
-- to this distinguished list of WISE discoveries within the canonical 8-pc sample. 
The motion measurements provided by AllWISE will likely provide 
researchers many new gifts over the years to come.

\acknowledgments

This publication makes use of data products from WISE, which is a 
joint project of the University of California, Los Angeles, and the Jet Propulsion Laboratory (JPL)/California 
Institute of Technology (Caltech), funded by the National Aeronautics and Space Administration (NASA). 
This research has made use of the NASA/IPAC Infrared Science Archive,
which is operated by JPL/Caltech, under contract with NASA. 
We are indebted to the SIMBAD database and the VizieR catalog access tool, provided by CDS, Strasbourg, France, and we
acknowledge use of the Database of Ultracool Parallaxes maintained by Trent Dupuy.
We thank our referee, whose critique of the original draft resulted in a clearer, more complete paper.
We thank Fiona Harrison, George Djorgovski, Brian Mazur, and Barry Madore for PI-ing telescope time used for 
spectroscopic follow-up and are grateful for the time alloted by Caltech, NASA/IRTF, and the Carnegie Observatories.
We also thank Nicolas Lodieu, John Gizis, and Sebastien L{\'e}pine for providing published spectra of subdwarfs.

Facilities: \facility{WISE}, \facility{Hale(Double Spectrograph)}, \facility{Keck:I(LRIS)}, \facility{Keck:II(DEIMOS, NIRSPEC)}, 
\facility{IRTF(SpeX)}, \facility{Magellan:Baade(IMACS)}.

\clearpage
\begin{center}
\clearpage
\begin{longtable*}{cccccl}
\tabletypesize{\tiny}
\tablenum{2}
\tablecaption{Common-proper-motion Binaries Noted During AllWISE Quality Checks\label{cpm_pairs}}
\tablehead{
\colhead{WISEA Designation} &                          
\colhead{W1} &  
\colhead{W2} &
\colhead{AllWISE} &     
\colhead{AllWISE} &
\colhead{Note} \\
\colhead{} &
\colhead{(mag)} &
\colhead{(mag)} &
\colhead{RA Motion} &
\colhead{Dec Motion} &
\colhead{} \\
\colhead{} &
\colhead{} &
\colhead{} &
\colhead{(mas yr$^{-1}$)} &
\colhead{(mas yr$^{-1}$)} &
\colhead{} \\
\colhead{(1)} &                          
\colhead{(2)} &  
\colhead{(3)} &     
\colhead{(4)} &
\colhead{(5)} &
\colhead{(6)}   
}
%\startdata
J012655.16+120022.0&   9.872$\pm$0.022&   9.886$\pm$0.021&    -83$\pm$36&  -364$\pm$33&   NLTT 4817, LSPM J0126+1200N \\                                     
J012654.11+120002.9&  12.830$\pm$0.023&  12.669$\pm$0.027&   -151$\pm$58&  -444$\pm$56&   NLTT 4814, LSPM J0126+1200S  (sep 24$\farcs$5)\\
\\                                                                                     
J013054.96+524442.3&   7.915$\pm$0.027&   7.977$\pm$0.020&   -222$\pm$26&    22$\pm$24&   BD+51 318A  \\
J013056.36+524500.8&   9.760$\pm$0.024&   9.658$\pm$0.020&   -189$\pm$25&   -40$\pm$25&   BD+51 318B (sep 22$\farcs$5) \\
\\                                                                                     
J015329.15+732940.2&   8.109$\pm$0.023&   8.058$\pm$0.020&    420$\pm$31&  -208$\pm$30&   LP 29-233, ``LHS 6036''  \\                                               
J015326.94+733015.4&   9.124$\pm$0.023&   8.939$\pm$0.020&    459$\pm$32&  -130$\pm$31&   LP 29-232, ``LHS 6035'' (sep 36$\farcs$4) \\
\\ 
J021157.87+042140.9&   6.955$\pm$0.054&   7.020$\pm$0.020&   -155$\pm$26&   -11$\pm$25&   BD+03 301, LSPM J0211+0421 \\
J021159.43+042150.7&   9.184$\pm$0.023&   9.080$\pm$0.020&   -145$\pm$26&   -80$\pm$26&   {\it new cpm companion}\tablenotemark{a} (sep 25$\farcs$3) \\
\\                                                                                    
J022104.69+365258.7&   8.358$\pm$0.024&   8.180$\pm$0.021&    768$\pm$26&  -562$\pm$26&   GJ 1047AB, LHS 1393 \\                                      
J022102.54+365241.5&   9.256$\pm$0.023&   9.060$\pm$0.020&    777$\pm$27&  -549$\pm$26&   GJ 1047C, LHS 1392 (sep 31$\farcs$0) \\
\\                                                                                     
J025051.72+033035.6&   8.852$\pm$0.022&   8.904$\pm$0.019&    175$\pm$26&   -83$\pm$25&   BD+02 436A, LSPM J0250+0330N \\                     
J025051.32+033006.7&   9.023$\pm$0.022&   9.077$\pm$0.019&    185$\pm$26&   -56$\pm$25&   BD+02 436B, LSPM J0250+0330S (sep 29$\farcs$5) \\
\\                                                                                     
J025403.21-355418.7&   5.853$\pm$0.119&   5.731$\pm$0.047&    523$\pm$26&  -150$\pm$26&   LHS 1466 \\                   
J025402.74-355454.7&   8.179$\pm$0.022&   8.066$\pm$0.019&    479$\pm$24&  -143$\pm$24&   LHS 1467 (sep 36$\farcs$4) \\
\\                                                                                     
J030531.36+114952.8&   7.997$\pm$0.023&   7.863$\pm$0.019&    165$\pm$26&  -254$\pm$26&   G 5-13, LSPM J0305+1149N   \\                                  
J030530.76+114932.1&   8.346$\pm$0.023&   8.249$\pm$0.020&    157$\pm$26&  -254$\pm$26&   G 5-12, LSPM J0305+1149S (sep 22$\farcs$6)  \\
\\                                                                                     
J031407.66+083319.0&   8.391$\pm$0.023&   8.433$\pm$0.021&    303$\pm$38&   -14$\pm$36&   G 79-15, LSPM J0314+0833S \\           
J031408.29+083345.6&   8.707$\pm$0.022&   8.702$\pm$0.021&    275$\pm$37&    14$\pm$36&   G 79-16, LSPM J0314+0833N (sep 28$\farcs$2) \\
\\                                                                                 
J031608.21$-$374158.4&  9.565$\pm$0.023&  9.624$\pm$0.020&     65$\pm$24&    51$\pm$24&   TYC 7561-77-1 \\
J031606.54$-$374215.4& 13.552$\pm$0.024& 13.341$\pm$0.027&     55$\pm$42&     6$\pm$42&   {\it new cpm companion}\tablenotemark{b} (sep 26$\farcs$1) \\
\\    
J033528.69-321806.2&   9.007$\pm$0.023&   8.941$\pm$0.020&   -292$\pm$25&  -331$\pm$24&   LP 888-32, LEHPM 3410 \\                                                 
J033530.11-321829.1&  10.333$\pm$0.023&  10.146$\pm$0.021&   -303$\pm$26&  -306$\pm$25&   LP 888-33, LEHPM 3411 (sep 29$\farcs$1) \\
\\                                                                                     
J033936.37+252814.8&   7.777$\pm$0.024&   7.698$\pm$0.020&    377$\pm$36&  -497$\pm$34&   Wolf 204, LHS 1573  \\
J033940.66+252842.2&   8.007$\pm$0.023&   7.919$\pm$0.021&    319$\pm$36&  -517$\pm$35&   Wolf 205, LHS 1574 (sep 64$\farcs$4) \\                             
\\                                                                                     
J035228.33-315026.6&  12.741$\pm$0.023&  12.499$\pm$0.023&    231$\pm$32&  -531$\pm$32&   LHS 1609  \\                                      
J035227.56-315105.2&  13.508$\pm$0.023&  13.254$\pm$0.026&    189$\pm$41&  -506$\pm$42&   {\it new cpm companion}\tablenotemark{c} (sep 39$\farcs$8) \\ 
\\                                                                                     
J043942.59+095215.6&   6.464$\pm$0.086&   6.406$\pm$0.023&    -24$\pm$37&  -456$\pm$35&   G 83-28   \\                                                  
J043943.24+095142.9&   9.238$\pm$0.023&   9.077$\pm$0.020&     14$\pm$39&  -456$\pm$37&   G 83-29 (sep 34$\farcs$1)  \\
\\                                                                                     
J044349.59-481934.0&   5.706$\pm$0.130&   5.558$\pm$0.046&     63$\pm$37&   347$\pm$36&   LTT 2077  \\                                                  
J044350.89-482003.8&   9.517$\pm$0.023&   9.363$\pm$0.020&     63$\pm$31&   297$\pm$30&   {\it new cpm companion}\tablenotemark{d} (sep 32$\farcs$5)  \\
\\                                                                                  
J051729.18$-$345346.0& $>$1.155       &   $>$1.308       &    712$\pm$42&   627$\pm$39&   $o$ Columbae, CD-35 2214 \\
J051723.88$-$345121.8& 9.724$\pm$0.023&   9.538$\pm$0.019&    253$\pm$34&  -292$\pm$33&   {\it new cpm companion}\tablenotemark{e} (sep 158$\farcs$2)  \\
\\
J065438.74+131038.5&   3.694$\pm$0.394&   3.369$\pm$0.283&    2033$\pm$57& -1035$\pm$64&  38 Geminorum AB, HR 2564 \\
J065448.82+131002.5&   9.840$\pm$0.023&   9.664$\pm$0.020&     250$\pm$38&   -21$\pm$36&  {\it new cpm companion}\tablenotemark{f} (sep 151$\farcs$6)  \\                
\\
J074020.22-172451.5&   9.064$\pm$0.022&   8.831$\pm$0.020&   1563$\pm$34&  -628$\pm$33&   LHS 235 \\                                       
J074021.62-172454.8&  12.440$\pm$0.023&  12.432$\pm$0.025&   1928$\pm$53&  -726$\pm$54&   LHS 234\tablenotemark{g} (sep 20$\farcs$3) \\
\\                                                                                     
J075812.35+875734.7&   8.850$\pm$0.022&   8.681$\pm$0.020&   -222$\pm$28&  -536$\pm$28&   LHS 1962   \\                                                 
J075909.31+875742.9&  10.207$\pm$0.023&  10.001$\pm$0.020&   -238$\pm$30&  -556$\pm$30&   LHS 1965 (sep 31$\farcs$5)  \\
\\ 
J082854.03$-$243541.8& 9.308$\pm$0.022&   9.185$\pm$0.021&    200$\pm$36&  -335$\pm$36&   WT 1549 \\                                                                                    
J082852.10$-$243536.5& 11.881$\pm$0.023& 11.665$\pm$0.022&    297$\pm$44&  -186$\pm$46&   {\it new cpm companion}\tablenotemark{h} (sep 26$\farcs$9)  \\          
\\
J103734.02+295955.5&  10.092$\pm$0.023&   9.939$\pm$0.021&    284$\pm$38&   -84$\pm$38&   LP 316-395, LSPM J1037+2959N  \\                              
J103734.43+295937.9&  11.153$\pm$0.022&  10.967$\pm$0.022&    327$\pm$42&   -98$\pm$42&   LP 316-396, LSPM J1037+2959S (sep 18$\farcs$4) \\
\\ 
J105607.88$-$575041.1& 10.309$\pm$0.022& 10.164$\pm$0.020&   -182$\pm$25&   125$\pm$25&   UPM J1056-5750 \\
J105536.09$-$575042.1& 10.366$\pm$0.023& 10.227$\pm$0.020&   -160$\pm$24&   152$\pm$25&   {\it new cpm companion}\tablenotemark{i} (sep 254$\farcs$8)  \\             
\\                                                                                    
J110604.58+425246.3&  10.323$\pm$0.023&  10.151$\pm$0.020&   -156$\pm$35&  -384$\pm$36&   G 176-13  \\                                             
J110605.07+425303.3&  11.361$\pm$0.023&  11.111$\pm$0.021&    -84$\pm$39&  -325$\pm$39&   G 176-14 (sep 17$\farcs$9) \\
\\ 
J113333.55$-$413954.4& 8.675$\pm$0.023&   8.712$\pm$0.020&   -140$\pm$25&  -134$\pm$25&   CD-40 6796 \\
J113333.67$-$414016.7& 10.589$\pm$0.024& 10.484$\pm$0.020&   -186$\pm$26&  -118$\pm$26&   {\it new cpm companion}\tablenotemark{j} (sep 22$\farcs$3)  \\       
\\
J120907.21+473602.2&   9.166$\pm$0.023&   8.975$\pm$0.021&    713$\pm$35&  -349$\pm$35&   LHS 2516  \\                                                    
J120907.24+473531.8&  10.036$\pm$0.024&   9.856$\pm$0.020&    804$\pm$35&  -353$\pm$35&   LHS 2517 (sep 30$\farcs$4) \\
\\ 
J121058.01$-$461917.3& 6.042$\pm$0.097&   5.997$\pm$0.040&     -4$\pm$26&   169$\pm$25&   CD-45 7595, LTT 4560 \\                                                                                 
J121058.26$-$461204.5& 8.822$\pm$0.022&   8.711$\pm$0.020&    103$\pm$25&   129$\pm$25&   {\it new cpm companion}\tablenotemark{k} (sep 432$\farcs$8)  \\                  
\\
J124007.18+204828.9&   6.801$\pm$0.066&   6.837$\pm$0.020&    250$\pm$35&  -449$\pm$35&   G 59-32, BD+21 2442 \\	
J124014.80+204752.7&  13.020$\pm$0.024&  12.803$\pm$0.026&    426$\pm$63&  -381$\pm$67&   {\it new cpm companion}\tablenotemark{l} (sep 112$\farcs$7)  \\               
\\
J124725.86$-$434353.2&  7.850$\pm$0.023&  7.893$\pm$0.019&   -165$\pm$24&  -162$\pm$24&   LTT 4892, CD-43 7881 \\
J124726.75$-$434441.8& 12.192$\pm$0.023& 11.997$\pm$0.021&   -220$\pm$30&  -138$\pm$30&   {\it new cpm companion}\tablenotemark{m} (sep 49$\farcs$5)  \\             
\\
J124737.85-274637.3&   9.715$\pm$0.022&   9.568$\pm$0.020&   -220$\pm$27&   -15$\pm$26&   LP 853-37 \\                                                               
J124739.94-274642.7&   9.849$\pm$0.022&   9.700$\pm$0.020&   -154$\pm$28&   -60$\pm$27&   LP 853-38 (sep 28$\farcs$4) \\
\\                                                                                     
J125433.93-381122.7&   7.951$\pm$0.024&   8.016$\pm$0.020&    208$\pm$26&   -68$\pm$26&   TYC 7772-1225-1  \\                      
J125435.54-381111.7&   8.175$\pm$0.022&   8.209$\pm$0.019&    161$\pm$26&   -78$\pm$26&   {\it possible new cpm companion}\tablenotemark{n}  (sep 22$\farcs$0) \\
\\                                                                                     
J131407.63+061814.3&  11.804$\pm$0.022&  11.790$\pm$0.022&   -206$\pm$33&  -294$\pm$33&   G 62-19, LSPM J1314+0618S \\                                    
J131408.36+061828.3&  12.291$\pm$0.022&  12.231$\pm$0.022&   -152$\pm$36&  -240$\pm$36&   G 62-20, LSPM J1314+0618N (sep 17$\farcs$8) \\
\\                                                                                     
J134454.85-453518.5&   7.547$\pm$0.025&   7.570$\pm$0.019&   -304$\pm$33&    67$\pm$32&   LTT 5330   \\           
J134457.17-453537.2&   9.343$\pm$0.022&   9.160$\pm$0.019&   -403$\pm$35&    68$\pm$34&   {\it new cpm companion}\tablenotemark{o} (sep 30$\farcs$7) \\
\\
J140400.30$-$592400.5& 9.363$\pm$0.021&   9.179$\pm$0.020&    -44$\pm$34&  -418$\pm$34&   L 197-165, NLTT 36089 \\
J140350.20$-$592348.0& 9.371$\pm$0.022&   9.194$\pm$0.020&    -76$\pm$35&  -396$\pm$34&   {\it new cpm companion}\tablenotemark{p} (sep 78$\farcs$1) \\   
\\
J140926.79$-$305551.4& 7.772$\pm$0.026&   7.819$\pm$0.021&   -496$\pm$29&  -271$\pm$29&   LHS 2807 \\
J140924.56$-$305534.1& 8.728$\pm$0.023&   8.643$\pm$0.020&   -492$\pm$26&  -285$\pm$25&   {\it new cpm companion}\tablenotemark{q} (sep 33$\farcs$4) \\
\\                                                                                     
J141509.28+221537.3&   8.877$\pm$0.022&   8.891$\pm$0.020&   -212$\pm$26&   129$\pm$26&   NLTT 36714, LSPM J1415+2215S  \\                                 
J141509.80+221559.8&   8.966$\pm$0.022&   9.042$\pm$0.019&   -191$\pm$26&   110$\pm$26&   NLTT 36716, LSPM J1415+2215N (sep 23$\farcs$7) \\
\\ 
J142051.70$-$045805.5& 8.348$\pm$0.032&   8.393$\pm$0.029&    -76$\pm$25&   -75$\pm$26&   BD-04 3668 \\
J142053.60$-$050137.8& 11.003$\pm$0.023& 10.877$\pm$0.020&   -185$\pm$29&   -56$\pm$29&   {\it new cpm companion}\tablenotemark{r} (sep 214$\farcs$2) \\                
\\                                                                                    
J153519.64+174245.2&   7.762$\pm$0.022&   7.597$\pm$0.019&  -1455$\pm$38&    34$\pm$34&   Ross 513A, LHS 399  \\                                                  
J153519.45+174302.6&   9.316$\pm$0.023&   9.081$\pm$0.020&  -1478$\pm$36&   -82$\pm$33&   Ross 513B, LHS 400 (sep 17$\farcs$6)  \\
\\                                                                                     
J154850.51+175056.1&   8.855$\pm$0.023&   8.781$\pm$0.020&   -268$\pm$36&   165$\pm$36&   G 137-55, LSPM J1548+1750E   \\                           
J154848.52+175050.0&   9.673$\pm$0.022&   9.525$\pm$0.020&   -288$\pm$37&   157$\pm$38&   G 137-54, LSPM J1548+1750W (sep 29$\farcs$1)  \\
\\                                                                                     
J155306.56+344508.8&   6.966$\pm$0.034&   6.877$\pm$0.020&    158$\pm$32&  -470$\pm$29&   LHS 3129  \\                                                                
J155306.84+344442.4&   8.009$\pm$0.023&   7.884$\pm$0.020&    112$\pm$32&  -479$\pm$31&   LHS 3130 (sep 26$\farcs$6) \\
\\                                                                                     
J155831.76+572242.3&   7.597$\pm$0.030&   7.674$\pm$0.020&    133$\pm$24&  -241$\pm$25&   G 225-39  \\                                                 
J155831.67+572309.1&  10.000$\pm$0.023&   9.823$\pm$0.020&    143$\pm$24&  -209$\pm$24&   G 225-38 (sep 26$\farcs$8) \\
\\                                                                                     
J164836.25+550743.4&   7.273$\pm$0.030&   7.325$\pm$0.020&    162$\pm$25&  -269$\pm$25&   G 226-28  \\                                                  
J164835.74+550809.7&   8.379$\pm$0.023&   8.461$\pm$0.020&    124$\pm$26&  -258$\pm$25&   G 226-27 (sep 26$\farcs$6) \\
\\                                                                                     
J165524.66-081930.5&   6.588$\pm$0.061&   6.374$\pm$0.022&  -1207$\pm$42&  -764$\pm$40&   LHS 427  \\
J165534.68-082349.7&   8.619$\pm$0.023&   8.393$\pm$0.020&  -1388$\pm$37&  -781$\pm$35&   LHS 429, vB 8 (sep 298$\farcs$9) \\                               
\\                                                                                     
J170428.76+034342.7&   8.570$\pm$0.022&   8.586$\pm$0.020&   -408$\pm$37&  -196$\pm$36&   G 19-10  \\
J170428.57+034422.7&   8.587$\pm$0.023&   8.648$\pm$0.020&   -346$\pm$37&  -170$\pm$37&   G 19-11 (sep 40$\farcs$1)  \\                                        
\\ 
J171828.99$-$224630.2& 9.217$\pm$0.024&   9.056$\pm$0.020&   -270$\pm$43&  -138$\pm$39&   {\it new cpm system}\tablenotemark{s} \\             
J171826.98$-$224543.5& 9.424$\pm$0.023&   9.313$\pm$0.019&   -327$\pm$42&   -52$\pm$40&   {\it new cpm system}\tablenotemark{s} (sep 54$\farcs$3)  \\      
\\
J172230.07$-$695119.2& 8.336$\pm$0.023&   8.243$\pm$0.020&   -236$\pm$40&  -221$\pm$36&   {\it possible new cpm system}\tablenotemark{t} \\      
J172237.14$-$695112.2& 11.426$\pm$0.023& 11.259$\pm$0.021&    -81$\pm$51&  -308$\pm$50&   {\it possible new cpm system}\tablenotemark{t}  (sep 54$\farcs$3)  \\    
\\                                                                                    
J182000.63-522138.6&   9.244$\pm$0.023&   9.170$\pm$0.020&    125$\pm$39&  -308$\pm$38&   L 272-87, NLTT 46256  \\                          
J182001.39-522200.7&   9.512$\pm$0.023&   9.399$\pm$0.020&    151$\pm$38&  -385$\pm$38&   L 272-88, NLTT 46257 (sep 23$\farcs$1) \\
\\                                                                                    
J182038.54+404836.0&  11.156$\pm$0.023&  11.047$\pm$0.020&   -202$\pm$39&  -264$\pm$42&   NLTT 46342, LSPM J1820+4048N \\                                              
J182038.57+404759.2&  11.774$\pm$0.023&  11.609$\pm$0.020&   -188$\pm$41&  -274$\pm$46&   NLTT 46343, LSPM J1820+4048S (sep 36$\farcs$9) \\
\\
J185252.01$-$570745.3& 7.473$\pm$0.032&   7.462$\pm$0.020&   -368$\pm$35&  -836$\pm$34&   LHS 3421 \\ 
J185257.45$-$570821.9& 10.804$\pm$0.024& 10.620$\pm$0.020&   -466$\pm$41&  -847$\pm$41&   2MASS J18525777$-$5708141 (sep 57$\farcs$4) \\ 
\\
J190250.67$-$755058.1& 10.928$\pm$0.023& 10.768$\pm$0.021&    -13$\pm$38&  -225$\pm$38&   SCR J1902-7550A \\
J190254.95$-$755110.8& 12.353$\pm$0.023& 12.135$\pm$0.023&     87$\pm$51&  -375$\pm$56&   SCR J1902-7550B (sep 22$\farcs$5) \\    
\\                                                                                   
J201605.71-115838.5&   8.778$\pm$0.022&   8.629$\pm$0.019&   -258$\pm$39&  -230$\pm$37&   {\it new cpm system}\tablenotemark{u}   \\         
J201605.96-115916.7&   8.897$\pm$0.022&   8.746$\pm$0.019&   -197$\pm$40&  -202$\pm$38&   {\it new cpm system}\tablenotemark{u} (sep 38$\farcs$3)  \\
\\                                                                                     
J205103.71-013357.8&   9.639$\pm$0.023&   9.611$\pm$0.019&   -285$\pm$41&  -168$\pm$40&   Wolf 885, NLTT 50000   \\                                                
J205103.50-013342.6&  10.248$\pm$0.023&  10.136$\pm$0.019&   -310$\pm$43&  -179$\pm$41&   Wolf 884, NLTT 49999 (sep 15$\farcs$6)  \\
\\
J210630.56$-$382606.1& 12.874$\pm$0.023& 12.599$\pm$0.026&    156$\pm$67&  -530$\pm$69&   {\it new cpm system}\tablenotemark{v}   \\ 
J210632.73$-$382640.4& 12.992$\pm$0.023& 12.717$\pm$0.025&     40$\pm$70&  -577$\pm$71&   {\it new cpm system}\tablenotemark{v} (sep 42$\farcs$7)  \\  
\\
J214737.18$-$135422.0&  9.809$\pm$0.023&  9.798$\pm$0.020&    156$\pm$41&  -260$\pm$42&   Ross 208 \\
J214736.41$-$135335.4& 11.575$\pm$0.023& 11.464$\pm$0.021&    164$\pm$51&  -468$\pm$50&   {\it new cpm companion}\tablenotemark{w} (sep 48$\farcs$0) \\  
\\
J230213.97+801412.0& 	6.683$\pm$0.071&  6.704$\pm$0.022&    -73$\pm$27&   168$\pm$28&   BD+79 762 \\
J230226.16+801241.8&    9.417$\pm$0.023&  9.278$\pm$0.020&   -108$\pm$29&   290$\pm$30&   {\it new cpm companion}\tablenotemark{x} (sep 95$\farcs$3) \\       
\\
J231049.97+453041.3&   5.245$\pm$0.174&    5.107$\pm$0.072&  -864$\pm$54&   331$\pm$55&   HD 218868, LTT 16813, WDS 23108+4531A \\
J231054.77+453043.5&   9.690$\pm$0.023&    9.515$\pm$0.021&  -200$\pm$36&  -396$\pm$35&   WDS 23108+4531C (sep 50$\farcs$5)  \\   
%\enddata
\end{longtable*}
\end{center}

\noindent $^a$ WISEA J021159.43+042150.7: Using the 2MASS-to-WISE time baseline we obtain a motion, ($\mu_\alpha$, 
$\mu_\delta$), of ($-140.5{\pm}10.2$, $-65.8{\pm}9.3$) mas yr$^{-1}$ for this object. The other component, BD+03 301, has 
a published motion of ($-144.70{\pm}1.64$, $-78.66{\pm}1.18$) mas yr$^{-1}$ (\citealt{vanleeuwen2007}). These motions are 
identical within their $2\sigma$ errors, so this is likely a common-proper-motion system.

\noindent $^b$ WISEA J031606.54$-$374215.4: Using the 2MASS-to-WISE time baseline we obtain a motion, ($\mu_\alpha$, 
$\mu_\delta$), of ($48.0{\pm}11.3$, $36.3{\pm}10.4$) mas yr$^{-1}$ for this object. The other component, TYC 7561-77-1, has 
a published motion of ($57.4{\pm}2.2$, $30.6{\pm}2.1$) mas yr$^{-1}$ (\citealt{hog2000}). Although small, these motions 
are identical within their 1$\sigma$ errors, so this is likely a common-proper-motion system.

\noindent $^c$ WISEA J035227.56$-$315105.2: Using the 2MASS-to-WISE time baseline we obtain a motion, ($\mu_\alpha$, 
$\mu_\delta$), of ($196.6{\pm}11.0$, $-531.9{\pm}10.4$) mas yr$^{-1}$ for this object. The other component, LHS 1609, 
has a published motion of ($219$, $-533$) mas yr$^{-1}$ (\citealt{luyten1979lhs}). These motions are identical to 
within 2$\sigma$, so this is likely a common-proper-motion binary.

\noindent $^d$ WISEA J044350.89$-$482003.8: Using the 2MASS-to-WISE time baseline we obtain a motion, ($\mu_\alpha$, 
$\mu_\delta$), of ($-5.5{\pm}9.0$, $263.4{\pm}8.8$) mas yr$^{-1}$ for this object. The other component, LTT 2077, has 
a published motion of ($4.75{\pm}0.37$, $276.85{\pm}0.43$) mas yr$^{-1}$ (\citealt{vanleeuwen2007}). These motions are 
identical to within 2$\sigma$, so this is likely a common-proper-motion binary.

\noindent $^e$ WISEA J051723.88$-$345121.8: Using the 2MASS-to-WISE time baseline we obtain a motion, ($\mu_\alpha$, 
$\mu_\delta$), of ($86.7{\pm}8.2$, $-340.4{\pm}8.0$) mas yr$^{-1}$ for this object. The other component, $o$ Columbae, 
has a published motion of ($92.67{\pm}0.14$, $-336.23{\pm}0.22$) mas yr$^{-1}$ (\citealt{vanleeuwen2007}). These motions 
are identical within their 1$\sigma$ errors, so this is very likely a common-proper-motion system.

\noindent $^f$ WISEA J065448.82+131002.5: Using the 2MASS-to-WISE time baseline we obtain a motion, ($\mu_\alpha$, 
$\mu_\delta$), of ($63.7{\pm}9.2$, $-86.2{\pm}8.8$) mas yr$^{-1}$ for this object. The other component, 38 Geminorum AB, 
has a published motion of ($68.48{\pm}1.78$, $-72.98{\pm}1.46$) mas yr$^{-1}$ (\citealt{vanleeuwen2007}). These motions 
are identical within their 2$\sigma$ errors, so this is likely a common-proper-motion system.

\noindent $^g$ WISEA J074021.62$-$172454.8: This object falls in a tile overlap region, so it was processed twice by 
AllWISE. The apparition chosen for inclusion in the AllWISE Source Catalog, whose values are listed in the table above, 
was selected over the other apparition because the former is further from the tile edge. The other apparition, which 
appears in the AllWISE Reject Table as WISEAR J074021.62$-$172454.8, has motion values of $1563\pm50$ mas yr$^{-1}$ 
and $-608\pm52$ mas yr$^{-1}$ in RA and Dec, respectively. The values for the AllWISE Reject source are $<1\sigma$ 
discrepant from either apparition of the primary, LHS 235. Because the same underlying frames are used for both tiles, 
it is unclear why the motion for one of the apparitions of LHS 234 is so wildly discrepant in RA.

\noindent $^h$ WISEA J082852.10$-$243536.5: Using the 2MASS-to-WISE time baseline we obtain a motion, ($\mu_\alpha$, 
$\mu_\delta$), of ($157.6{\pm}10.8$, $-137.5{\pm}10.7$) mas yr$^{-1}$ for this object. The other component, WT 1549, 
has a published motion of ($151{\pm}13$, $-140{\pm}13$) mas yr$^{-1}$ (\citealt{wroblewski1996}). These motions are 
identical within their 1$\sigma$ errors, so this is very likely a common-proper-motion system.

\noindent $^i$ WISEA J105536.09$-$575042.1: Using the 2MASS-to-WISE time baseline we obtain a motion, ($\mu_\alpha$, 
$\mu_\delta$), of ($-196.5{\pm}9.3$, $128.2{\pm}9.3$) mas yr$^{-1}$ for this object. The other component, UPM J1056-5750, 
has a published motion of ($-194.4{\pm}7.4$, $121.9{\pm}6.9$) mas yr$^{-1}$ (\citealt{finch2010}). These motions are 
identical within their 1$\sigma$ errors, so this is very likely a common-proper-motion system.

\noindent $^j$ WISEA J113333.67$-$414016.7: Using the 2MASS-to-WISE time baseline we obtain a motion, ($\mu_\alpha$, 
$\mu_\delta$), of ($-152.9{\pm}9.2$, $-120.1{\pm}8.5$) mas yr$^{-1}$ for this object. The other component, CD-40 6796, 
has a published motion of ($-169.7{\pm}1.9$, $-117.3{\pm}1.7$) mas yr$^{-1}$ (\citealt{hog2000}). These motions are 
identical within their 2$\sigma$ errors, so this is likely a common-proper-motion system.

\noindent $^k$ WISEA J121058.26$-$461204.5: Using the 2MASS-to-WISE time baseline we obtain a motion, ($\mu_\alpha$, 
$\mu_\delta$), of ($101.3{\pm}8.3$, $167.7{\pm}8.1$) mas yr$^{-1}$ for this object. The other component, CD-45 7595, 
has a published motion of ($89.11{\pm}0.40$, $168.15{\pm}0.47$) mas yr$^{-1}$ (\citealt{vanleeuwen2007}). These 
motions are identical within their 2$\sigma$ errors, so this is likely a common-proper-motion system.

\noindent $^l$ WISEA J124014.80+204752.7: Using the 2MASS-to-WISE time baseline we obtain a motion, ($\mu_\alpha$, 
$\mu_\delta$), of ($205.6{\pm}10.6$, $-369.1{\pm}10.8$) mas yr$^{-1}$ for this object. The other component, BD+21 2442, 
has a published motion of ($202.04{\pm}1.44$, $-368.45{\pm}0.99$) mas yr$^{-1}$ (\citealt{vanleeuwen2007}). These 
motions are identical within their 1$\sigma$ errors, so this is very likely a common-proper-motion system.

\noindent $^m$ WISEA J124726.75$-$434441.8: Using the 2MASS-to-WISE time baseline we obtain a motion, ($\mu_\alpha$, 
$\mu_\delta$), of ($-182.0{\pm}7.1$, $-95.4{\pm}7.1$) mas yr$^{-1}$ for this object. The other component, CD-43 7881, 
has a published motion of ($-196.60{\pm}0.80$, $-94.31{\pm}0.89$) mas yr$^{-1}$ (\citealt{vanleeuwen2007}). These
 motions are identical within their 2$\sigma$ errors, so this is likely a common-proper-motion system.

\noindent $^n$ WISEA J125435.54$-$381111.7: Using the 2MASS-to-WISE time baseline we obtain a motion, ($\mu_\alpha$, 
$\mu_\delta$), of ($161.1{\pm}7.9$, $-25.4{\pm}7.8$) mas yr$^{-1}$ for this object. The other component, TYC 7772-1225-1, 
has a published motion of ($149.6{\pm}2.8$, $-40.1{\pm}2.7$) mas yr$^{-1}$ (\citealt{hog2000}). These motions are 
identical to within 2$\sigma$, so this could be a common-proper-motion binary. In the UCAC4 Catalog this pair has 
listed motions of ($147.6{\pm}0.7$, $-38.2{\pm}1.0$) mas yr$^{-1}$ for the brighter component and ($144.5{\pm}0.8$, 
$-29.0{\pm}0.8$) mas yr$^{-1}$ for the fainter component. Although the 10$\sigma$ discrepancy in Dec would seem to 
rule these out as a cpm pair, UCAC4 measurements can be unreliable and thus we still list this pair as a possible 
physical system.

\noindent $^o$ WISEA J134457.17$-$453537.2: Using the 2MASS-to-WISE time baseline we obtain a motion, ($\mu_\alpha$, 
$\mu_\delta$), of ($-237.8{\pm}8.0$, $35.7{\pm}8.0$) mas yr$^{-1}$ for this object. The other component, LTT 5330, has 
a published motion of ($-240.51{\pm}13.1$, $25.99{\pm}13.1$) mas yr$^{-1}$ (\citealt{roser2008}). These motions are 
identical within their 1$\sigma$ errors, so this is very likely a common-proper-motion system.

\noindent $^p$ WISEA J140350.20$-$592348.0: Using the 2MASS-to-WISE time baseline we obtain a motion, ($\mu_\alpha$, 
$\mu_\delta$), of ($30.5{\pm}8.7$, $-494.3{\pm}8.5$) mas yr$^{-1}$ for this object. The other component, L 197-165, has 
a published motion of ($33$, $-469$) mas yr$^{-1}$ (\citealt{luyten1979nltt}). The motion of the unpublished WISE object 
is identical within its 1$\sigma$ errors to the RA motion of the Luyten object but discrepant by $\sim$3$\sigma$ 
from the Luyten Dec measurement. Given the large motion in Dec and the proximity of the sources, this is still likely 
a common-proper-motion system.

\noindent $^q$ WISEA J140924.56$-$305534.1: Using the 2MASS-to-WISE time baseline we obtain a motion, ($\mu_\alpha$, 
$\mu_\delta$), of ($-456.4{\pm}8.1$, $-213.8{\pm}8.0$) mas yr$^{-1}$ for this object. The other component, LHS 2807, 
has a published motion of ($-460.60{\pm}3.68$, $-219.82{\pm}3.02$) mas yr$^{-1}$ (\citealt{vanleeuwen2007}). These 
motions are identical to $<$1$\sigma$, so this is likely a common-proper-motion system.

\noindent $^r$ WISEA J142053.60$-$050137.8: Using the 2MASS-to-WISE time baseline we obtain a motion, ($\mu_\alpha$, 
$\mu_\delta$), of ($-131.2{\pm}10.2$, $-55.4{\pm}10.0$) mas yr$^{-1}$ for this object. The other component, BD-04 3668, 
has a published motion of ($-137.5{\pm}3.5$, $-58.0{\pm}3.2$) mas yr$^{-1}$ (\citealt{roeser1988}). These motions 
are identical to $<$1$\sigma$, so this is likely a common-proper-motion system.

\noindent $^s$ WISEA J171828.99$-$224630.2 and WISEA J171826.98$-$224543.5: Using the 2MASS-to-WISE time baseline 
we obtain a motion, ($\mu_\alpha$, $\mu_\delta$), of ($-149.3{\pm}8.9$, $-149.9{\pm}8.4$) mas yr$^{-1}$ for the 
brighter object and ($-148.7{\pm}8.8$, $-148.8{\pm}8.3$) mas yr$^{-1}$ for the fainter one. These motions are 
identical to $<$1$\sigma$, so this is very likely a common-proper-motion pair.

\noindent $^t$ WISEA J172230.07$-$695119.2 and WISEA J172237.14$-$695112.2: Using the 2MASS-to-WISE time baseline 
we obtain a motion, ($\mu_\alpha$, $\mu_\delta$), of ($-145.6{\pm}10.4$, $-217.6{\pm}8.7$) mas yr$^{-1}$ for the 
brighter object and ($-37.6{\pm}11.5$, $-221.9{\pm}9.8$) mas yr$^{-1}$ for the fainter one. Although the (larger) 
Dec motions are identical to $<1\sigma$, the RA motions are discrepant by $>9\sigma$. A visual check of the system 
more strongly supports common proper motion than do the actual measurements. We suspect that the WISE astrometry 
for the fainter source in the pair may be corrupted by a background object -- seen at nearly equal magnitude in 
the earlier DSS2 images -- that lies near its position in the WISE data.

\noindent $^u$ WISEA J201605.71$-$115838.5 and WISEA J201605.96$-$115916.7: Both of these objects are listed in 
the UCAC4 Catalog, but no other literature is available. Using the 2MASS-to-WISE time baseline we obtain motions, 
($\mu_\alpha$, $\mu_\delta$), of ($-72.4{\pm}8.9$, $-141.1{\pm}7.5$) mas yr$^{-1}$ for the brighter component and
 ($-69.4{\pm}9.0$, $-134.4{\pm}7.7$) mas yr$^{-1}$ for the dimmer component, which are identical within their 
1$\sigma$ errors. In the UCAC4 Catalog this pair has listed motions of ($-70.9{\pm}2.1$, $-135.6{\pm}2.2$) mas 
yr$^{-1}$ for the brighter component and ($-57.3{\pm}2.1$, $-130.5{\pm}2.1$) mas yr$^{-1}$ for the fainter component. 
Although the 6.5$\sigma$ discrepancy in RA would seem to rule these out as a cpm pair, the secondary appears to be 
blended with another source of nearly equal magnitude in the DSS2 images. Because visual checks strongly suggest 
this is a co-moving system, we still list this as a possible physical pair.

\noindent $^v$ WISEA J210630.56$-$382606.1 and WISEA J210632.73$-$382640.4: Using the 2MASS-to-WISE time 
baseline we obtain a motion, ($\mu_\alpha$, $\mu_\delta$), of ($174.9{\pm}10.1$, $-323.4{\pm}10.1$) mas yr$^{-1}$ 
for the brighter object and ($187.1{\pm}10.3$, $-323.3{\pm}10.2$) mas yr$^{-1}$ for the fainter one. The motions 
are identical to $<2\sigma$, so these are likely a common-proper-motion pair.

\noindent $^w$ WISEA J214736.41$-$135335.4: Using the 2MASS-to-WISE time baseline we obtain a motion, 
($\mu_\alpha$, $\mu_\delta$), of ($151.6{\pm}10.0$, $-300.2{\pm}9.3$) mas yr$^{-1}$ for this object. The other 
component, Ross 208, has a published motion of ($152{\pm}5$, $-296{\pm}5$) mas yr$^{-1}$ (\citealt{salim2003}). 
These motions are identical within $<1\sigma$, so this is very likely a common-proper-motion system.

\noindent $^x$ WISEA J230226.16+801241.8: Using the 2MASS-to-WISE time baseline we obtain a motion, 
($\mu_\alpha$, $\mu_\delta$), of ($-68.1{\pm}9.9$, $114.1{\pm}9.9$) mas yr$^{-1}$ for this object. The other 
component, BD+79 762, has a published motion of ($-53.2{\pm}1.2$, $127.2{\pm}1.2$) mas yr$^{-1}$ (\citealt{hog2000}). 
These motions are identical within their $2\sigma$ errors, so this is likely a common-proper-motion system.

\end{document}